\documentclass[a4paper, 11pt]{article}
\pdfoutput=1 
\usepackage{jheppub} 
\usepackage[T1]{fontenc} 
\usepackage{caption}
\usepackage{subcaption}
\usepackage{hyperref}
\usepackage{revsymb}
\usepackage{indentfirst}
\usepackage{mwe}
\usepackage{simpler-wick}
\usepackage{amsmath}
\usepackage{amsthm}
\usepackage{tikz}
\usepackage{tikz}
\usetikzlibrary{decorations.pathmorphing, decorations.markings}

\tikzset{
    photon/.style={decorate, decoration={snake}, draw=black},
    electron/.style={draw=black, postaction={decorate},
        decoration={markings,mark=at position .55 with {\arrow[black]{>}}}},
    vertex/.style={draw,fill=black,circle,minimum size=1.5pt,inner sep=0pt},
    cross/.style={path picture={
        \draw[black]
        (path picture bounding box.south east) -- (path picture bounding box.north west)
        (path picture bounding box.south west) -- (path picture bounding box.north east);
    }},
}

\newtheorem{thm}{Theorem}

\title{\boldmath Associativity is enough: an all-orders 2d chiral algebra for 4d form factors}
 \author{V{\'i}ctor E. Fern{\'a}ndez, }
 \author{Natalie M. Paquette$^*$}
 \affiliation{Department of Physics, University of Washington, Seattle, WA, USA}
 \affiliation{{$^*$} Corresponding author}
\emailAdd{vfer@uw.edu}
\emailAdd{npaquett@uw.edu}
\abstract{There is a special set of massless four-dimensional gauge theories which admit local and gauge-anomaly-free uplifts to twistor space; we call such theories twistorial. In twistorial theories, generalized towers of soft modes (including states of both helicities) form a 2d chiral algebra \textit{even at the quantum level}. The 2d OPE limit of this chiral algebra coincides with the holomorphic collinear limit in 4d. This is true, in particular, for self-dual Yang-Mills (SDYM) theory coupled to special choices of matter, the latter being required to make the theory twistorial. Costello and the second author recently proposed that form factors of such twistorial 4d theories could be computed as 2d chiral algebra correlators. In turn, there exist form factors of self-dual theories, with insertions of appropriate local operators, that compute a subclass of observables in full (i.e. non-self-dual) QCD, coupled to appropriate matter. For example, the n-point two-loop all-plus QCD amplitude has recently been computed using the 1-loop chiral algebra OPEs determined thus far, but higher orders of the quantum-deformed chiral algebra must be determined to continue the ``chiral algebra bootstrap'' program for higher-loop-level form factors of these twistorial theories. \\ 

In this paper, using only elementary constraints from symmetries and associativity, we obtain closed-form expressions for the extended chiral algebra to arbitrary loop-order. This can be viewed as providing an all-loop result for a subset of collinear splitting functions in non-supersymmetric, massless QCD coupled to special choices of matter.}
\begin{document} 
\maketitle
\flushbottom

\section{Introduction \& Conclusions}

In this paper, we will fix the (singular part of the) OPEs of several chiral algebras, which are of interest in the study of scattering amplitudes and flat space holography, to all orders in $\hbar$. The methods to fix the chiral algebra are elementary, although their implementation is rather involved. For the rest of this introduction, therefore, we will motivate this undertaking by explaining its application, to be presented in follow-up work, to the computation of loop-level form factors in non-supersymmetric quantum field theories. \\

\subsection{Background}

Computing loop-level scattering amplitudes of non-supersymmetric quantum field theories remains a challenge. While there has been astonishing progress in the case of  (planar) 4d $\mathcal{N}=4$ super-Yang Mills (SYM) (e.g. \cite{Arkani-Hamed:2022rwr} and references therein), going beyond two-loop computations in ordinary, non-supersymmetric Yang-Mills theory has proven difficult---though see e.g. \cite{Bern:2000dn, Badger:2013gxa, Dunbar:2016cxp,Dunbar:2016gjb, Dunbar:2017nfy, Kosower:2022bfv, Kosower:2022iju} for impressive computations of two-loop all-plus helicity amplitudes\footnote{Our references are far from exhaustive, and there have been some selected results at 3-loops and higher. For example, see \cite{Caola:2022dfa} for a recent 3-loop result for 2-to-2 scattering in QCD; our interest will rather be on obtaining analytic $n$-point results for special external helicity configurations, and with special matter content, particularly at 3-loops and higher.}. \\

In \cite{CP}, the authors proposed a novel perturbative approach, hereafter called the \textit{chiral algebra bootstrap}, to computing form factors for a special (but not necessarily supersymmetric!) class of 4d theories. This approach in principle transmutes difficult higher-loop computations into elementary---albeit potentially tedious---algebraic manipulations. Let us briefly recapitulate how and why this works. At the end of this section, we will sketch several applications of the chiral algebra bootstrap to higher-loop QCD (coupled to very particular matter content) observables, which serve as primary motivation for this note. \\

The chiral algebra bootstrap, in brief, equates 4d form factors for so-called \textit{twistorial} theories with correlation functions of a 2d chiral algebra \cite{CP}. This allows us to trade the computation of higher-loop quantities in this restricted class of non-supersymmetric gauge theories, for algebraic manipulations in a holomorphic 2d algebra. Twistorial theories are quantum field theories that lift to local, holomorphic quantum field theories on twistor space, which has six real dimensions. The existence of such a lift has long been understood to be an avatar of integrability of the 4d theory. A necessary condition for a 4d theory to lift to a holomorphic theory on twistor space is that the corresponding 6d theory is itself well-defined and free from gauge anomalies; a gauge anomaly on twistor space can therefore be viewed as an obstruction to quantum integrability of a classically integrable 4d field theory. \\

Recently, perturbative gauge anomalies for holomorphic theories on twistor space have been studied \cite{CosAnomaly} and interesting families of anomaly-free theories have emerged. These take the form of self-dual gauge theories \cite{CosAnomaly} and gravity \cite{BSS}, coupled to additional \textit{massless} matter. The anomaly-cancellation condition for gauge theory, which constrains the possible matter content, is given by the equation\footnote{See \cite{okubo2} for the general mathematical identities underpinning these formulas.}
\begin{equation}\label{eq:anomaly}
    \textrm{Tr}_{\textrm{adj}}(X^4) - \textrm{Tr}_R(X^4) = \lambda_{\mathfrak{g}, R}^2 \textrm{Tr}_{\textrm{fun}}(X^2)^2,
\end{equation} for some proportionality constant $\lambda_{\mathfrak{g}, R}^2$ depending on the Lie algebra and matter representation. Here, $X \in \mathfrak{g}$ and the traces are respectively, in the adjoint, matter representation $R$, and fundamental representation. Throughout the text, $R$ is taken to be a real representation (to also ensure cancellation of the twistorial uplift of the ABJ anomaly). Scalar matter can also be incorporated, as in certain supersymmetric examples, with a similar trace contribution on the left hand side with an overall positive-sign. \\

For certain choices of Lie algebras (delineated below), the anomaly can be cancelled without ordinary matter, so that the second term in \eqref{eq:anomaly} vanishes, and the anomaly-cancellation condition takes the form of a Green-Schwarz mechanism with proportionality constant $\lambda_{\mathfrak{g}}^2 = {10 ({h^{\vee}})^2 \over \textrm{dim} \mathfrak{g} + 2}$ \cite{okubo}, wherein tree exchange of a scalar field cancels the one-loop box diagram in 6d \footnote{In fact, this condition can be understood as the holomorphic twist of the anomaly cancellation condition for 6d $\mathcal{N}=(1, 0)$ theories with a tensor multiplet.}. On the other hand, there are certain choices of $\mathfrak{g}, R$ for which $\lambda_{\mathfrak{g}, R}^2 = 0$ and the anomaly can be cancelled purely by ordinary massless matter in $R$. \\ 

We refer to complete descriptions of these 6d theories, and the derivation of their reductions to 4d, to \cite{CosAnomaly}. Instead, we content ourselves by listing some interesting anomaly-free twistorial theories (see \cite{CosAnomaly, CosQCD, Bittleston:2024efo} for more details and examples, and \cite{Bittleston:2024efo} for twisted holographic interpretations in of some of these systems): 
\begin{enumerate}
    \item The 4d WZW model of \cite{LMNS} with $G=SO(8)$ coupled to an additional scalar field with a fourth-order kinetic term. This example has a known uplift to a holomorphic twist of the Type I string, which was exploited in \cite{burnsShort, burns} to create a top-down holographic dual for an asymptotically flat spacetime. \\
    \item Self-dual Yang-Mills theory (SDYM) can be supersymmetrized, which cancels the anomaly. The case of $\mathcal{N}=4$ SDYM in particular has been well-studied by many other authors following \cite{WittenTwistor}. \\
    \item SDYM with gauge algebra $\mathfrak{g} = sl_2, sl_3, so(8), e_{6, 7, 8}$ coupled to a periodic scalar field (which we call henceforth ``the axion'', since it couples to an $F \wedge F$ term in 4d) with a fourth-order kinetic term. \\
    \item Self-dual general relativity (SDGR), also coupled to a  certain fourth-order scalar field \cite{BSS}. We call the scalar-coupled theory ``extended self-dual GR'', and its corresponding algebra the ``extended celestial chiral algebra'', following earlier work \cite{roland}. \\
    \item SDYM with gauge group $SU(N)$ coupled to the ``axion'' as well as additional $N$ fundamental Dirac fermions\footnote{Therefore, in terms of Weyl representations we have $N_f$ fundamentals and $N_f$ antifundamentals.} (i.e. with $N_f = N_c$). In this case, $\lambda_{\mathfrak{g}, R}^2= 6.$ \\
     \item SDYM with gauge group $SO(N)$ coupled to the ``axion'' and $N_f = N-8$ fundamental fermions. Here, $\lambda_{\mathfrak{g}, R}^2= 3.$ \\
    \item SDYM with $SU(N)$ gauge group, 8 fundamentals, and one two-index antisymmetric tensor field. This representation is often expressed more formally in terms of Weyl fermions as $8 F \oplus 8 F^{\vee} \oplus \wedge^2 F \oplus \wedge^2 F^{\vee}$ \cite{CosQCD}. Here $F$ is the fundamental and $F^{\vee}$ the anti-fundamental, and $\wedge^2$ is the exterior square (i.e. the projection onto the antisymmetric part of $F \otimes F$). 
    
    The simplest examples in this sequence, and perhaps the most interesting, are $SU(2)$ SDYM with $N_f = 8$ and $SU(3)$ SDYM with $N_f = 9$. 
    \end{enumerate}

We emphasize that in the last item, the anomaly is cancelled \textit{purely by ordinary matter}; the ``axion'' of the Green-Schwarz mechanism decouples completely for $\lambda_{\mathfrak{g}, R}^2 = 0$. \\

In this note, we will focus on the twistorial theories given by SDYM coupled to either the axion, ordinary matter, or both (in all cases, as subject to the anomaly cancellation condition), but our methods should readily generalize to all twistorial theories\footnote{The OPE expressions for SDGR are slightly more complicated than for the SDYM theories. We sketch results for the SDGR throughout the text separately. See also \cite{roland}.}. \\

With twistorial theories in hand, we can now state the main theorem of \cite{CP}. The theorem concerns form factors, which are matrix elements between a local operator acting on the vacuum, and asymptotic scattering states $\langle 0| \mathcal{O} | p_1 \ldots p_n \rangle$\footnote{We have the operator acting on the left vacuum to more immediately match with notational conventions from twistor space.}. In the self-dual gauge theories, in the convention where all asymptotic states are taken to be incoming, the helicities of all external states are positive. In short, we propose
\begin{equation}
\langle 0| \mathcal{O} | p_1 \ldots p_n \rangle = \langle \mathcal{O} | \mathcal{V}_1[\tilde{\lambda}_1](z_1) \ldots \mathcal{V}_n[\tilde{\lambda}_n](z_n) \rangle_{2d}
\end{equation} where the angle brackets on the right-hand side denote a 2d correlation function with a choice of conformal block denoted by $\langle \mathcal{O}|$ \footnote{All of our correlation functions are computed on a 2d sphere; our ``celestial'' chiral algebra, with operators of mostly-negative integer conformal dimensions, admits an infinite family of conformal blocks even on $\mathbb{P}^1$.} and vertex operator insertions $\mathcal{V}_n$ labeled by spinor helicity variables for on-shell massless momenta $p_{\alpha \dot{\alpha}} = \lambda_{\alpha}\tilde{\lambda}_{\dot{\alpha}}, \  p^2 = 0$. We fix the scaling ambiguity of the spinors, per usual, such that $\lambda_{\alpha} = (1, z)$ with $z$ the holomorphic coordinate at null infinity. The chiral algebra is therefore often called a celestial chiral algebra, because it is supported on the celestial sphere or, more precisely, on the twistor sphere associated to the origin of $\mathbb{R}^4$. Since the chiral algebra is a holomorphic object, the dependence on the two spinor helicity variables is asymmetric, with rotations of the $\tilde{\lambda}$ variables corresponding to a 2d $SU(2)$ flavor symmetry. This is immediate in the twistor space formulation, since the latter naturally geometrizes analytically continued momenta; the $SU(2)$ flavor symmetry rotates the fibres of twistor space. \\

We will compute form factors in these 4d theories, with the insertion of a single local operator at a fixed location (the origin in position space). Note that ordinary S-matrix elements in twistorial theories vanish, i.e. twistorial theories are in a sense integrable, but form factors do not. Since we leave the operator fixed at the origin and do not integrate over the operator's position (although one may choose to subsequently do this by hand, after multiplying by $e^{i \sum_j p_j \cdot x}$), this quantity was called the \textit{form factor integrand} in \cite{CP}.
\begin{thm}[Costello-Paquette \cite{CP}]
Form factors are equivalent to 2d correlation functions in the chiral algebra of conformally soft modes of the 4d theory (often called the celestial chiral algebra \cite{Pate:2019lpp, Guevara:2021abz, Strominger:2021mtt}). \footnote{An alternative, but classically equivalent, derivation of the chiral algebra from the 6d perspective is also given in \cite{CP}.} \\ \\
The dictionary between 4d and 2d quantities is:
\begin{itemize}
    \item \textbf{Local operators in 4d are in bijective correspondence with conformal blocks of the chiral algebra.} Note that the chiral algebra is non-unitary, and has an infinite family of conformal blocks even on the sphere. This result was established abstractly in \cite{CP}, and the authors computed several simple examples which were essentially fixed by symmetry. Additional constructive methods for deriving the correspondence using the 6d uplift of these local operators in combination with the methods of \cite{CP} can be found in \cite{BuCasali}, and further generalizations applicable to scattering in the presence of self-dual (magnetically charged) line defects are obtained in \cite{GP}. \\
    \item \textbf{States of the twistorial theory in a boost eigenbasis are in bijective correspondence with local operators in the 2d chiral algebra.} It is often helpful to work with the 4d Mellin-transformed states and the corresponding 2d local operators, i.e. conformally soft modes, since the latter have simple OPEs. On the other hand, for form factor computations, we will typically resum the soft modes into hard states $\mathcal{V}[\tilde{\lambda}](z)$ using appropriate powers of the energy and components of $\tilde{\lambda}_{\dot{\alpha}}$ (i.e. the anti-holomorphic coordinates of the chiral algebra plane in Euclidean signature), leading in 4d to the usual asymptotic scattering states in the momentum eigenbasis. \\
    \item \textbf{Holomorphic collinear limits in the 4d twistorial theory correspond to the OPE limit of the celestial chiral algebra}. The OPE singularities include the usual singularities arising from collinear-splitting functions of the self-dual gauge theory, which are one-loop exact and appear in the SDYM S-matrix. There are, however, additional subleading singularities which can appear in form factors. From the OPE point of view, these singularities multiply normal-ordered products of operators and comprise non-factorizing contributions to the holomorphic collinear limit of the twistorial theory. Without these contributions, the chiral algebra OPE based on 2-particle collinear splitting functions fails to be associative on its own.
\end{itemize}
\end{thm}
In \cite{CPassoc}, it was explained that twistorial theories have well-defined, and in particular associative, celestial chiral algebras even at the quantum level; the cancellation of the 6d gauge anomaly restores a naive failure of the Jacobi identity that appears at the one-loop level \footnote{See \cite{Ren:2022sws, Kapec:2022hih, Ball:2023qim} for other explorations and proposed solutions for failures of associativity at the quantum level in celestial chiral algebras more generally. In particular, see \cite{Ball:2022bgg, Ball:2023sdz} for proposed repairs of associativity involving contributions from multiple simultaneous collinear limits, which requires weakening the locality axiom of the algebra.}. Thus, the chiral algebra bootstrap is well-defined for loop-level form factors. \\

An important property of form factors computed in twistorial theories is that they are \textit{rational} functions with certain allowed poles. Poles can only appear when two 2d local operators approach each other $z_i - z_j \rightarrow 0$ (in holomorphic coordinates on the celestial sphere, i.e. the support of the chiral algebra). In 4d notation, this amounts to poles only in $\langle ij \rangle \propto \epsilon_{\alpha \beta}\lambda_i^{\alpha}\lambda_j^{\beta}$. This is a consequence of the fact that correlation functions in a twistorial theory (which, unlike S-matrix elements, do not vanish) are entire analytic functions with singularities only on the analytically continued lightcone, $||x_i - x_j||^2 = 0$, for $x_i \in \mathbb{C}^4$. In brief, this is because points in 4d spacetime lift to copies of $\mathbb{CP}^1$ on twistor space, which do not intersect each other unless the condition above is satisfied. In particular, this disallows theories with divergences of the form $\textrm{log}||x||$. For example, in \cite{CosAnomaly} it was demonstrated that the logarithmic divergence in the two-point function of $F_{-}^2$, the anti-self-dual part of the field strength, in SDYM is cancelled by a logarithmic divergence arising from exchange of the fourth-order anomaly-cancelling axion. \\

Let us now consider form factors for 4d SDYM gauge theory, which we will shortly couple to additional matter to cancel the anomaly. SDYM has the action \cite{Chalmers:1996rq}
\begin{equation}\label{eq:BF}
   \int \textrm{tr}\left(B \wedge F(A)_- \right) 
\end{equation} where $A$ is the gauge field, $F_-$ is the anti-self-dual part of the field strength, and $B$ is a $\mathfrak{g}$-valued anti-self-dual two-form field. It is well-known that if we add to this theory the term ${1 \over 2} g_{YM}^2 \int \textrm{tr}(B \wedge B)$ and integrate out $B$, we obtain a theory that is perturbatively equivalent to ordinary Yang-Mills theory\footnote{There is a nontrivial theta-angle, which we ignore for the purposes of this note. We will only do perturbative computations.}. This deformation immediately tells us that form factors of SDYM with a single insertion of the operator $\textrm{tr}(B^2)$ are equivalent to (a subset of) ordinary Yang-Mills amplitudes. However, the twistorial SDYM theory is coupled to the axion (and/or other anomaly-cancelling matter), so that such form factors yield amplitudes of Yang-Mills theory also coupled to such special matter. This however suggests one natural avenue of progress: focus on the form factors of the theories for which the axion decouples, to obtain quantities in QCD coupled to a fine-tuned number of fundamentals, but no fourth-order axion. Although this is a different amount of (massless) matter from the real-world, it is a way to obtain loop-level results in certain non-supersymmetric QFTs without any fourth-order exotica. By twistoriality, these quantities will be rational. \\
 
The key ingredient to running the chiral algebra bootstrap is knowledge of the chiral algebra OPEs to, ideally, arbitrary loop order. This is because products of chiral algebra operators in the corresponding correlation function can be reduced iteratively to more elementary form factors with the OPE (either at lower loop order, or with fewer 2d operator insertions/4d external states). This is independent of the choice of conformal block/4d local operator. For example, a negative-helicity gluon operator becoming coincident with a positive-helicity counterpart can, according to the OPE, be replaced with a normally-ordered product of some number of negative-helicity gluons (depending on the loop order), multiplied by a known function of $\langle ij \rangle ^{-1} $ (coming from the OPE pole) and $[ij]$ (coming from the repackaging of soft modes into hard eigenstates).  \\

In \cite{CPassoc, CosQCD}, the chiral algebra OPEs for certain terms up to one-loop order, and for various choices of anomaly-cancelling matter, were determined. The known pieces of those chiral algebras were sufficient to run the chiral algebra bootstrap for certain interesting tree-level \cite{CP}, one-loop \cite{CPassoc}, and two-loops \cite{CosQCD} form factors\footnote{More precisely, \cite{CosQCD} computed the single-trace terms of the n-point all-+ amplitude, while \cite{DM} used the chiral algebra bootstrap to obtain the double-trace terms as well. Moreover, \cite{DM} verified the agreement of the 4-point specialization of this quantity with a more standard computation in QCD using a mass regulator.}. In particular, the form factors in our twistorial SDYM theories with a $\textrm{tr}(B^2)$ insertion, arbitrary numbers of positive helicity gluons, and $2-l$ negative helicity gluons for $l=0, 1, 2$ loops are equivalent to the corresponding amplitudes in full Yang-Mills theory coupled to same matter content. The tree-level result for the axion-coupled theory, in particular, reproduces the famous Parke-Taylor formula. Happily, the soft modes that give nonvanishing contributions to these form factors were those whose OPEs were completely determined by the known one-loop chiral algebra. \\

(As a brief aside, note that in twistorial theories, the one-loop all $+$ amplitudes vanish; this is one way to account for the simplicity of these theories. Recently, \cite{DM} used this fact to reproduce and explain known relations among color-ordered QCD amplitudes. Nonetheless, a tree-level chiral algebra computation involving an insertion of the conformal block corresponding to $(\Delta \rho)^2$ with $\rho$ the axion, rather than $\textrm{tr}(B^2)$, can recover the SDYM or YM one-loop all $+$ amplitude \cite{CP}. At four points, this is immediately apparent since tree-level axion exchange is introduced precisely to cancel the one-loop gauge box-diagram, but the chiral algebra bootstrap can be used to compute the complete $n$-point result of \cite{Bern:1993qk, Mahlon:1993fe}). \\

\subsection{Main results}
In this paper, we obtain closed-form expressions for the chiral algebra OPEs associated to self-dual Yang-Mills (SDYM) coupled to twistorial matter. The expressions we obtain hold for any choice of matter satisfying equation \eqref{eq:anomalymatter}. 

First, we use 1. symmetries of the holomorphic theory on twistor space, and 2. the Koszul duality construction of the celestial chiral algebra \cite{CP} to fix the form of the OPEs up to a set of undetermined coefficients. We decompose these coefficients into representation theoretic factors, which depend on the choice of gauge group and matter, and numerical factors. These formulas are available in \S \ref{sec:SDYMOPE}. 

Next, we 3. show that we can impose associativity to solve for the undetermined numerical factors \footnote{In particular, we obtain closed form solutions for all but one coefficient, $c$. We explain in Appendix \ref{app:examples} how the associativity equations involving $c$ can be used to solve for $c$ at a given loop order, and do so explicitly at one and two loops; we believe it should be possible to obtain a closed-form solution to $c$ as well, without any additional input.}. First, we use the constraints of associativity to rewrite each numerical coefficient in terms of one particular coefficient we call $f$, which appears in the OPE of the strong generators of the chiral algebra (physically, $f$ determines the OPE between (modes of) a positive helicity gluon and a negative helicity gluon); these expressions are in \S \ref{sec:coeffs}. Then, we again use associativity to obtain a recursive formula that expresses $f$ at $m$-loops in terms of $f$ at 1-loop and at $(m-1)$-loops (Equation \eqref{eq:frecursion}). The closed-form solution to this equation has been obtained using properties of the holomorphic twistorial uplift \cite{Zeng}, and is presented in Equations \eqref{eq:fcoefficient}, \eqref{eq:ffromm}. 

Since the closed form expression for $f$ is a bit unwieldy, we find it easier, in practice, to use the recursion relation with the one-loop solution as input to generate $f$ to the desired loop order. An explicit, conveinent expression for the 1-loop $f$-coefficient is given in Equations \eqref{eq:main_m1}.

Note that $\mathcal{N}=1, 2$ SDYM are captured by the same formulas presented in the text, since they are also given by holomorphic BF theories for certain (super) Lie groups that are extended by new generators corresponding to fermionic matter (see \S \ref{sec:matter}). See e.g. \cite{Boels:2006ir} for more on these theories. In fact, the same trick also works for $\mathcal{N}=4$ SDYM with slight modification: one needs to also include additional bosonic generators corresponding to the adjoint-valued scalar superpartners. The local holomorphic lift of 4d $\mathcal{N}=4$ SDYM is well-known to be a holomorphic Chern-Simons theory on super-twistor space $\mathbb{CP}^{3|4}$ \cite{WittenTwistor}, but integrating out the superspace directions reproduces a holomorphic BF theory on ordinary twistor space for an extended super Lie algebra, with a Yukawa term furnishing an extension of the bilinear form. Although we only explicate theories with purely fermionic matter (and possibly the axion) in this note, we expect 4d $\mathcal{N}=4$ SDYM to be technically simpler; many of the traces multiplying numerical coefficients vanish for this specialization of matter, and only ladder-type diagrams can contribute, cf. \S \ref{sec:diagrams} \footnote{In all the supersymmetric examples, one can set the $E, F$ generators associated to the axion field to zero and $\lambda_{\mathfrak{g}, R}=0$.}. 

Along the way, we also complete steps 1. and 2. for the chiral algebra associated to self-dual gravity coupled to the anomaly-cancelling matter of \cite{BSS}; see \S \ref{sec:SDGR}. It is reasonable to conjecture that the same application of associativity in step 3. can be used to solve for the numerical coefficients of these OPEs, since the theorem of \cite{CP} guarantees that the chiral algebra for any local holomorphic theory on the twistor space associated to flat spacetime exists and is associative once the one-loop anomaly on twistor space is cancelled. We leave a formal proof, or full calculation for SDGR, to future work.

It would also be interesting to explore whether twistorial symmetries plus associativity can determine the OPEs for twistorial theories in more general backgrounds or in the presence of various deformations \cite{Bu:2022iak, Bittleston:2023bzp, Bogna:2024gnt, Bittleston:2024efo} (see also \cite{Bogna:2023bbd, Adamo:2025fqt, Garner:2024tis} for form factor computations in nontrivial self-dual backgrounds using twistor methods), up to a small number of parameters governing the background deformation\footnote{There is also recent work in the context of AdS/CFT that studies recursive constraints from associativity on a certain supersymmetric chiral algebra \cite{Gaberdiel:2025eaf}. They find that the algebra is determined up to a central charge. Note that twistorial theories on 4d flat backgrounds have central charge exactly zero by Koszul duality, but a central charge can be generated by, e.g. the presence of a logarithmic axion profile (see \S 12 of \cite{CP}, \cite{Bittleston:2024efo}).}.

\subsection{Future directions}

\begin{center}
\begin{figure}
    \centering
    \includegraphics[width=0.8\linewidth]{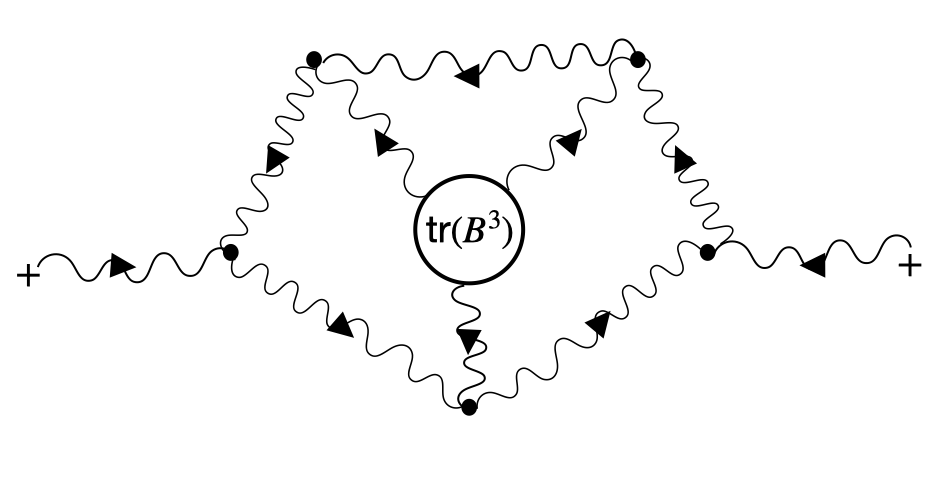}
    \caption{The 3-loop form factor for $tr(B^3)$ in the self-dual theory which can be used to obtain the corresponding $tr(F_-^3)$ form factor in  the ordinary (i.e. non-self-dual) gauge theory.}
    \label{fig:diagram}
\end{figure}
\end{center}

Now that we have the all-orders chiral algebra OPEs in hand, there are several directions one can explore with the chiral algebra bootstrap, and which we hope to report on in follow-up work. These include the following.
\begin{enumerate}
    \item One can study higher-loop form factors in our self-dual theories with certain helicity configurations which agree with the corresponding quantities in full gauge theory. The simplest example of such a quantity is the $l$-loop term in the form factor with an insertion of $\textrm{tr}(B^l)$, and arbitrary numbers of positive helicity gluons. The $l=2$ case was computed in \cite{CosQCD}. More generally, form factors of $tr(B^n)$ agree with massless QCD form factors of $tr(F_-^n)$ at $l$-loops, with external states of $n-l$ negative helicity gluons and arbitrary numbers of positive helicity gluons. We plan to report on the $n=3$ case, illustrated in Figure \ref{fig:diagram}, in a future publication; this, to our knowledge, would be the first such form factor computation in a non-supersymmetric theory beyond 2-loops\footnote{See \cite{DGK, BDDD} for tree and one-loop level  $tr(F_-^2)$ form factor computations in various theories using other methods.}.

    \item One can study the $SU(N)$ self-dual gauge theories coupled simultaneously to standard matter ($N$ fundamental flavors) and an axion anomaly-cancelling sector. Form factors for these theories are rational, but $l$-loop axion exchanges contribute to $l+1$-loop QCD contributions, wherein they cancel non-rational contributions order by order. As pointed out in \cite{CosQCD}, this suggests one can extract transcendental pieces of ordinary QCD quantities by computing lower-point and lower-order axion exchanges. This is a complementary application of twistorial theories to the chiral algebra bootstrap we focus on here.
    On the other hand, the computations of \cite{CosQCD} showed that leading terms in an expansion in $N$ are independent of the axion, so that these terms can be computed with the chiral algebra bootstrap and matched with the corresponding quantity in QCD at higher loops.

    \item The complete theorem of \cite{CP} did not restrict to a single insertion of a 4d local operator. In fact, multiple local operator insertions can be considered. In that case, the form factor integrand takes the form of a sum of products of 2d chiral algebra correlators with appropriate conformal blocks and 4d OPE coefficients. The 4d OPE coefficients and the conformal blocks arise by taking the OPE of the 4d local operators. The 4d OPE coefficients will produce more general functions of the spacetime coordinates than in the single-operator case, but should still be constrained by crossing symmetry in the 4d sense. Nevertheless, they need to be specified in addition to the 2d OPE in order for this more general bootstrap program to be run. Only a tree-level example of this procedure has been computed in \cite{CP}, leading to an expression closely related to the CSW formula \cite{CSW}. While the single-operator form factors provide a bootstrap approach rather analogous to BCFW recursion relations, this more general form factor integrand should provide a twistorial analogue of CSW rules. However, the precise relationship of these more general form factors with QCD amplitudes/form factors has yet to be fully understood.
    
\item Related to the previous point, a natural arena to benchmark our chiral bootstrap methods is in the case of self-dual $\mathcal{N}=4$ super Yang-Mills, which is of course anomaly-free. There have been a number of successful form factor calculations from a twistor space perspective in 4d $\mathcal{N}=4$ SYM, such as \cite{Brandhuber:2011tv, Chicherin:2016qsf, Koster:2016loo, Koster:2016fna}. We expect these results also closely match our prescription. Understanding the precise relationship between these approaches should be a fruitful way to make progress.  
\end{enumerate}

In this paper, we pave the way for these and other chiral algebra bootstrap developments by determining the complete, all-order 2d chiral algebra for the SDYM theory with arbitrary choices of anomaly-cancelling matter. The plan for the rest of the note is as follows. \\ 

We consider SDYM coupled to matter satisfying \ref{eq:anomaly}. In \S \ref{sec:symm} we introduce the chiral algebra and our notations, and then explain how to fix the form of the OPEs, up to undetermined coefficients, using a scaling symmetry of the twistorial Lagrangian. We also discuss the diagrammatic interpretation of our methods, using the Koszul duality interpretation of the chiral algebra as a defect theory coupled to the holomorphic `bulk' theory on twistor space. We then present the same results for SDGR coupled to the fourth-order scalar. \\

In \S \ref{sec:assoc}, we present the key formulas for associativity that we use to fix the coefficients of the chiral algebra OPEs. We display the results of this analysis, and present several explicit examples of such computations for illustration in Appendix \ref{app:examples}. The twistorial SDYM algebra has the nice property that at the $m$th order in the loop expansion, all coefficients can be fixed in terms of a single OPE coefficient that we call $\overset{(m)}{f}$. \\

We then use associativity to derive a simple and solvable recursion relation for $\overset{(m)}{f}$, thereby completing the determination of the algebra. We sketch how one could perform a similar analysis for SDGR, but leave its complete determination to future work. Several technical results and proofs are relegated to appendices. 

%In \S \ref{sec:matter}, we generalize our results to the case where SDYM is coupled to ordinary matter (consistent with anomaly-cancellation). 

\section{Form of OPE Corrections from Symmetry}\label{sec:symm}
Let us first establish our notation and background, following \cite{CP}, to which we refer for more details and derivations.
\subsection{Background}
The uplift of self-dual Yang Mills (SDYM) to twistor space $\mathbb{PT}$ is realized by a holomorphic BF-type action given by\footnote{Note that we normalize the integral such that the (extended) chiral algebra tree-level OPEs have no factors of $(2 \pi i)^{-1}$. A different choice was made in \cite{CPassoc} to match the standard one-loop collinear splitting function. Such overall constants can be absorbed by a redefinition of $\mathcal{B}$.}
\begin{equation}
   S[\mathcal{A}, \mathcal{B}] = \bigg(\frac{1}{2 \pi i}\bigg) \underset{\mathbb{PT}}{\int} \text{Tr}(\mathcal{B} F^{0,2}(\mathcal{A})) = \bigg(\frac{1}{2 \pi i}\bigg)\underset{\mathbb{PT}}{\int} \text{Tr}(\mathcal{B} \overline{\partial} \mathcal{A}+\frac{1}{2} \mathcal{B} [\mathcal{A}, \mathcal{A}])
\end{equation}
where the holomorphic gauge field is $\mathcal{A} \in \Omega^{0,1}(\mathbb{PT}, \mathfrak{g})$ and the other  adjoint-valued field\footnote{In general, $\mathcal{B}$ takes values in the dual Lie algebra $\mathfrak{g}^{\vee}$, and the action employs the canonical pairing between $\mathfrak{g}, \mathfrak{g}^{\vee}$.} is $\mathcal{B} \in \Omega^{3,1}(\mathbb{PT}, \mathfrak{g})$, for $\mathfrak{g}$ a complex semi-simple Lie algebra. Note that $\mathcal{B}$ also transforms nontrivially under 6d gauge transformation (e.g. \cite{Mason, Boels:2006ir}). Upon reduction to 4d, the reduction of $\mathcal{B}$, called $B$, gives negative helicity gluons in the 4d BF theory eq.(\ref{eq:BF}); the 4d gauge field $A$ contains the positive-helicity degrees of freedom as usual. \\

This theory suffers from an anomaly which \cite{CosAnomaly}, if the gauge group is either $\mathfrak{sl}_2$, $\mathfrak{sl}_3$, $\mathfrak{so}_8$ or one of the exceptional algebras, can be canceled by a Green-Schwarz mechanism. This works by coupling our theory to a field $\eta \in \Omega^{2, 1}(\mathbb{PT})$ which self-interacts via the free-limit of the BCOV \cite{BCOV} action: \\
\begin{equation}
    S_{\eta} = \frac{1}{2} \bigg( \frac{1}{2 \pi i} \bigg) \underset{\mathbb{PT}}{\int} \partial^{-1} \eta \overline{\partial} \eta.
\end{equation}
The full anomaly-free action is then
\begin{equation}\label{eq:twistoraction}
    S[\mathcal{A},\mathcal{B},\eta] = \bigg( \frac{1}{2 \pi i} \bigg) \underset{\mathbb{PT}}{\int} \bigg(\text{Tr}(\mathcal{B} F^{0,2}(\mathcal{A})) + \frac{1}{2} \partial^{-1} \eta \overline{\partial} \eta + \frac{1}{2} \hat{\lambda}_{\mathfrak{g}} \sqrt{\hbar} \eta \text{Tr}(\mathcal{A}\partial \mathcal{A}) \bigg)
\end{equation}
where we have defined\footnote{Recall that $\lambda_{\mathfrak{g}}$ is defined by the anomaly-cancellation condition in the absence of additional matter, $\text{Tr}(X^4)_{\text{adj}} =  \lambda_{\mathfrak{g}}^2 \text{Tr}(X^2)^2_{\text{fun}}$.}
\begin{equation}\label{eq:hatlambdag}
    \hat{\lambda}_{\mathfrak{g}} \equiv \frac{\lambda_{\mathfrak{g}}}{(2 \pi i)\sqrt{12}}
\end{equation}
and inserted a factor of $\sqrt{\hbar}$ in the interaction term to ensure that the theory remains invariant under a rescaling of $\hbar$. \\

To see this, note that holomorphic BF theory is invariant under simultaneous rescalings by an arbitrary constant $\alpha$:
\begin{equation}
    \hbar \rightarrow \alpha \hbar \quad \quad \mathcal{B} \rightarrow \alpha \mathcal{B} 
\end{equation} upon restoring the usual factor of ${1 \over \hbar}$ which multiplies the (Euclidean) action eq.(\ref{eq:BF}). 
Extending this to the $\eta$-coupled theory, we find that the theory is invariant under simultaneous rescalings:
\begin{equation}
    \hbar \rightarrow \alpha \hbar \quad \quad \mathcal{B} \rightarrow \alpha\mathcal{B} \quad \quad \eta \rightarrow \sqrt{\alpha} \eta
\end{equation}
with the added factor of $\sqrt{\hbar}$ in the interaction. In this work, we will treat $\hbar$ as a formal real parameter deforming the chiral algebra, and enabling us to keep track of the corresponding loop order in the 4d theory. \\

We note that upon reduction to 4d, we obtain the usual BF action eq.(\ref{eq:BF}) plus the following terms:
\begin{equation}
    S[A, \rho] = \int {1 \over 2}(\Delta \rho)^2 + {\lambda_{\mathfrak{g}} \over \sqrt{6} (2 \pi i)} \rho (F \wedge F),
\end{equation} where we call $\rho$ the axion due to its coupling with the gauge sector. We will refer to this 4d theory as the twistorial theory, due to the existence of the uplift eq.(\ref{eq:twistoraction}) to twistor space at the quantum level.\\

To extract the chiral algebra of conformally soft modes of the 4d twistorial theory, consider a defect along a holomorphic plane $\mathbb{C} \subset \mathbb{C}^3$ and choose coordinates $z$, $v^1$, $v^2$ so that the plane is located at $v^i = 0$.\footnote{Working with this local model for $\mathbb{CP}^1\subset \mathbb{PT}$ is sufficient since anomalies are local.} Coupling this defect theory to our axion-coupled theory as an order-type holomorphic defect \cite{CPsugra} gives rise to four towers of operators which the chiral algebra comprises. For example, one such (tower of) defect operator(s) is the current $J$ which couples to the 6d theory by inserting the following expression (where we leave the usual trace over gauge indices implicit) into the path integral:
\begin{equation}\label{eq:coupling}
  \sum_{k=0}^{\infty}{1 \over k!} \int_{z_1, \ldots z_k \in \mathbb{CP}^1}\prod_{i=1}^k\left(\int d^2 z_i {1 \over n!}{1 \over m!}J_a[m, n](z_i) \partial_{v_1}^m \partial_{v_2}^n \mathcal{A}^a_{\bar{z}}(z_i)  \right).
\end{equation} This may be viewed as a holomorphic analogue of inserting an order-type defect, such as an ordinary Wilson line, into the path integral as a path-ordered exponential. Notice that, in contrast to the classical action of an ordinary current/gauge field coupling, in a holomorphic theory we must include holomorphic modes of the gauge field in the transverse directions to obtain the most general possible expression \cite{CPsugra}. \\

To summarize, we list the operators, their quantum numbers, and the fields to which they couple via these bulk-defect couplings, in Table \ref{table1}. Since the axion has a fourth-order kinetic term, and hence twice the degrees of freedom of a scalar with an ordinary second-order kinetic term, it couples to two towers of chiral algebra operators. Note that charge under combined dilatations tells us that the $\tilde{J} \tilde{J}$, $E \tilde{J}$, and $E E$ OPEs are non-singular ($\frac{1}{z}$ has combined dilatation 1).
\begin{table}[t]
    \centering
    \begin{tabular}{|c|c|c|c|c|c|}
    \hline
         Generator&Field&Scaling Dimension&Spin&Combined Dilatation&Weight  \\
         \hline
         $J[t_1,t_2]$, $t_i \geq$ 0&$\mathcal{A}$&$-(t_1+t_2)$&$1-\frac{t_1+t_2}{2}$&$1$&$0$ \\
         \hline
         $\tilde{J}[t_1,t_2]$, $t_i \geq$ 0&$\mathcal{B}$&$-(t_1+t_2+2)$&$-1-\frac{t_1+t_2}{2}$&$0$&$1$ \\
         \hline
          $E[t_1,t_2]$, $t_1+t_2 \geq$ 1&$\eta$&$-(t_1+t_2)$&$-\frac{t_1+t_2}{2}$&$0$&$1/2$\\
         \hline
          $F[t_1,t_2]$, $t_i \geq$ 0&$\eta$&$-(t_1+t_2+2)$&$-\frac{t_1+t_2}{2}$&$1$&$1/2$\\
         \hline
    \end{tabular}
    \caption{Local operators of the 2d chiral algebra, the 6d fields they source, and their quantum numbers. By scaling dimension, we mean charge under scaling of Euclidean 4d spacetime $\mathbb{R}^4$. Here, spin refers to the holomorphic 2d conformal weight, and combined dilatation corresponds to the charge of the operator under simultaneous dilatations $z \rightarrow \frac{z}{r}$ on the celestial sphere and $x \rightarrow \sqrt{r}x$ on 4d spacetime. Finally, weight describes how the operator transforms under a rescaling of $\hbar$. }
    \label{table1}
\end{table}
\\ 

Requiring that the coupling to the defect theory be invariant under rescaling of $\hbar$, we find that the defect operators must transform non-trivially:
\begin{equation}\label{eq:scalingsymm}
    \hbar \rightarrow \alpha \hbar \quad \quad \tilde{J} \rightarrow \frac{\tilde{J}}{\alpha} \quad \quad E \rightarrow \frac{E}{\sqrt{\alpha}} \quad \quad F \rightarrow \frac{F}{\sqrt{\alpha}}.
\end{equation} 
We say that $J$ has weight 0, $E$ and $F$ have weight $\frac{1}{2}$, and $\tilde{J}$ has weight 1. 

\subsection{Including Ordinary Matter}\label{sec:matter}
So far, we have set up the problem when the anomaly is cancelled purely the axion field. Happily, it is very straightforward to augment the previous subsection when the anomaly is cancelled by a combination of the axion and additional ordinary matter, as in \cite{CosQCD}. 

Recall that matter content in a (real) representation $R$ of $\mathfrak{g}$ can cancel the gauge anomaly more generally if the following trace identity holds \cite{CosQCD}:
\begin{equation}\label{eq:anomalymatter}
   \text{Tr}_{\textrm{adj}}(X^4)-\text{Tr}_{R}(X^4) =  \lambda_{\mathfrak{g},R}^2 \text{Tr}_{\text{fun}}(X^2)^2
\end{equation} Analogously to \S \ref{eq:hatlambdag}, we define $\hat{\lambda}_{\mathfrak{g},R} \equiv {\lambda_{\mathfrak{g},R} \over (2 \pi i) \sqrt{12}}$, and we simply make the following replacement in the ``axion'' coupling to the gauge sector in eq.(\ref{eq:twistoraction}): $\hat{\lambda}_{\mathfrak{g}} \mapsto \hat{\lambda}_{\mathfrak{g},R}$. \\

In some special cases with fundamental matter, like $SU(2)$ with $N_f = 8$ or $SU(3)$ with $N_f=9$, the axion decouples completely ($\lambda_{\mathfrak{g},R}=0$) and anomaly cancellation is satisfied simply via $\text{Tr}_{\textrm{adj}}(X^4) = \text{Tr}_{R}(X^4)$. \\

To add in the real matter, we follow \cite{CosQCD}, whose notation we will also closely follow, by using a superfield formalism that simply replaces the Lie algebra with a ``superalgebra'' (which here simply means a $\mathbb{Z}/2\mathbb{Z}$-graded Lie algebra, with the extra mod-2 grading understood as fermion parity) $\mathfrak{g} \mapsto \mathfrak{g}_{R} \equiv \mathfrak{g}\oplus \Pi R$, where $\Pi$ denotes a fermionic parity shift on the matter. In practice, this means that traces over the $R$ indices of $\mathfrak{g}_{R}$ come with an overall minus sign: $\textrm{Tr}_{\mathfrak{g}_R} = \textrm{Tr}_{\mathfrak{g}} - \textrm{Tr}_{R}$. \\ 

The boon of this formalism is that we can simply replace the usual Lie algebra structure constants with structure constants of $\mathfrak{g}_R$ in our formulas; $a, b$ will continue to denote the gauge indices of $\mathfrak{g}$, while indices $i, j$ span a basis of $R$; in particular, the new, additional structure constants $g^j_{ia}$ encode the action of $\mathfrak{g}$ on $R$. More precisely, let $e_i$ denote a basis of $R$ and $e^i$ a basis for the dual. We write $g^{j}_{ia} = - \langle e^j, t_a e_i \rangle$, so that $f^c_{ab}$ and $g^j_{ia}$ are the structure constants of $\mathfrak{g}_R$. \\

The Killing form $K^{ab}$ with indices $a, b, \ldots$ continues to refer to the invariant pairing on $\mathfrak{g}$ only. Since the matter representation $R$ is real, it enjoys a symmetric invariant pairing, which can be used to raise and lower the matter indices of $g^j_{ia}$ as well. We sometimes find it convenient to denote by $R^{ij}$ a formal extension of the Killing form to the matter generators of $\mathfrak{g}_R$, to make certain expressions more symmetrical between $f, g$.\\

This formalism follows from the twistorial uplift of the 4d self-dual gauge theory minimally coupled to matter, which again takes the form of a holomorphic BF theory, but now valued in the Lie superalgebra $\mathfrak{g}_R(-1)$, where the $(-1)$ factor denotes a twist of $R$ by the line bundle $\mathcal{O}(-1)$. In particular, the 6d fields include positive-helicity fermions $\psi_i \in \Omega^{0, 1}(\mathbb{PT}, R \otimes \mathcal{O}(-1))$ and negative-helicity fermions $\psi^i \in \Omega^{0, 1}(\mathbb{PT}, \mathfrak{g}\otimes \mathcal{O}(-3))$, whose action is:
\begin{equation}\label{eq:twistormatter}
S[\psi, \tilde{\psi}, \mathcal{A}] = \left(1 \over 2 \pi i \right)\int_{\mathbb{PT}}\textrm{Tr} (\tilde{\psi}\overline{\partial}\psi - \tilde{\psi}\mathcal{A} \psi),
\end{equation} which we add to eq.(\ref{eq:twistoraction}) (with its $\mathfrak{g}_R$-appropriate coupling constant). Notice that in the classical action the matter and axion fields do not couple directly to each other.\\

The inclusion of ordinary matter means that our 2d chiral algebra acquires two additional towers of operators $M_i[t], \tilde{M}_j[r]$. We present their quantum numbers in Table \ref{table3}. We will find it more convenient in various computations to lower the $R$-index of the defect $\tilde{M}_j$, in a slight departure from the conventions of \cite{CosQCD}.  
\begin{table}[t]
    \centering
    \begin{tabular}{|c|c|c|c|c|c|}
    \hline
         Generator&Field&Scaling Dimension&Spin&Combined Dilatation&Weight  \\
         \hline
         $J[t_1,t_2]$, $t_i \geq$ 0&$\mathcal{A}$&$-(t_1+t_2)$&$1-\frac{t_1+t_2}{2}$&$1$&$0$ \\
         \hline
         $\tilde{J}[t_1,t_2]$, $t_i \geq$ 0&$\mathcal{B}$&$-(t_1+t_2+2)$&$-1-\frac{t_1+t_2}{2}$&$0$&$1$ \\
         \hline
          $E[t_1,t_2]$, $t_1+t_2 \geq$ 1&$\eta$&$-(t_1+t_2)$&$-\frac{t_1+t_2}{2}$&$0$&$1/2$\\
         \hline
          $F[t_1,t_2]$, $t_i \geq$ 0&$\eta$&$-(t_1+t_2+2)$&$-\frac{t_1+t_2}{2}$&$1$&$1/2$\\
          \hline
         $M_i[t_1,t_2]$, $t_i \geq$ 0&$\psi_i$&$-(t_1+t_2+1/2)$&$\frac{1-t_1-t_2}{2}$&$3/4$&$1/2$ \\
         \hline
         $\tilde{M}_j[t_1,t_2]$, $t_i \geq$ 0&$\psi^i$&$-(t_1+t_2+3/2)$&$-\frac{1+t_1+t_2}{2}$&$1/4$&$1/2$ \\
         \hline
    \end{tabular}
    \caption{Extended chiral algebra operators when coupled to matter. We display the 6d fields to which they couple and their quantum numbers, in the same notation as in Table \ref{table1}.}
    \label{table3}
\end{table}

\subsection{Tree-Level OPEs}\label{sec:tree}
First, we recall the tree-level OPEs, which can be determined by computing the BRST-variation of the classical bulk-defect actions (such as the action in \eqref{eq:coupling}) \cite{CP, CosQCD}. One can easily check that the classical OPEs (i.e. $\mathcal{O}(\hbar^0)$) furnished by $J, \tilde{J}, M, \tilde{M}$ OPEs are associative by themselves\footnote{The classical $J, \tilde{J}$ OPEs are also associative in the absence of matter.}. \\

\noindent For notational brevity, we make the following definitions:
    \begin{equation*}
        t \equiv (t^1,t^2) \quad \quad r \equiv (r^1,r^2) \quad \quad s \equiv (s^1,s^2)
\end{equation*}
We also introduce the following notation:
  \begin{equation*}
      n \in \mathbb{Z}^{+} \quad  (n,n) \equiv n
\end{equation*}
The tree-level OPEs are then:
\begingroup \allowdisplaybreaks \begin{align}
      J_a[t](z) J_b[r](0) &\sim \frac{1}{z}f_{ab}^{d_1}J_{d_1}[t+r](0)+\frac{1}{z} \hat{\lambda}_{\mathfrak{g}, R} \sqrt{\hbar} K_{ab}(r^1t^2-r^2t^1)F[t+r-1](0) \\ \notag \\ &-\frac{1}{z} \hat{\lambda}_{\mathfrak{g}, R} \sqrt{\hbar} K_{ab}(t^1+t^2)\partial E[t+r](0) \notag \\ \notag \\&-\frac{1}{z^2}\hat{\lambda}_{\mathfrak{g}, R} \sqrt{\hbar} K_{ab}(t^1+t^2+r^1+r^2) E[t+r](0). \notag \\ \notag \\
        \tilde{J}_a[t](z) J_b[r](0) &\sim \frac{1}{z} f_{ab}^{d_1} \tilde{J}_{d_1} [t+r](0). \\ \notag \\
        E[t](z) J_a[r](0) &\sim \frac{1}{z}\bigg(\frac{t^1r^2-t^2r^1}{t^1+t^2}\bigg) \hat{\lambda}_{\mathfrak{g}, R} \sqrt{\hbar} \tilde{J}_{a} [t+r-1](0). \\ \notag \\ 
        F[t](z) J_a[r](0) &\sim -\frac{1}{z} \hat{\lambda}_{\mathfrak{g}, R} \sqrt{\hbar} \partial \tilde{J}_a [t+r](0) \\ \notag \\&\quad - \frac{1}{z^2} \hat{\lambda}_{\mathfrak{g}, R} \sqrt{\hbar} \bigg(1+\frac{r^1+r^2}{t^1+t^2+2} \bigg)\tilde{J}_a[t+r](0) \notag \\ \notag \\
        M_i[t](z) J_a[r](0) &\sim \frac{1}{z} g^j_{ia} M_j[t+r](0) \\ \notag \\
    \tilde{M}_i[t](z) J_a[r](0) &\sim \frac{1}{z} g^j_{ia} \tilde{M}_j[t+r](0) \\ \notag \\
    \tilde{M}_i[t](z) M_j[r](0) &\sim \frac{1}{z} g^{d_1}_{ij} \tilde{J}_{d_1} [t+r](0) 
\end{align} \endgroup
where $K_{ab}$ is the Killing form on $\mathfrak{g}$ and $f_{bc}^a, g^i_{aj}$ are the structure constants of the Lie superalgebra $\mathfrak{g}_R$, as described in the previous section. \\

Here, we have chosen to present axion OPEs of order $\sqrt{\hbar}$ as part of our ``tree-level'' expressions because they arise from cancelling the gauge variation of a diagram with the topology of a tree: one cubic vertex connecting two external legs to a single defect operator. The form of the remaining corrections determined in \S \ref{sec:diagrams} are best organized by diagram topology, starting with box topologies, and will be displayed below. In our conventions, associativity is determined order by order in $\hbar$ (i.e. so that associativity at order $\hbar^{m}$ is determined by all terms of the selfsame order plus the lower orders).
\subsection{Feynman Diagrams \& the OPE} \label{sec:diagrams}
\noindent The OPEs between our 2d chiral algebra operators are determined by the requirement that the coupling between the defect theory (as in eq.(\ref{eq:coupling}))\footnote{There is a subtlety for the operators associated to the fourth-order scalar field, which in 6d couple to a constrained BCOV field $\eta$, satisfying $\partial \eta = 0$. In the Koszul duality approach of \cite{CP}, `off-shell' currents which couple to the unconstrained field components were introduced via couplings as in eq.(\ref{eq:coupling}) and the constraint was applied at the end to convert their OPEs into OPEs for the `on-shell' operators $E, F$ using the linear relation eq.(7.2.5) in \cite{CP}. One can run the arguments of this section directly with the off-shell fields and reach the same conclusion. This is guaranteed to work since our arguments (matching quantum numbers, cubic interactions in the Lagrangian, and how the diagram topology affects the form of the OPE corrections) apply to both the off-shell and on-shell degrees of freedom. To directly compute holomorphic Feynman integrals (which we will be able to circumvent in this note) it is easier to work with the off-shell modes.} and the 6d twistorial uplifted theory (eq.(\ref{eq:twistoraction}) plus eq.(\ref{eq:twistormatter})) be gauge invariant. This is analogous to how gauge-invariance of an ordinary Wilson line tells us that the $J$'s satisfy the current algebra of the gauge algebra. In (holomorphic) theories with a 2d holomorphic defect, we compute chiral algebra OPEs rather than ordinary commutators. Moreover, by expanding the exponential of the coupling action in the path integral, we find contributions to the OPE at each order in $\hbar$, whereas no such deformations to the current algebra are possible for an ordinary Wilson line in a theory with compact gauge group. The mathematical formulation of this principle is called Koszul duality; see \cite{PW} for a review and other references therein. In other words, the OPEs are defined such that, order-by-order in pertubation theory, all non-vanishing BRST-variations of Feynman diagrams (which in this case are Witten-like diagrams, since we focus on the interactions from bulk/defect couplings) at that loop order must cancel. The diagrammatic perspective will be useful for us in what follows. \\

\begin{figure}[t]
    \centering
    \begin{subfigure}[b]{0.3\textwidth}
    \centering
    \includegraphics[scale=0.40]{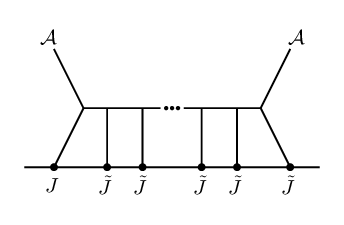}
    \end{subfigure}
    \begin{subfigure}[b]{0.3\textwidth}
\centering
         \includegraphics[scale=0.40]{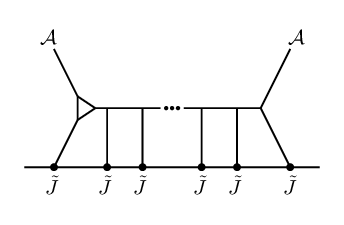}
    \end{subfigure}
    \begin{subfigure}[b]{0.3\textwidth}
\centering
         \includegraphics[scale=0.40]{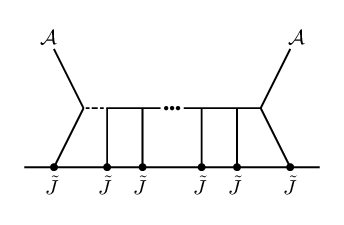}
    \end{subfigure}
    \begin{subfigure}[b]{0.3\textwidth}
\centering
         \includegraphics[scale=0.40]{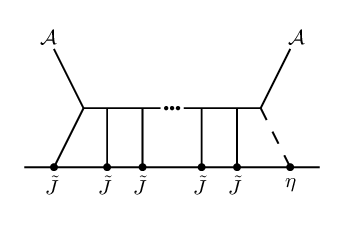}
    \end{subfigure}
    \begin{subfigure}[b]{0.3\textwidth}
\centering
         \includegraphics[scale=0.40]{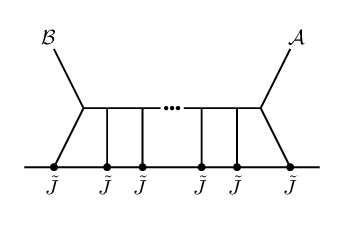}
    \end{subfigure}
    \begin{subfigure}[b]{0.3\textwidth}
\centering
         \includegraphics[scale=0.40]{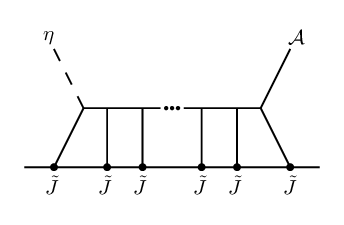}
    \end{subfigure}
     \caption{Requiring that the gauge anomaly of these diagrams cancel lead to non-trivial higher-order corrections to the OPEs of our defect operators.}
     \label{figure_1}
\end{figure}

Schematically, the contributions from BRST non-invariant diagrams consist of a normal-ordered product of the operators on the defect, and contractions between these operators. To determine which diagrams contribute to a given OPE, we will restrict the structural diagrammatic possibilities using a.) the fact that the interactions in the twistorial Lagrangian are only cubic in form, and b.) enforcement of invariance under rescaling of $\hbar$.
This will allow us to determine the allowed operators on the defect. Let us see this explicitly. \\

We will start by considering the case when the anomaly is cancelled by the axion alone, and then generalize to include fermionic matter. \\

Suppose that we are interested in the OPE of two defect operators which couple to the fields $\varphi_1$ and $\varphi_2$, $l$ of which are $\eta$ ($2 \geq l \geq 0$), and assume that the combined weight of such operators (in the sense of Table \ref{table1}) is $\omega_{12}$. Consider an $m$-loop diagram with external legs $\varphi_1$ and $\varphi_2$, and $N$ operators on the defect with combined weight $\omega$. Due to the cubic nature of the interactions, the lowest-order diagram with $N$ operators on the defect and two external legs is an $(N-1)$th-order diagram, therefore $N-m \leq 1$. Suppose now that out of the $N$ operators, $n_1$ are $J$, $n_2$ are $\tilde{J}$, $n_3$ are $E$, and $n_4$ are $F$. The assumption that the combined weight is $\omega$ gives us the following equation
\begin{equation}\label{eq:weight1}
    \omega = n_2 +\frac{n_3+n_4}{2}.
\end{equation}

In the standard loop-expansion, an $m$-loop diagram comes with an explicit factor of $\hbar^{m}$. If this diagram also has $s$ vertices of the form $ \hat{\lambda}_{\mathfrak{g}} \sqrt{\hbar} \eta \mathcal{A} \mathcal{A}$, then the normal-ordered product OPE correction comes with a factor of $\hat{\lambda}_{\mathfrak{g}}^{\frac{s}{2}} \hbar^{m+\frac{s}{2}}$. Note that since this is the only interaction involving $\eta$ and it is linear in $\eta$, all axion external legs and axion defect operators necessarily come with a factor of $\hat{\lambda}_{\mathfrak{g}} \sqrt{\hbar}$. This gives us the inequality $s \geq n_3+n_4+l$. \\

We now impose invariance under rescaling of $\hbar$ by requiring that both sides of the OPE have matching weight. This results in the following equation: 
\begin{equation}\label{eq:weight2}
    \omega = \omega_{12}+m+\frac{s}{2}.
\end{equation}
Using these equations, we find the following inequalities:
\begin{equation}\label{eq:inequality}
   1 \geq N-m \geq n_2+n_3+n_4-m = \omega_{12}+\frac{s+n_3+n_4}{2} \geq \omega_{12}+\frac{l}{2}+n_3+n_4 \geq 0.
\end{equation}
 From this, we immediately learn that there are no anomalous Feynman diagrams that contribute non-trivially to the $F \tilde{J}$, $EF$, and $FF$ OPEs since $\omega_{12}+\frac{l}{2} > 1$. Therefore, the only OPE corrections are to the $JJ$, $ \tilde{J} J$, $EJ$, and $FJ$ OPEs, and considering the various possible cases in eq.(\ref{eq:inequality}) and using eq.(\ref{eq:weight1}), eq.(\ref{eq:weight2}), we find that only the (BRST variations of) Feynman diagrams in Figure \ref{figure_1} contribute to the OPE. \\
 
 In other words, these inequalities are sufficiently strong to fix the allowed operator content on the defect, or equivalently, the operators that can appear on the right-hand-side of the OPE. In particular, eq.(\ref{eq:inequality}) precludes an arbitrary number of $J$'s from appearing on the defect. To see how this works, fix $s,n_3,n_4$ such that $1 \geq \omega_{12}+\frac{s+n_3+n_4}{2} \geq 0$. Then eq.(\ref{eq:inequality}) fixes $N$ to be either $m$ or $m+1$ ($m$ is only allowed if $\omega_{12}+\frac{s+n_3+n_4}{2}$ equals zero), and eq.(\ref{eq:weight1}) fixes $n_2$ in terms of $n_3,n_4,\omega$, the last of which is determined by eq.(\ref{eq:weight2}). Using $n_1 = N-n_2-n_3-n_4$, we see that the number of $J$'s is fixed as well. As an example consider $\omega_{12}=1, l=0$. The only allowed $s,n_3,n_4$ are $s=n_3=n_4=0$. The inequality in eq.(\ref{eq:inequality}) reduces to $1 \geq N-m \geq 1$ which fixes $N=m+1$, and eq.(\ref{eq:weight1}), eq.(\ref{eq:weight2}) give us the following equality $\omega = m+1=n_2$. This determines $n_1=0$.

\subsubsection{Adding Matter}
\begin{figure}[t] 
    \centering
    \begin{subfigure}[b]{0.3\textwidth}
    \centering
    \includegraphics[scale=0.40]{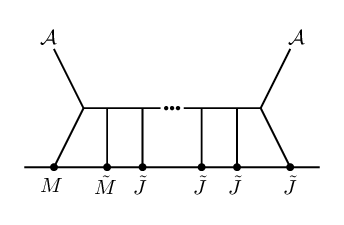}
    \end{subfigure}
    \begin{subfigure}[b]{0.3\textwidth}
\centering
         \includegraphics[scale=0.40]{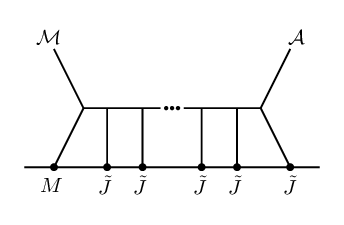}
    \end{subfigure}
    \begin{subfigure}[b]{0.3\textwidth}
\centering
         \includegraphics[scale=0.40]{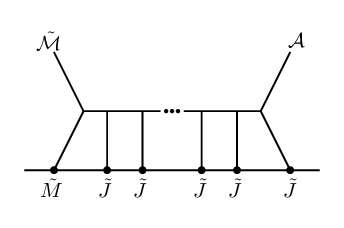}
    \end{subfigure}
      \begin{subfigure}[b]{0.3\textwidth}
\centering
         \includegraphics[scale=0.40]{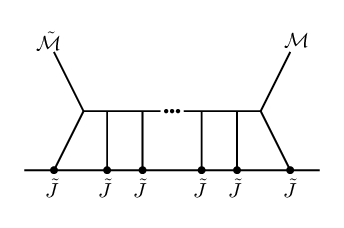}
    \end{subfigure}
     \caption{ The cancellation of the gauge anomaly of these diagrams results in additional
corrections to the OPEs of the operators in the chiral algebra with matter}
     \label{figure_matter}
\end{figure}
The results of \S \ref{sec:diagrams} are still valid here (upon replacement with $\hat{\lambda}_{\mathfrak{g}, R}$, which can be zero for certain $R$), and can be simply generalized. \\

We now let $n_5$ be the number of $M$ defect operators, and $n_6$ the number of $\tilde{M}$ operators. We modify all equations by $n_3+n_4 \to n_3+n_4+n_5+n_6$, except for the constraint on $s$. Explicitly,
\begin{equation}
    \omega = n_2+\frac{n_3+n_4+n_5+n_6}{2} \quad \quad 1 \geq N-m \geq \omega_{12}+\frac{s+n_3+n_4+n_5+n_6}{2} \geq 0.
\end{equation}

Considering all allowed cases, we find that the only new diagrams are those of Figure \ref{figure_matter}. In particular, only the $JJ$ OPE gets deformed by the inclusion of matter. Note that combined dilatation rules out some of the resulting cases.

\subsection{General Form of The All-Orders OPE Corrections}\label{sec:SDYMOPE}
The diagrams in Figures \ref{figure_1} and \ref{figure_matter} lead to all higher-order (i.e. beyond tree-level, in the sense of \S \ref{sec:tree}) OPE corrections, which can be expressed as an expansion in $\hbar \in \mathbb{R}_{\geq 1}$. We display these OPEs below. The operator content on the right-hand-side of the OPE was determined by the arguments of \S \ref{sec:diagrams}. By matching the quantum numbers under combined dilatations (see Table \ref{table1}) we can further fix the order of the poles they multiply, since $\frac{1}{z}$ and $\partial_z$ each have combined dilatation $=1$. In the subsequent sections, we will use associativity to determine the unknown coefficients that appear below. \\ \\
For notational convenience we define $p_m = t+r-m$.
\begingroup \allowdisplaybreaks
\begin{align*}
      &J_a[t](z)J_b[r](0)  \sim \\ &  \quad \quad \frac{1}{z} \overset{\sum k_j = p_m}{\sum_{ m \geq 1 \quad k_j^i \geq 0}}  \hbar^m \overset{(m)}{\underset{(t,r)}{a}}[k_1,...,k_{m+1}]^{d_1 \cdot \cdot \cdot d_{m+1}}_{ab}:J_{d_1}[k_1] \prod_{j=2}^{m+1} \tilde{J}_{d_j}[k_j]: 
      \\ 
      & \quad \quad +\overset{\sum k_j = p_m}{\sum_{ m \geq 1 \quad k_j^i \geq 0}} \hbar^m \bigg( \frac{1}{z^2}\overset{(m)}{\underset{(t,r)}{b}}[k_1,...,k_{m}]^{d_1 \cdot \cdot \cdot d_m}_{ab}+\frac{1}{z}\overset{(m)}{\underset{(t,r)}{c}}[k_1,...,k_{m}]^{d_1 \cdot \cdot \cdot d_m}_{ab} \hat{\partial_1} \bigg):\prod_{j=1}^{m} \tilde{J}_{d_j}[k_j]:  
      \\  
      & \quad \quad + \overset{k_j = p_m}{\sum_{ m \geq 2 \quad k_j^i \geq 0}}  \hat{\lambda}_{\mathfrak{g}, R}^2 \hbar^{m} \bigg( \frac{1}{z^2}\overset{(m)}{\underset{(t,r)}{d}}[k_1,...,k_m]^{d_1 \dotsm d_m}_{ab}+\frac{1}{z}\overset{(m)}{\underset{(t,r)}{e}}[k_1,...,k_m]^{d_1 \dotsm d_m}_{ab}\hat{\partial_1} \bigg):\prod_{j=1}^{m} \tilde{J}_{d_j}[k_j]:
      \\  
      & \quad \quad + \overset{\sum k_j = p_{m+1}}{\sum_{ m \geq 1 \quad k_j^i \geq 0}}  \hat{\lambda}_{\mathfrak{g}, R} \hbar^{m+\frac{1}{2}} \frac{1}{z}\overset{(m)}{\underset{(t,r)}{i}}[k_1,...,k_{m+1}]^{d_2 \dotsm d_{m+1}}_{ab}:F[k_1]\prod_{j=2}^{m+1} \tilde{J}_{d_j}[k_j]:
      \\  
      & \quad \quad + \overset{\sum k_j = p_m}{\sum_{ m \geq 1 \quad k_j^i \geq 0}}  \hat{\lambda}_{\mathfrak{g}, R} \hbar^{m+\frac{1}{2}} \bigg( \frac{1}{z^2}\overset{(m)}{\underset{(t,r)}{g}}[k_1,...,k_{m+1}]^{d_2 \dotsm d_{m+1}}_{ab} 
      \\
      & \quad \quad \quad \quad \quad \quad \quad \quad \quad \quad \quad \quad \quad \quad \quad +\frac{1}{z}\sum_{l=1}^{2}\overset{(m)}{\underset{(t,r)}{h_l}}[k_1,...,k_{m+1}]^{d_2 \dotsm d_{m+1}}_{ab} \hat{\partial_l} \bigg):E[k_1]\prod_{j=2}^{m+1} \tilde{J}_{d_j}[k_j]: 
      \\
      & \quad \quad + \frac{1}{z} \overset{\sum k_j = p_m}{\sum_{ m \geq 1 \quad k_j^i \geq 0}} \hbar^m  \overset{(m)}{\underset{(t,r)}{a_*}}[k_1,...,k_{m+1}]^{d_3 \cdot \cdot \cdot d_{m+1},ij}_{ab}:M_i[k_1] \tilde{M}_j[k_2] \prod_{j=3}^{m+1} \tilde{J}_{d_j}[k_j]:
      \\ 
      \\ 
      &\tilde{J}_a[t](z)J_b[r](0)  \sim \quad \frac{1}{z}  \overset{\sum k_j = p_m}{\sum_{ m \geq 1 \quad k_j^i \geq 0}}   \hbar^m \overset{(m)}{\underset{(t,r)}{f}}[k_1,...,k_{m+1}]^{d_1 \cdot \cdot \cdot d_{m+1}}_{ab}: \prod_{j=1}^{m+1} \tilde{J}_{d_j}[k_j]:  
      \\ 
      \\ 
      &E[t](z)J_b[r](0)  \sim \quad \frac{1}{z}  \overset{\sum k_j = p_{m+1}}{\sum_{ m \geq 1 \quad k_j^i \geq 0}}  \hat{\lambda}_{\mathfrak{g}, R} \hbar^{m+\frac{1}{2}}  \overset{(m)}{\underset{(t,r)}{j}}[k_1,...,k_{m+1}]^{d_1 \cdot \cdot \cdot d_{m+1}}_{b}: \prod_{j=1}^{m+1} \tilde{J}_{d_j}[k_j]: 
      \\ 
      \\ 
      &F[t](z)J_b[r](0)  \sim \\ &  \quad \quad \overset{\sum k_j = p_m}{\sum_{ m \geq 1 \quad k_j^i \geq 0}} \hat{\lambda}_{\mathfrak{g}, R} \hbar^{m+\frac{1}{2}} \bigg( \frac{1}{z^2}\overset{(m)}{\underset{(t,r)}{k}}[k_1,...,k_{m+1}]^{d_1 \cdot \cdot \cdot d_{m+1}}_{b} +\frac{1}{z}\overset{(m)}{\underset{(t,r)}{l}}[k_1,...,k_{m+1}]^{d_1 \cdot \cdot \cdot d_{m+1}}_{b} \hat{\partial_1} \bigg):\prod_{j=1}^{m+1} \tilde{J}_{d_j}[k_j]:
      \\ 
      \\ 
      &M_i[t](z)J_b[r](0) \sim \quad \frac{1}{z} \overset{\sum k_j = p_m}{\sum_{ m \geq 1 \quad k_j^i \geq 0}} \hbar^m  \overset{(m)}{\underset{(t,r)}{d_*}}[k_1,...,k_{m+1}]^{d_2 \cdot \cdot \cdot d_{m+1},j}_{b,i}: M_j[k_1]\prod_{j=2}^{m+1} \tilde{J}_{d_j}[k_j]:
      \\  
      \\
      &\tilde{M}_i[t](z)J_b[r](0) \sim \quad  \frac{1}{z} \overset{\sum k_j = p_m}{\sum_{ m \geq 1 \quad k_j^i \geq 0}} \hbar^m  \overset{(m)}{\underset{(t,r)}{e_*}}[k_1,...,k_{m+1}]^{d_2 \cdot \cdot \cdot d_{m+1},j}_{b,i}: \tilde{M}_j[k_1]\prod_{j=2}^{m+1} \tilde{J}_{d_j}[k_j]: 
      \\  
      \\
      &\tilde{M}_i[t](z)M_j[r](0) \sim \quad \frac{1}{z} \overset{\sum k_j = p_m}{\sum_{ m \geq 1 \quad k_j^i \geq 0}} \hbar^m  \overset{(m)}{\underset{(t,r)}{h_*}}[k_1,...,k_{m+1}]^{d_1 \cdot \cdot \cdot d_{m+1}}_{ij}: \prod_{j=1}^{m+1} \tilde{J}_{d_j}[k_j]:
\end{align*} 
\endgroup
where $k_j^i$ is the i-th component of $k_j$, and $\hat{\partial_j} \prod_{i=1}^{m+1} \theta_{i} \equiv (\prod_{i=1}^{j-1} \theta_{i}) \partial \theta_{j}(\prod_{i=j+1}^{m+1} \theta_{i})$, where the $\theta_i$ denote chiral algebra operators. As always, $: \ldots  :$ denotes normal-ordered products:
\begin{equation}
    :\phi_i \phi_j:(w) = \lim_{z \to w} \bigg(\phi_i(z) \phi_j(w)-\wick{\c\phi_i(z) \c\phi_j(w)} \bigg).
\end{equation}
The sum over the $k_j$ has been constrained by several symmetries. The first is simply matching scaling dimension (in the sense of Table \ref{table1}), i.e. the holomorphic conformal dimension must be the same on both sides of the OPE. This leads to the expression $\sum_{i=1}^2 \sum_{j=1}^{m+1} k_j^i = \sum_{i=1}^2 (t^i + r^i - m)$. Next, we write the complexified 4d Lorentz algebra as $\mathfrak{sl}_2(\mathbb{C})_- \times \mathfrak{sl}_2(\mathbb{C})_+$. The eigenvalue of the former factor is spin (holomorphic conformal weight) we just accounted for, and which acts on the celestial sphere by M{\"o}bius transformations. The second factor $\mathfrak{sl}_2(\mathbb{C})_{+}$, which we will use to further constrain the sum, mixes the twistor fibre coordinates, which are the components of $\tilde{\lambda}^{\dot{\alpha}}$ and hence is a flavor symmetry from the point of view of the chiral algebra. The operators $\theta[m, n]$, from either matter or gluon towers, live in a representation of highest weight $(m+n)/2$ (with $\theta[m+n, 0]$ furnishing the highest weight state, etc.) and have eigenvalue $(m-n)/2$ under the Cartan of $\mathfrak{sl}(2)_+$.  Imposing invariance under this flavor symmetry restricts the sum further, such that the components each satisfy $\sum_j k_j^i = t^i + r^i - m$. \\

We remark that, at one-loop, some of these OPE coefficients were determined in \cite{CPassoc, CosQCD} with associativity. Those were the OPEs for the strong generators of the chiral algebra: $J[1], \tilde{J}[1]$. We will reproduce and extend these results. In \S \ref{sec:solve_1}, we will see explicitly that all OPE coefficients can be determined, via associativity, in terms of the OPE coefficients appearing in the one-loop OPE of the strong generators, as conjectured in \cite{CPassoc}. Those, and some additional, one-loop OPEs were also reproduced in \cite{victor} with the direct Koszul duality methods; we review this approach in Appendix \ref{app:koszul}. \\

Additionally, the coefficients of the terms with single poles, for the non-axion-coupled chiral algebras (i.e. the chiral algebras generated by $J, \tilde{J}$ and $w, \tilde{w}$ only), were determined in \cite{Zeng} using the (quite different) method of homotopy transfer. The latter technique uses the presence of a homotopical algebra that is inherent in twisted theories of cohomological type (see \cite{TASI} for pedagogical lectures), like the holomorphic theories on twistor space that appear in this note. Because of the difference in methods and perspective, the expressions of \cite{Zeng} look at first sight quite different from the expressions we find\footnote{Indeed, one of our goals for this note is to provide workmanlike formulas for the OPE coefficients that can be easily plugged in to Mathematica for use in form factor computations.}. We have nonetheless checked that our overlapping results agree beautifully. In particular, the recursion relation we derive from associativity in \S \ref{sec:recursion} gives the same solution as the $A_{\infty}$ relation satisfied by the higher homotopical operations $m_n$ of \cite{Zeng} for the corresponding OPE coefficient.

\subsection{Extended Self-Dual Gravity}
The arguments from the previous sections can be repeated \textit{mutis mutandis} for the case of self-dual gravity and its twistorial uplift, which is Poisson-BF theory. We will simply present the results of this analysis. To consistently quantize the theory, one must deform it just as in the case of self-dual Yang-Mills. Again, a Green-Schwarz mechanism can be applied, introducing a fourth-order pseudoscalar field \cite{BSS}, to which we refer the reader for details. We call this anomaly-cancelled theory ``extended'' self-dual GR. Analogously to the gauge theory case \cite{CPassoc, victor}, part of the one-loop OPE for this theory has been fixed via associativity and Koszul duality analyses \cite{roland}. In particular, our presentation here of the chiral algebra and its bulk-defect coupling elide several subtleties which are treated carefully in \cite{roland}, but which do not affect the present analysis. \\

\begin{table}[t]
    \centering
    \begin{tabular}{|c| c| c|c|c|c|}
    \hline
         Generator& Field & Scaling Dimension&Spin&Combined Dilatation&Weight  \\
         \hline
         $w[t_1,t_2]$, $t_i \geq$ 0& $\mathcal{H}$ &$2-(t_1+t_2)$&$2-\frac{t_1+t_2}{2}$&$1$&$0$ \\
         \hline
         $\tilde{w}[t_1,t_2]$, $t_i \geq$ 0& $\mathcal{G}$&$-(t_1+t_2+4)$&$-2-\frac{t_1+t_2}{2}$&$0$&$1$ \\
         \hline
          $e[t_1,t_2]$, $t_1+t_2 \geq$ 1& $\eta$ &$-(t_1+t_2)$&$-\frac{t_1+t_2}{2}$&$0$&$1/2$\\
         \hline
          $f[t_1,t_2]$, $t_i \geq$ 0& $\eta$ &$-(t_1+t_2+2)$&$-\frac{t_1+t_2}{2}$&$1$&$1/2$\\
         \hline
    \end{tabular}
    \caption{Local operators of the extended chiral algebra for axion-coupled SDGR, the 6d fields to which they couple, and the quantum numbers labeled as in Table \ref{table1}.}
    \label{table2}
\end{table}
\noindent The extended Poisson-BF theory is
\begin{equation*}
    S[\mathcal{H},\mathcal{G},\eta]=\bigg( \frac{1}{2 \pi i} \bigg) \underset{\mathbb{PT}}{\int} \bigg(\mathcal{G}T^{0,2}(\mathcal{H})+\frac{1}{2}\partial^{-1} \eta \overline{\partial} \eta + \eta \partial^{\dot{\alpha}}\mathcal{H} \partial_{\dot{\alpha}}\text{ $\lrcorner$ } \eta + {1 \over 2} \mu \sqrt{\hbar} \eta \partial^{\dot{\beta}} \partial_{\dot{\alpha}} \mathcal{H} \partial^{\dot{\alpha}} \partial_{\dot{\beta}} \partial \mathcal{H} \bigg).
\end{equation*}
Here, $\mathcal{G} \in \Omega^{3, 1}(\mathbb{PT}, \mathcal{O}(-2))$, $\mathcal{H} \in \Omega^{0, 2}(\mathbb{PT}, \mathcal{O}(2))$ is the ``Hamiltonian field'' and $T^{0, 2}(\mathcal{H}) = \overline{\partial}\mathcal{H} + \left\lbrace \mathcal{H}, \mathcal{H} \right\rbrace$, using the Poisson bracket of weight $-2$ on the fibres of twistor space. Like $\lambda_{\mathfrak{g}}$, $\mu$ is fixed by the anomaly-cancellation condition to be $\mu^2 = {1 \over 5!}({i \over 2 \pi})^2$. \\ \\
The tree-level OPEs for extended SDGR are as follows \cite{CP, AMS, roland}:
\begingroup
\allowdisplaybreaks
\begin{align}
    w[t](z)w[r](0) &\sim \frac{1}{z}(t^1r^2-t^2r^1) w[t+r-1](0)- \frac{1}{z} \mu \sqrt{\hbar} R_3(t,r) f[t+r-3](0)  \\ \notag \\
   &+\frac{1}{z^2} \mu \sqrt{\hbar} R_2(t,r) e[t+r-2](0) \notag \\ \notag \\ &+\frac{1}{z} \mu \sqrt{\hbar} \bigg( \frac{t^1+t^2-2}{t^1+t^2+r^1+r^2-4} \bigg) R_2(t,r) \partial e[t+r-2](0). \notag \\ \notag \\
    \tilde{w}[t](z)w[r](0) &\sim \frac{1}{z}(t^1r^2-t^2r^1) \tilde{w}[t+r-1](0). \\ \notag \\
    e[t](z)w[r](0) &\sim \frac{1}{z}(t^1r^2-t^2r^1) e[t+r-1](0)+\frac{1}{z} \mu \sqrt{\hbar} R_3(r,t) \tilde{w}[t+r-3](0).  \\ \notag \\
     f[t](z)w[r](0) &\sim \frac{1}{z}(t^1r^2-t^2r^1) f[t+r-1](0) \\ \notag \\
     &+\frac{1}{z^2} \bigg(\frac{r^1+r^2}{t^1+t^2+2}\bigg) e[t+r](0)+\bigg(\frac{r^1+r^2}{t^1+t^2+r^1+r^2}\bigg)\partial e[t+r](0) \notag \\ \notag \\
     &-\frac{1}{z^2} \mu \sqrt{\hbar} R_2(r,t) \bigg( \frac{t^1+t^2+r^1+r^2}{t^1+t^2+2}\bigg) \tilde{w}[t+r-2](0) \notag \\ \notag \\ 
     &-\frac{1}{z} \mu \sqrt{\hbar} R_2(r,t) \bigg( \frac{r^1+r^2-2}{t^1+t^2+2}\bigg) \partial \tilde{w}[t+r-2](0).
     \notag \\ \notag \\
     f[t](z) f[r](0) &\sim \frac{1}{z^2} \bigg(2+\frac{t^1+t^2}{r^1+r^2+2}+\frac{r^1+r^2}{t^1+t2+2}\bigg) \tilde{w}[t+r](0) \\ \notag \\
     &+\frac{1}{z}\bigg(1+\frac{t^1+t^2}{r^1+r^2+2}\bigg) \partial \tilde{w}[t+r](0) \notag \\ \notag \\
    e[t](z)f[r](0) &\sim \frac{1}{z}(t^1r^2-t^2r^1) \tilde{w}[t+r-1](0) 
\end{align}
\endgroup
\\ 
where the $R_n(t,r)$ are defined as follows:
\begin{equation}
    R_n(t,r)=\sum_{j=0}^n (-1)^j {n \choose j} \frac{t^1!t^2!r^1!r^2!}{(t^1+j-n)!(t^2-j)!(r^1-j)!(r^2+j-n)!}.
\end{equation}
Note that, as in the case of SDYM, we label the corrections of order $\sqrt{\hbar}$ as tree-level because they arise from cancelling the gauge variation of a tree-topology diagram. \\ 

This theory is invariant under simultaneous rescalings:
\begin{equation}
    \hbar \rightarrow \alpha \hbar \quad \quad \mathcal{G} \rightarrow \alpha \mathcal{G} \quad \quad \eta \rightarrow \sqrt{\alpha} \eta.
\end{equation}
Again, invariance of the coupling to the defect theory tells us that the defect operators transform non-trivially:
\begin{equation}
    \tilde{w} \rightarrow \frac{\tilde{w}}{\alpha} \quad \quad e \rightarrow \frac{e}{\sqrt{\alpha}} \quad \quad f \rightarrow \frac{f}{\sqrt{\alpha}}.
\end{equation} 
We say that $w$ has weight 0, $e$ and $f$ have weight $\frac{1}{2}$, and $\tilde{w}$ has weight 1. 
\\ 

Exactly as in SDYM, $w, \tilde{w}$ are conformally soft graviton modes of positive and negative helicities, respectively, while $e, f$ are matter currents which couple to the fourth-order scalar. See \cite{roland} for the explicit bulk/defect couplings. \\ 

\begin{figure}[t]
    \centering
    \begin{subfigure}[b]{0.3\textwidth}
\centering
         \includegraphics[scale=0.40]{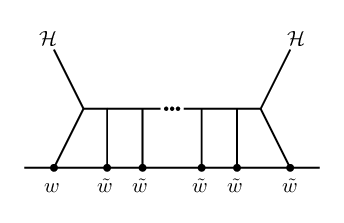}
    \end{subfigure}
    \begin{subfigure}[b]{0.3\textwidth}
    \centering
    \includegraphics[scale=0.40]{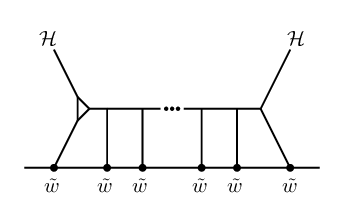}
    \end{subfigure}
     \begin{subfigure}[b]{0.3\textwidth}
\centering
         \includegraphics[scale=0.40]{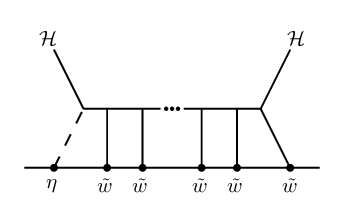}
    \end{subfigure}
    \begin{subfigure}[b]{0.3\textwidth}
\centering
         \includegraphics[scale=0.40]{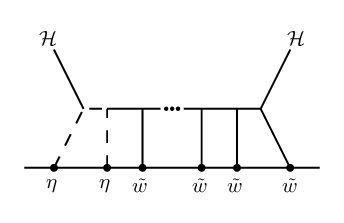}
    \end{subfigure}
        \begin{subfigure}[b]{0.3\textwidth}
\centering
         \includegraphics[scale=0.40]{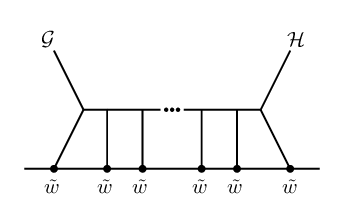}
    \end{subfigure}
    \begin{subfigure}[b]{0.3\textwidth}
\centering
         \includegraphics[scale=0.40]{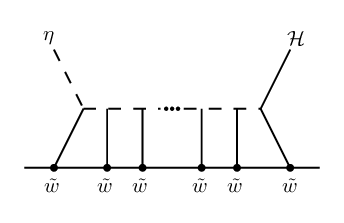}
    \end{subfigure}
    \begin{subfigure}[b]{0.3\textwidth}
\centering
         \includegraphics[scale=0.40]{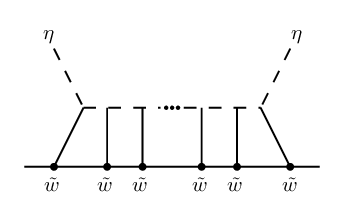}
    \end{subfigure}
     \caption{Requiring that the gauge anomaly of these diagrams---and all possible variations of these diagrams that arise from permuting the positions of the defect operators---cancel lead to non-trivial higher-order corrections to the extended SDGR chiral algebra OPEs.}
     \label{figure_SDGR}
\end{figure}

Noting that the interactions in this theory are also only cubic, we can use our results from before: 
\begin{equation}
     1 \geq N-m \geq \omega_{12}+\frac{s+n_3+n_4}{2} \geq 0.
\end{equation}
together with eq.(\ref{eq:weight1}), eq.(\ref{eq:weight2}). Note that we no longer have the constraint on $s$. This is because, unlike their SDYM counterparts $E, F$, the towers coupling to the axion $e, f$ are no longer required to appear with the factor of $\sqrt{\hbar}$ from the $\eta \mathcal{H} \mathcal{H}$ vertex: SDGR has an additional $\eta \eta \mathcal{H}$ vertex which scales like $\hbar$. This leads to the possibility of additional diagrams with external $\eta$-legs; see Figure \ref{figure_SDGR}. Note that the case $\omega_{12}=0,n_3=1,n_4=0,s=1$ is ruled out by that the assumption that $m$ is an integer. 

\newpage
\subsection{General Form of The OPE For Extended SDGR}\label{sec:SDGR}
\noindent We use the same notation as in \S \ref{sec:SDYMOPE}, but we now define $p_m = t+r-2m$. \\ \\
Using the same logic as before, the general OPE corrections at higher orders in $\hbar$ are determined to be: 
\begingroup \allowdisplaybreaks
\begin{align*}
&w[t](z)w[r](0)  \sim \\ & \quad \quad \frac{1}{z} \overset{\sum k_j = p_m-1}{\sum_{ m \geq 1 \quad k_j^i \geq 0}}  \hbar^m \overset{(m)}{\underset{(t,r)}{A}}(k_1,...,k_{m+1}):w[k_1] \prod_{j=2}^{m+1} \tilde{w}[k_j]: 
      \\ 
&\quad \quad + \overset{\sum k_j = p_{m+1}}{\sum_{ m \geq 1 \quad k_j^i \geq 0}} \hbar^m \bigg( \frac{1}{z^2}\overset{(m)}{\underset{(t,r)}{B}}(k_1,...,k_{m})+\frac{1}{z}\overset{(m)}{\underset{(t,r)}{C}}(k_1,...,k_{m}) \hat{\partial_1} \bigg):\prod_{j=1}^{m} \tilde{w}[k_j]:  \\  
&\quad \quad + \overset{\sum k_j = p_{m+1}}{\sum_{ m \geq 2 \quad k_j^i \geq 0}} \mu^2 \hbar^{m} \bigg( \frac{1}{z^2}\overset{(m)}{\underset{(t,r)}{D}}(k_1,...,k_m)+\frac{1}{z}\overset{(m)}{\underset{(t,r)}{E}}(k_1,...,k_m)\hat{\partial_1} \bigg):\prod_{j=1}^{m} \tilde{w}[k_j]:
      \\  
&\quad \quad + \overset{\sum k_j = p_{m+1}-1}{\sum_{ m \geq 1 \quad k_j^i \geq 0}} \mu \hbar^{m+\frac{1}{2}} \frac{1}{z}\overset{(m)}{\underset{(t,r)}{I}}(k_1,...,k_{m+1}):f[k_1]\prod_{j=2}^{m+1} \tilde{w}[k_j]: 
      \\  
&\quad \quad + \overset{\sum k_j = p_{m+1}}{\sum_{ m \geq 1 \quad k_j^i \geq 0}}  \mu \hbar^{m+\frac{1}{2}} \bigg( \frac{1}{z^2}\overset{(m)}{\underset{(t,r)}{G}}(k_1,...,k_{m+1})+\frac{1}{z}\sum_{k=1}^{2}\overset{(m)}{\underset{(t,r)}{H_k}}(k_1,...,k_{m+1}) \hat{\partial_k} \bigg):e[k_1]\prod_{j=2}^{m+1} \tilde{w}[k_j]:
      \\
&\quad \quad + \overset{\sum k_j = p_m-1}{\sum_{ m \geq 1 \quad k_j^i \geq 0}}  \hbar^{m} \frac{1}{z}\overset{(m)}{\underset{(t,r)}{M}}(k_1,...,k_{m+1}):f[k_1] e[k_2] \prod_{j=3}^{m+1} \tilde{w}[k_j]: 
      \\
      &\quad \quad + \overset{\sum k_j = p_m}{\sum_{ m \geq 1 \quad k_j^i \geq 0}}   \hbar^{m} \bigg( \frac{1}{z^2}\overset{(m)}{\underset{(t,r)}{N}}(k_1,...,k_{m+1})+\frac{1}{z}\underset{k \neq 2}{\sum_{k=1}^{3}} \overset{(m)}{\underset{(t,r)}{O_k}}(k_1,...,k_{m+1}) \hat{\partial_k} \bigg):e[k_1] e[k_2]\prod_{j=3}^{m+1} \tilde{w}[k_j]:
      \\ 
      \\ 
      &\tilde{w}[t](z)w[r](0)  \sim \quad \frac{1}{z}  \overset{\sum k_j = p_m-1}{\sum_{ m \geq 1 \quad k_j^i \geq 0}}   \hbar^m \overset{(m)}{\underset{(t,r)}{F}}(k_1,...,k_{m+1}): \prod_{j=1}^{m+1} \tilde{w}[k_j]:  
      \\ 
      \\ 
      &e[t](z)w[r](0)  \sim  \\ & \quad \quad \frac{1}{z}  \overset{\sum k_j = p_{m+1}-1}{\sum_{ m \geq 1 \quad k_j^i \geq 0}}  \mu \hbar^{m+\frac{1}{2}}  \overset{(m)}{\underset{(t,r)}{J}}(k_1,...,k_{m+1}): \prod_{j=1}^{m+1} \tilde{w}[k_j]: \\
      & \quad \quad + \frac{1}{z}  \overset{\sum k_j = p_m-1}{\sum_{ m \geq 1 \quad k_j^i \geq 0}} \hbar^{m}  \overset{(m)}{\underset{(t,r)}{P}}(k_1,...,k_{m+1}):e[k_1]\prod_{j=2}^{m+1} \tilde{w}[k_j]:
      \\ 
      \\ 
      &f[t](z)w[r](0)  \sim \\ & \quad \quad \overset{\sum k_j = p_{m+1}}{\sum_{ m \geq 1 \quad k_j^i \geq 0}} \mu \hbar^{m+\frac{1}{2}} \bigg( \frac{1}{z^2}\overset{(m)}{\underset{(t,r)}{K}}(k_1,...,k_{m+1})+\frac{1}{z}\overset{(m)}{\underset{(t,r)}{L}}(k_1,...,k_{m+1})\hat{\partial_1} \bigg):\prod_{j=1}^{m+1} \tilde{w}[k_j]: 
      \\
      & \quad \quad +\frac{1}{z} \overset{\sum k_j = p_m-1}{\sum_{ m \geq 1 \quad k_j^i \geq 0}}  \hbar^m \overset{(m)}{\underset{(t,r)}{Q}}(k_1,...,k_{m+1}): f[k_1]\prod_{j=2}^{m+1} \tilde{w}[k_j]:
      \\
      & \quad \quad +\overset{\sum k_j = p_m}{\sum_{ m \geq 1 \quad k_j^i \geq 0}} \hbar^{m} \bigg( \frac{1}{z^2}\overset{(m)}{\underset{(t,r)}{R}}(k_1,...,k_{m+1})+\frac{1}{z}\overset{(m)}{\underset{(t,r)}{S}}(k_1,...,k_{m+1})\hat{\partial_1} \bigg):e[k_1]\prod_{j=2}^{m+1} \tilde{w}[k_j]: 
      \\ 
      \\
       &f[t](z)f[r](0) \sim \quad  \overset{\sum k_j = p_m}{\sum_{ m \geq 1 \quad k_j^i \geq 0}} \hbar^m \bigg( \frac{1}{z^2}\overset{(m)}{\underset{(t,r)}{T}}(k_1,...,k_{m})+\frac{1}{z}\overset{(m)}{\underset{(t,r)}{U}}(k_1,...,k_{m}) \hat{\partial_1} \bigg):\prod_{j=1}^{m+1} \tilde{w}[k_j]: 
       \\ 
       \\ 
       &e[t](z)f[r](0)  \sim \quad \frac{1}{z}  \overset{\sum_{j=1}^{m+1} k_j = p_m-1}{\sum_{ m \geq 1 \quad k_j^i \geq 0}}   \hbar^m \overset{(m)}{\underset{(t,r)}{V}}(k_1,...,k_{m+1}): \prod_{j=1}^{m+1} \tilde{w}[k_j]:
\end{align*}  
\endgroup

In the case of SDYM, we labeled the OPE coefficients as $a^{(m)} \ldots l^{(m)}$ in alphabetical order. To ease comparison with that example, we label the OPE coefficients for SDGR of the same structural type as their SDYM counterparts (with obvious relabelings of the positive helicity, negative helicity, and matter operators) with the corresponding capital letters $A^{(m)} \ldots L^{(m)}$. The coefficients from $M^{(m)} \ldots V^{(m)}$ are unique to extended SDGR for the reasons explicated in the previous section. \\

Also notice that, although the OPEs for both (extended) SDYM and SDGR involve an infinite sum of terms in an $\hbar$ expansion, the chiral algebras are extremely simple: the high degree of symmetry constrains the OPEs to have only single or double poles.  

\section{Associativity is Enough}\label{sec:assoc}
\noindent We now proceed to determine the coefficients of the OPEs in \S \ref{sec:SDYMOPE}. It will turn out to be convenient to express all of the coefficients as elementary functions of $\overset{(m)}{f}$, and then to find the exact solution for $\overset{(m)}{f}$. 
\subsection{Compilation of Key Formulas}
\noindent Our strategy is simple. We enforce associativity of the OPEs order-by-order in $\hbar$ by demanding that the following identity holds for all choices of operators $\phi_1,\phi_2,\phi_3$: 
\begingroup \allowdisplaybreaks
\begin{align*}
        \oint_{\lvert z_2 \rvert = 2} dz_2 z_2^l\bigg( \oint_{\lvert z_{12} \rvert = 1} \phi_1(z_1) \phi_2(z_2)\bigg)\phi_3(0) &= \oint_{\lvert z_{1} \rvert = 2} dz_1 \phi_1(z_1)\bigg(\oint_{\lvert z_{2} \rvert = 1} dz_2 z_2^l \phi_2(z_2) \phi_3(0) \bigg) \\ \\  &- (-1)^{F_1 F_2}\oint_{\lvert z_{2} \rvert = 2} dz_2 z_2^l \phi_2(z_2) \bigg(\oint_{\lvert z_{1} \rvert = 1} dz_1 \phi_1(z_1) \phi_3(0)\bigg)
\end{align*}
\endgroup
where $l \in \mathbb{Z}_{\geq 0}$, and $F_i$ is $1$ if the operator $\phi_i$ is fermionic and 0 otherwise, so the factor $(-1)^{F_1 F_2}$ is relevant when we couple to fermionic matter in representation $R$. This will fix the unknown coefficients of our OPE corrections. \\

Let us start by considering the case where the anomaly is cancelled by the axion only, and then we will generalize to the inclusion of fermionic matter.
\\ \\
Writing the OPEs formally as
\begin{equation*}
    \phi_i(z) \phi_j(w) \sim \sum_{n \geq 1} \frac{\{ \phi_i \phi_j \}_n}{(z-w)^n}
\end{equation*}
we find that, after performing the contour integrals, this equality reduces to:
\begin{equation}\label{eq:assoc}
    \{\{ \phi_1 \phi_2 \}_1 \phi_3 \}_{l+1} = \{\phi_1 \{\phi_2 \phi_3 \}_{l+1}\}_1 - (-1)^{F_1 F_2}\{\phi_2 \{\phi_1 \phi_3\}_1\}_{l+1}.
\end{equation}
It will also be convenient, from time to time, to rewrite the left-hand-side of this expression using the identity:
\begin{equation}
    \sum_{n \geq l+1} \frac{(-1)^n \partial^{n-l-1} }{(n-l-1)!} \{\phi_3 \{ \phi_1 \phi_2 \}_1 \}_{n} = \{\phi_1 \{\phi_2 \phi_3 \}_{l+1}\}_1 -  (-1)^{F_1 F_2} \{\phi_2 \{\phi_1 \phi_3\}_1\}_{l+1}.
\end{equation}
This is because we can use Wick's theorem to easily evaluate OPEs between an operator and a normal-ordered product (in that order): 
\begingroup \allowdisplaybreaks
    \begin{align*}
         &\{\phi_1 :\phi_2 \phi_3:\}_l(w) = \\ &\bigg( \frac{1}{2 \pi i} \bigg)^2 \oint_{w} dz (z-w)^{l-1} \oint_{w} \frac{dx}{x-w} \bigg( \wick{\c\phi_1(z)\c\phi_2(x)} \phi_3(w)+(-1)^{F_1 F_2}\phi_2(x)\wick{\c\phi_1(z) \c\phi_3(w)}\bigg) \\ \\ = \bigg(& \frac{1}{2 \pi i} \bigg)^2  \sum_{n \geq 1} \oint_{w} dz (z-w)^{l-1} \oint_{w} \frac{dx}{x-w} \bigg( \frac{\{\phi_1 \phi_2\}_n(x)\phi_3(w)}{(z-x)^n}+(-1)^{F_1 F_2}\frac{\phi_2(x)\{\phi_1\phi_3\}_n(w)}{(z-w)^n}\bigg) \\ = { }:&\{\phi_1 \phi_2\}_{l}\phi_3:(w)+
    (-1)^{F_1 F_2}:\phi_2\{\phi_1\phi_3\}_{l}:(w)+\sum_{n=1}^{l-1} {l-1 \choose n-1}:{\{\{\phi_1 \phi_2\}}_n\phi_3\}_{l-n}:(w).
    \end{align*} 
\endgroup
This expression is particularly useful because charge under combined dilatation tells us that $l$ is at most equal to $3$. In practice, we will only need to consider $l=1,2$, as we will see shortly. \\

It turns out that only the following choice of operators $\phi_1$,$\phi_2$,$\phi_3$ yield a non-trivial equality in eq.(\ref{eq:assoc}): 
\begin{equation*}
    \phi_1 = \phi_2 =\phi_3 = J  \quad \quad \quad \phi_i = \phi_j = J \quad \phi_k \in \{\tilde{J}, E, F \}.
\end{equation*}
To see that this is the case note that the only singular OPEs involving $\phi_k$ are those with $J$, ie. $\phi_k(z) J(0)$. This means that non-trivial associativity requirements can only be obtained from at least one of the operators being $J$. The results of \S \ref{sec:symm} obtained from the rescaling symmetry tells us that the OPE $\phi_k(z) J(0)$ results in a normal-ordered product of $\tilde{J}$, which in turn only has singular OPEs with $J$. Therefore, we need at least two of the operators to be $J$ for a non-trivial equality. \\ 

We prove in Appendix \ref{app:minimal} that it is sufficient to consider the following associativity conditions: 
\begingroup \allowdisplaybreaks
\begin{align}
  \{ \tilde{J}_a [t] \{ J_b[r] J_c[s] \}_1 \}_{1} &= \{J_b[r] \{\tilde{J}_a[t] J_c[s]\}_{1}\}_1 - \{J_c[s] \{\tilde{J}_a[t] J_b[r] \}_1\}_{1} \label{eq:minimalset01} \\ \notag \\
           \{\{ J_a[t] J_b[r] \}_1 J_c[s] \}_{1} &=  \{J_a[t] \{J_b[r] J_c[s] \}_{1}\}_1  - \{J_b[r] \{J_a[t] J_c[s]\}_1\}_{1} \label{eq:minimalset02} \\ \notag \\
         \{ E [t] \{ J_b[r] J_c[s] \}_1 \}_{1} &= \{J_b[r] \{E[t] J_c[s]\}_{1}\}_1 - \{J_c[s] \{E[t] J_b[r] \}_1\}_{1} \label{eq:minimalset03}
         \\ \notag \\   
          \{F[t] \{J_b[r] J_c[s] \}_{1}\}_1  &=  \{\{ F[t] J_b[r] \}_1 J_c[s] \}_{1}+ \{J_b[r] \{F[t] J_c[s]\}_1\}_{1}  \label{eq:minimalset04} \\ \notag \\
           \{F[t] \{J_b[r] J_c[s]\}_1\}_{2}   &= \{J_b[r] \{F[t] J_c[s] \}_{2}\}_1 - \{ J_c[s] \{ J_b[r] F[t] \}_1  \}_{2}. \label{eq:minimalset05}
    \end{align}
\endgroup
To understand why these five equations can fix our thirteen unknown coefficients, it is important to note that some of these associativity conditions result in more than one meaningful equality. This is because the coefficients of distinct operators cannot cancel one another, and therefore terms in the associativity equations with different operator content should satisfy these relations independently of one another. \\

The generalization to incorporate nontrivial matter is again immediate. The minimal set of non-trivial associativity conditions that need to be satisfied is expanded by the emergence of: 
\begingroup \allowdisplaybreaks
\begin{align}
      \label{eq:matter1} \{M_i[s] \{ J_a[t] J_b[r] \}_1 \}_{1} &= \{J_a[t] \{M_i[s] J_b[r]  \}_{1}\}_1-\{J_b[r] \{ M_i[s] J_a[t]\}_1\}_{1}   \\ \notag \\
   \label{eq:matter2} \{\tilde{M}_i[s] \{ J_a[t] J_b[r] \}_1 \}_{1} &= \{J_a[t] \{ \tilde{M}_i[s] J_b[r]  \}_{1}\}_1-\{J_b[r] \{\tilde{M}_i[s] J_a[t] \}_1\}_{1}  \\ \notag \\
     \label{eq:matter3} \{ J_c[s] \{ \tilde{M}_j[t] M_i[r] \}_1 \}_1 &=- \{ \tilde{M}_j[t] \{ M_i[r] J_c [s] \}_1 \}_1- \{  M_i[r] \{ \tilde{M}_j[t] J_c [s] \}_1 \}_1
\end{align} \endgroup
where we have used the same arguments as in Appendix \ref{app:minimal} to reduce the number of equations. \\

The fact that none of these associativity conditions couple the fermionic matter fields to the axion towers readily follow from the general OPE corrections. In particular, $M$ and $\tilde{M}$ have trivial OPEs with $\tilde{J}, E$ and $F$. However, the associativity conditions eq.(\ref{eq:minimalset02}) pick up contributions from the deformation of the $JJ$ OPE. Notice in particular that the coefficients $c$ and $b$ depend on all anomaly-cancelling matter contributions present in the theory and are nonzero even in the special cases where $\lambda_{g, R}^2 = 0$ and we can set $E, F$ to zero, or when the $M, \tilde{M}$ generators are set to zero and the axion coupling simplifies to $\lambda_{\mathfrak{g}}$. \\

The operator content of the associativity conditions and the coefficients they fix are as follows:
\begingroup \allowdisplaybreaks
\begin{align*}
    \ref{eq:minimalset01} &\rightarrow : \underset{j}{\prod} \tilde{J}_{d_j}: { } \Rightarrow \overset{(m)}{f} \text{ Associativity Equation} \\ \\
    \ref{eq:minimalset02} &\rightarrow 
        \begin{cases}
        : J_{d_1} \underset{j}{\prod} \tilde{J}_{d_j}: { } \Rightarrow \overset{(m)}{a} \quad \quad : \underset{j}{\prod} \tilde{J}_{d_j}: { } \Rightarrow \overset{(m)}{c_{\lambda}}   \\
        : F \underset{j}{\prod} \tilde{J}_{d_j}: { } \Rightarrow \overset{(m)}{i} \quad \quad : \partial E \prod_{j}^{} \tilde{J}_{d_j}: { } \Rightarrow \overset{(m)}{h_1} \quad \quad   : E \partial J_{d_1} \underset{j}{\prod} \tilde{J}_{d_j}: { } \Rightarrow \overset{(m)}{h_2}  \\
        \end{cases}
        \longrightarrow \overset{(m)}{f}
        \\ \\
    \ref{eq:minimalset03} &\rightarrow : \underset{j}{\prod} \tilde{J}_{d_j}: { } \Rightarrow \overset{(m)}{j} \rightarrow \overset{(m)}{f} \quad \quad 
     \ref{eq:minimalset04} \rightarrow : \underset{j}{\prod} \tilde{J}_{d_j}: { } \Rightarrow \overset{(m)}{l} \rightarrow \overset{(m)}{f} \quad \quad
      \ref{eq:minimalset05} \rightarrow : \underset{j}{\prod} \tilde{J}_{d_j}: { } \Rightarrow \overset{(m)}{k} \rightarrow \overset{(m)}{f} 
\end{align*}
\endgroup
where we have grouped $c$ and $e$ into a single coefficient $c_{\lambda}$. We can similarly combine $b$ and $d$ into a coefficient called $b_{\lambda}$. It is convenient to keep the coefficients separate in some intermediate calculations, since they represent contributions from different diagrams in Figure \ref{figure_1}, although they contribute to the same singular term in the OPE. Later, however, we will find that $d, e$ are actually zero, so that  $c_{\lambda} = c$ (and likewise for $b_{\lambda}$) after all.  \\

We can fix the $b_{\lambda}$ and $g$ using the invariance of the left-hand-side of the OPE under the exchange $z \leftrightarrow w$ followed by the swap $(a,t) \leftrightarrow (b,r)$. Doing this, we find an expression relating $b_{\lambda}$ to $c_{\lambda}$:
\begingroup \allowdisplaybreaks
\begin{align} \label{eq:b}
     &\overset{\sum k_j = p_m}{\sum_{ m \geq 1 \quad k_j^i \geq 0}} \bigg(\underset{(t,r)}{\overset{(m)}{c_{\lambda}}} [k_1,...,k_{m}]_{ab}^{d_1 \cdot \cdot \cdot d_{m}} + \underset{(r,t)}{\overset{(m)}{c_{\lambda}}} [k_1,...,k_{m}]_{ba}^{d_1 \cdot \cdot \cdot d_{m}}\bigg) :\partial J_{d_1}[k_1] \overset{m}{\underset{j=2}{\prod}} \tilde{J}_{d_j}[k_j]: { } =  \\ & \quad \quad \overset{\sum k_j = p_m}{\sum_{ m \geq 1 \quad k_j^i \geq 0}}\bigg(\underset{(t,r)}{\overset{(m)}{b_{\lambda}}} [k_1,...,k_{m}]_{ab}^{d_1 \cdot \cdot \cdot d_{m}} + ... + \underset{(t,r)}{\overset{(m)}{b_{\lambda}}} [k_2,...,k_{m},k_1]_{ab}^{d_2 \cdot \cdot \cdot d_{m} d_1} \bigg) :\partial J_{d_1}[k_1] \overset{m}{\underset{j=2}{\prod}} \tilde{J}_{d_j}[k_j]: \notag
\end{align} \endgroup \\
and an expression relating $g^{(m)}$ to $h_1^{(m)}$:
\begin{equation}
    \underset{(t,r)}{\overset{(m)}{g}}[k_1,...,k_{m+1}]_{ab}^{d_2 \cdot \cdot \cdot d_{m+1}} = \underset{(t,r)}{\overset{(m)}{h_1}}[k_1,...,k_{m+1}]_{ab}^{d_2 \cdot \cdot \cdot d_{m+1}}+\underset{(r,t)}{\overset{(m)}{h_1}}[k_1,...,k_{m+1}]_{ba}^{d_2 \cdot \cdot \cdot d_{m+1}}.
\end{equation} 
These equalities turn out to be sufficient to obtain $b_{\lambda}$ and $g$. Thus, all the unknown coefficients can be accounted for with these relations. \\

Similar arguments hold for the coefficients in the general fermionic-matter-coupled OPEs. We can set $s=0$ in \eqref{eq:matter1} \eqref{eq:matter2}, and set $r=0$ in \eqref{eq:matter3}. Then, the equation \eqref{eq:matter2} results in the same expression as in \eqref{eq:matter1}, immediately giving us the equality: 
\begin{equation}
    \overset{(m)}{\underset{(t,r)}{e_*}}[k_1,...,k_{m+1}]^{d_2 \cdot \cdot \cdot d_{m+1},j}_{b,i} =  \overset{(m)}{\underset{(t,r)}{d_*}}[k_1,...,k_{m+1}]^{d_2 \cdot \cdot \cdot d_{m+1},j}_{b,i}.
\end{equation}

We present some illustrative worked examples of the manipulations that determine these coefficients in Appendix \ref{app:examples}. In the next section, we will present the results of the derivations.

\subsection{OPE Coefficients in Terms of $f$} \label{sec:coeffs}
\noindent We make the following definitions using the Killing form and structure constants of $\mathfrak{g}$:
\begin{equation}\label{eq:Kdef}
\quad \quad K^{d_1 \cdot \cdot \cdot d_{m+1}}_{ab} = -f^{d_1}_{a e_1} K^{e_1 e_2} f^{d_2}_{e_2 e_3} \cdot \cdot \cdot f^{d_m}_{e_{2m-2} e_{2m-1}} K^{e_{2m-1} e_{2m}} f^{i_{m+1}}_{e_{2m} b} 
\end{equation} 
\begin{equation}\label{eq:alphabeta}
     \alpha(t,k)=t^2(k^1+1)-t^1(k^2+1) \quad \quad \beta(t) = t^1+t^2
\end{equation}  
Recall $R^{ij}$, $i, j \in R$, is the formal extension of $K^{ab}$ to all of $\mathfrak{g}_R$. We work with conventions such that the Kronecker delta $\delta^{\alpha \beta}$ (with $\alpha, \beta$ denoting both indices either in $\mathfrak{g}$ or both in $R$), when it appears in explicit diagonalizations of the Killing form, is to be understood as relabeling, but not raising, the corresponding index (and similarly for $\delta_{\alpha \beta}$). \\

We need to define the structure $G^{ij d_3 \cdot \cdot \cdot d_{m+1}}_{ab}$, very similar to $K^{i_1 \cdot \cdot \cdot i_{m+1}}_{ab}$ but including both $f, g$ structure constants when the theory is coupled to matter. The easiest way to determine the composition of $f, g$ that constitute $G^{ij d_3 \cdot \cdot \cdot d_{m+1}}_{ab}$ is by going left to right starting with the first two indices: the first upper index and the first bottom index. These two indices determine which structure constant appears first. The remaining unfixed index of that structure constant and the second upper index of $G$ determines the next structure constant, and so on. Then, add as many $K$’s or $R$'s as you need to contract the unfixed indices, in the same way as in $K^{i_1 \cdot \cdot \cdot i_{m+1}}_{ab}$. Carrying this out fully fixes the definition of $G$. To illustrate, we present a few explicit examples: 
\begingroup \allowdisplaybreaks
\begin{align*}
    G^{ij}_{ab} &= R^{j_1 j_2} g^{i}_{a j_1} g^{j}_{b j_2} \quad \quad \quad \quad \quad \quad \quad \quad \quad
    G^{j d_2}_{ib} = K^{e_1 e_2} g^j_{i e_1} f^{d_2}_{b e_2} 
    \\ \\
    G^{ijd_3}_{ab} &= R^{j_1 j_2} K^{e_1 e_2} g^i_{a j_1} g^{j}_{j_2 e_1} f^{d_3}_{b e_2} \quad \quad \quad \quad
    G^{j d_3 i}_{ab} = R^{j_1 j_2} R^{j_3 j_4} g^{j}_{a j_1} g^{d_3}_{j_2 j_3} g^{i}_{b j_4}
\end{align*} \\
Performing computations as in Appendix \ref{app:examples} for each equation in the set eqs.(\ref{eq:minimalset01} - \ref{eq:minimalset05}) and eqs.(\ref{eq:matter1} - \ref{eq:matter3}) (except one, which we will solve in the next subsection for $\overset{(m)}{f}$ itself) fixes the coefficients in terms of the $\underset{(t,r)}{\overset{(m)}{f}}$  as follows:
\begingroup \allowdisplaybreaks \begin{align*}
    &\underset{(t,r)}{\overset{(m)}{f}}[k_1,...,k_{m+1}]_{ab}^{d_1 \cdot \cdot \cdot d_{m+1}} = \underset{(t,r)}{\overset{(m)}{f}}(k_1,...,k_{m+1}) K^{d_1 \cdot \cdot \cdot d_{m+1}}_{ab} 
    \\ \\
    &\underset{(t,r)}{\overset{(m)}{a}}[k_1,...,k_{m+1}]_{ab}^{d_1 \cdot \cdot \cdot d_{m+1}} = \\& \quad \underset{(t,r)}{\overset{(m)}{f}}(k_1,...,k_{m+1}) K^{d_1 \cdot \cdot \cdot d_{m+1}}_{ab} +\underset{(t,r)}{\overset{(m)}{f}}(k_2,k_1,...,k_{m+1}) K^{d_2 d_1 \cdot \cdot \cdot d_{m+1}}_{ab}+ 
    ...+\underset{(t,r)}{\overset{(m)}{f}}(k_2,...,k_{m+1},k_1) K^{d_2 \cdot \cdot \cdot d_{m+1} d_1}_{ab} 
    \\ \\
    &\underset{(t,r)}{\overset{(m)}{b}}[k_1,...,k_{m}]_{ab}^{d_1 \cdot \cdot \cdot d_{m}} = (2 h^\vee - T(R))\underset{(t,r)}{\overset{(m)}{b}}(k_1,...,k_{m}) K_{ab}^{d_1 \cdot \cdot \cdot d_{m}} 
    \\ \\
    &\underset{(t,r)}{\overset{(m)}{b}}(k_1,...,k_{m}) = \\& \quad \quad \bigg( \frac{1}{96 \pi^2} \bigg) \bigg(\underset{(t,r)}{\overset{(m-1)}{i}}(k_{m},...,k_1)+\underset{(t,r)}{\overset{(m-1)}{i}}(k_{1},k_{m},k_{m-1},...,k_2)\bigg) +
    \\& \quad \quad +
    \bigg( \frac{1}{96\pi^2} \bigg) \beta(t) \underset{(t,r)}{\overset{(m-1)}{j}}(k_m,...,k_1) 
    -(-1)^m\bigg( \frac{1}{96\pi^2} \bigg) \beta(r) \underset{(r,t)}{\overset{(m-1)}{j}}(k_m,...,k_1)
    \\& \quad \quad
    -\bigg( \frac{1}{96\pi^2} \bigg) \beta(t) \underset{(t,r)}{\overset{(m-1)}{j}}(k_1,...,k_m) 
    +(-1)^m\bigg( \frac{1}{96\pi^2} \bigg) \beta(r) \underset{(r,t)}{\overset{(m-1)}{j}}(k_1,...,k_m)
    \\
    & \quad \quad + \underset{l_j^i \geq 0}{\overset{\sum l_j = k_m}{\sum}}  \underset{(t,r)}{\overset{(m)}{f}}(l_1,l_2,k_{m-1},...,k_{1})
     + \underset{l_j^i \geq 0}{\overset{\sum l_j = k_{m-1}}{\sum}}  \underset{(t,r)}{\overset{(m)}{f}}(k_{m},l_1,l_2,k_{m-2}...,k_{1}) + 
     \\
     & \quad \quad ...+ \underset{l_j^i \geq 0}{\overset{\sum l_j = k_1}{\sum}}  \underset{(t,r)}{\overset{(m)}{f}}(k_{m},...,k_{2}, l_1,l_2) 
     \\& \quad \quad
     +\frac{1}{2} \sum_{n = 1}^{m-1} \bigg(
     \underset{l_j^i \geq 0}{\overset{\sum l_j = q^{m}_{n}}{\sum}} \underset{(t,r)}{\overset{(n)}{f}}(l_1,l_2,k_{n-1},...,k_{1}) \underset{(l_2,l_1)}{\overset{(m-n)}{f}}(k_{m},...,k_{n}) + 
     \\
     & \quad \quad + \underset{l_j^i \geq 0}{\overset{\sum l_j = q^{m-1}_{n-1}}{\sum}}  \underset{(t,r)}{\overset{(n)}{f}}(k_m,l_1,l_2,k_{n-2},...,k_{1}) \underset{(l_2,l_1)}{\overset{(m-n)}{f}}(k_{m-1},...,k_{n-1})+...
     \\
     &\quad \quad +\underset{l_j^i \geq 0}{\overset{\sum l_j = q^{m-n+1}_{1}}{\sum}}  \underset{(t,r)}{\overset{(n)}{f}}(k_m,...,k_{m-n+2},l_1,l_2) \underset{(l_2,l_1)}{\overset{(m-n)}{f}}(k_{m-n+1},...,k_{1}) + (-1)^m (t \leftrightarrow r)\bigg)
    \\ \\
    &\underset{(t,r)}{\overset{(m)}{i}}[k_1,...,k_{m+1}]_{ab}^{d_2 \cdot \cdot \cdot d_{m+1}} = \\& \quad \quad \bigg(\alpha(t,k_1)\underset{(t,r)}{\overset{(m)}{f}}(k_1+1,...,k_{m+1}) +\alpha(t-k_2-1,k_1)\underset{(t,r)}{\overset{(m)}{f}}(k_2,k_1+1,...,k_{m+1})+
    \\
    & \quad \quad ...+ \alpha(t-m-\sum_{i=2}^{m+1}k_i,k_1)\underset{(t,r)}{\overset{(m)}{f}}(k_2,...,k_{m+1},k_1+1) \bigg) K^{d_2 \cdot \cdot \cdot d_{m+1}}_{ab} 
    \\ \\
    &\underset{(t,r)}{\overset{(m)}{g}}[k_1,...,k_{m+1}]_{ab}^{d_2 \cdot \cdot \cdot d_{m+1}} = \\& \quad -\beta(k_1) \bigg( \underset{(t,r)}{\overset{(m)}{f}}(k_1, k_2, \ldots ,k_{m+1})+\underset{(t,r)}{\overset{(m)}{f}}(k_2,k_1, k_3, \ldots ,k_{m+1})+...+\underset{(t,r)}{\overset{(m)}{f}}(k_2, k_3, \ldots ,k_{m+1},k_1) \bigg) K^{d_2 \cdot \cdot \cdot d_{m+1}}_{ab} 
    \\ \\
    &\underset{(t,r)}{\overset{(m)}{h_1}}[k_1,...,k_{m+1}]_{ab}^{d_2 \cdot \cdot \cdot d_{m+1}} = \\& \quad \bigg(\beta(-t)\underset{(t,r)}{\overset{(m)}{f}}(k_1,...,k_{m+1}) +\beta(k_2+1-t)\underset{(t,r)}{\overset{(m)}{f}}(k_2,k_1,...,k_{m+1})+ \\ & \quad \quad ...+\beta(\sum_{i=2}^{m+1}k_i+m-t)\underset{(t,r)}{\overset{(m)}{f}}(k_2,...,k_{m+1},k_1) \bigg)K^{d_2 \cdot \cdot \cdot d_{m+1}}_{ab}
    \\ \\
    &\underset{(t,r)}{\overset{(m)}{h_2}}[k_1,...,k_{m+1}]_{ab}^{d_2 \cdot \cdot \cdot d_{m+1}} = \\& \quad \quad -\beta(k_1)\bigg( \underset{(t,r)}{\overset{(m)}{f}}(k_2,k_1,...,k_{m+1})+...+\underset{(t,r)}{\overset{(m)}{f}}(k_2,...,k_{m+1},k_1) \bigg) K^{d_2 \cdot \cdot \cdot d_{m+1}}_{ab}-
    \\ & \quad \quad
    -\beta(k_1)\bigg( \underset{(t,r)}{\overset{(m)}{f}}(k_3,k_2,k_1,...,k_{m+1})+...+\underset{(t,r)}{\overset{(m)}{f}}(k_3,k_2,...,k_{m+1},k_1) \bigg) K^{d_3 d_2 \cdot \cdot \cdot d_{m+1}}_{ab}-
    \\&
    \quad \quad ...-\beta(k_1) \underset{(t,r)}{\overset{(m)}{f}}(k_3,...,k_{m+1},k_2,k_1)  K^{d_3 \cdot \cdot \cdot d_{m+1} d_2}_{ab} 
    \\ \\
    & \underset{(t,r)}{\overset{(m)}{j}}[k_1,...,k_{m+1}]_{b}^{d_1 \cdot \cdot \cdot d_{m+1}} =  - \bigg(\frac{\alpha(t,k_1)}{\beta(t)}\bigg) \underset{(t-1,r)}{\overset{(m)}{f}}(k_1,...,k_{m+1}) K^{d_1 e_1} K^{d_2 \cdot \cdot \cdot d_{m+1}}_{e_1 b} 
    \\ \\
    &\underset{(t,r)}{\overset{(m)}{k}}[k_1,...,k_{m+1}]_{b}^{d_1 \cdot \cdot \cdot d_{m+1}} = -\bigg(\frac{\beta(k_1+1)}{\beta(t+1)}\bigg) \underset{(t,r)}{\overset{(m)}{f}}(k_1,...,k_{m+1}) K^{d_1 e_1} K^{d_2 \cdot \cdot \cdot d_{m+1}}_{e_1 b}  
    \\ \\
    &\underset{(t,r)}{\overset{(m)}{l}}[k_1,...,k_{m+1}]_{b}^{d_1 \cdot \cdot \cdot d_{m+1}} = -\underset{(t,r)}{\overset{(m)}{f}}(k_1,...,k_{m+1}) K^{d_1 e_1} K^{d_2 \cdot \cdot \cdot d_{m+1}}_{e_1 b} 
    \\ \\
    &\overset{(m)}{\underset{(t,r)}{a_*}}[k_1,...,k_{m+1}]^{d_3 \cdot \cdot \cdot d_{m+1},ij}_{ab} = \\& \quad \quad \overset{(m)}{\underset{(t,r)}{f}}[k_1, k_2, \ldots,k_{m+1}] G^{ij d_3 \cdot \cdot \cdot d_{m+1}}_{ab}+...+\overset{(m)}{\underset{(t,r)}{f}}[k_1,...,k_{m+1},k_2] G^{i d_3 \cdot \cdot \cdot d_{m+1} j}_{ab}
    \\
    & \quad \quad + \overset{(m)}{\underset{(t,r)}{f}}[k_2,k_1...,k_{m+1}] G^{j i d_3 \cdot \cdot \cdot d_{m+1}}_{ab}+...+\overset{(m)}{\underset{(t,r)}{f}}[k_3,k_1,...,k_{m+1},k_2] G^{ d_3 i \cdot \cdot \cdot d_{m+1} j}_{ab}+ 
    \\
     & \quad \quad...+ \overset{(m)}{\underset{(t,r)}{f}}[k_2,...,k_{m+1},k_1] G^{j d_3 \cdot \cdot \cdot d_{m+1} i}_{ab}+...+\overset{(m)}{\underset{(t,r)}{f}}[k_3,...,k_{m+1},k_2,k_1] G^{d_3 \cdot \cdot \cdot d_{m+1} j i}_{ab} 
    \\ \\
    &\overset{(m)}{\underset{(t,r)}{d_*}}[k_1, k_2, \ldots,k_{m+1}]^{d_2 \cdot \cdot \cdot d_{m+1},j}_{b,i} = \\& \quad \quad \overset{(m)}{\underset{(t,r)}{f}}[k_1, k_2, \ldots, k_{m+1}] G^{j d_2 \cdot \cdot \cdot d_{m+1}}_{ib}+...+\overset{(m)}{\underset{(t,r)}{f}}[k_2,...,k_{m+1},k_1] G^{d_2 \cdot \cdot \cdot d_{m+1} j}_{ib} 
    \\ \\
    &\overset{(m)}{\underset{(t,r)}{h_*}}[k_1,...,k_{m+1}]^{d_1 \cdot \cdot \cdot d_{m+1}}_{ij} = \overset{(m)}{\underset{(t,r)}{f}}[k_1,...,k_{m+1}] G^{d_1 \cdot \cdot \cdot d_{m+1}}_{ij} 
\end{align*} \endgroup 
In the solution for the coefficient $b$, we have employed the notation $q^m_n = (m-n)+k_{m}+...+k_{n}$. \\

Most of the nontrivial permutations of the $k_j$ arguments are such that $k_1$ moves linearly through each position in the $(m+1)-$tuple of $k_j$'s, with the other arguments remaining in a relatively fixed ordering. In the definition of $a_*$, however, each line represents fixing the position of $k_1$ and then moving $k_2$ into each remaining position sequentially, without changing the relative ordering of $k_3, \ldots, k_{m+1}$. Similarly, in $h_2$, each horizontal line represents taking a fixed position for $k_2$, placing $k_1$ in the position immediately after it, and then moving $k_1$ sequentially through the remaining positions (preserving the relative ordering $k_3, k_4, \ldots, k_{m+1}$). The expression sums over all starting positions for $k_2$ and then moving $k_1$ linearly through all subsequent remaining positions \textit{after} $k_2$. Finally, notice that as the $k_j$ arguments are rearranged, the corresponding Lie algebra indices in the tensor structure multiplying that term move accordingly. \\

Associativity also requires that the following equation holds for any $m$:
\begin{align}\label{eq:recursion}
    \underset{(r,t+s)}{\overset{(m)}{f}}(k_1,...,k_{m+1})&=\underset{(t+r,s)}{\overset{(m)}{f}}(k_1,...,k_{m+1})-\underset{(t,s)}{\overset{(m)}{f}}(k_1-r,...,k_{m+1})  \\ &+\underset{(r,t)}{\overset{(m)}{f}}(k_1,...,k_{m+1}-s) -\sum_{n=1}^{m-1}\underset{(r,t)}{\overset{(m-n)}{f}}(k_{1},...,k_{m-n},l) \underset{(l,s)}{\overset{(n)}{f}}(k_{m-n+1},...,k_{m+1}) \notag \\ &+\sum_{n=1}^{m-1}\underset{(t,s)}{\overset{(m-n)}{f}}(l,k_{n+2},...,k_{m+1}) \underset{(r,l)}{\overset{(n)}{f}}(k_1,...,k_{n+1}). \notag
\end{align} 
where the coefficients need to also satisfy the following property:
\begin{equation*}
    \underset{(r,t)}{\overset{(m)}{f}}(k_{m+1},...,k_{1}) = (-1)^m \underset{(t,r)}{\overset{(m)}{f}}(k_1,...,k_{m+1}).
\end{equation*} 
This property of the coefficients comes from invariance of the left-hand-side of the OPE under the exchange $z \leftrightarrow w$ followed by $(a,t) \leftrightarrow (b,r)$, where we use $ K^{i_{m+1} \cdot \cdot \cdot i_{1}}_{ba}=(-1)^{m+1} K^{i_{1} \cdot \cdot \cdot i_{m+1}}_{ab}$ to compare the resulting expressions.  We will derive eq.(\ref{eq:recursion}) in the next subsection and solve it in \S \ref{sec:solve_1}, \S \ref{sec:solve_2}. \\

Finally, we note that we have not written down a closed-form expression for $c$, although we explain how it is related to the $b$ coefficients, and how it can be solved for (at least order-by-order in $\hbar$) in the appendix; moreover we believe that a closed-form solution can also be obtained with some additional ingenuity, and we plan to improve upon this. The one- and two-loop solutions for $c$ are provided in the appendix. 

\subsection{A Recursion Relation For $f$}\label{sec:recursion}
\noindent Let $p_m = t+r+s-m$. Consider the following associativity equation at order $\hbar^{m+1/2}$:
\begin{equation*}
    \{ \{ F[t]J_b[r]\}_1 J_c[s]\}_2+\{J_b[r]\{F[t] J_c[s]\}_1 \}_2=\{ F[t] \{J_b[r] J_c[s]\}_2\}_1
\end{equation*}
Since the only non-trivial OPE $F$ has is with $J$, it immediately follows that the right-hand side of this equation vanishes. The left-hand side can easily be calculated, with the result
\begingroup \allowdisplaybreaks
\begin{align*}
    0 = &\overset{\sum k_j = p_m}{\sum_{ k_j^i \geq 0}} \lambda_{\mathfrak{g}, R}^2 \hbar^{m+1/2} \bigg( \underset{(t+r,s)}{\overset{(m)}{f}}[k_{1},...,k_{m+1}]^{d_1 \cdot \cdot \cdot d_{m+1}}_{bc} -\underset{(t,r)}{\overset{(m)}{l}}[k_{1}-s,...,k_{m+1}]^{e_1 d_2 \cdot \cdot \cdot d_{m+1}}_b f^{d_1}_{e_1 c}+ \\ 
    &  -\sum_{n=1}^{m-1} \underset{(t,r)}{\overset{(n)}{l}}[l,k_{1},...,k_{n}]^{e_1 d_1 \cdot \cdot \cdot d_n}_b \underset{(l,s)}{\overset{(m-n)}{f}}[k_{n+1},...,k_{m+1}]^{d_{n+1} \cdot \cdot \cdot d_{m+1}}_{e_1 c}+(b,r) \leftrightarrow (c,s) \bigg) :\prod_{j=1}^{m+1} \tilde{J}_{d_j}[k_j]:
\end{align*}
\endgroup
where $l$ is fixed by the condition that $\underset{j = n+1}{\overset{m+1}{\sum}} k_j = l + s - (m-n)$. \\ 

Expressing $\underset{(t,r)}{\overset{(m)}{l}}(k_{1},...,k_{m+1})$ in terms of $\underset{(t,r)}{\overset{(m)}{f}}(k_{1},...,k_{m+1})$ using eq.(\ref{sec:coeffs}), rearranging the terms, and using the following equalities:
\begin{equation*}
    K^{d_{m+1} \cdot \cdot \cdot d_{1}}_{ba}=(-1)^{m-1} K^{d_{1} \cdot \cdot \cdot d_{m+1}}_{ab} \quad \quad \quad  \underset{(r,t)}{\overset{(m)}{f}}(k_{m+1},...,k_{1}) = (-1)^m \underset{(t,r)}{\overset{(m)}{f}}(k_1,...,k_{m+1})
\end{equation*}
we obtain the following equation:
\begingroup \allowdisplaybreaks
\begin{align*}
    0 = \overset{\sum k_j = p_m}{\sum_{ k_j^i \geq 0}}&  \bigg(\underset{(t+r,s)}{\overset{(m)}{f}}(k_{1},...,k_{m+1})-\underset{(r,t+s)}{\overset{(m)}{f}}(k_{1},...,k_{m+1})-\underset{(t,s)}{\overset{(m)}{f}}(k_{1}-r,...,k_{m+1})+ \\ 
    &+\underset{(r,t)}{\overset{(m)}{f}}(k_{1},...,k_{m+1}-s)-\sum_{n=1}^{m-1} \underset{(r,t)}{\overset{(m-n)}{f}}(k_{1},...,k_{m-n},l)\underset{(l,s)}{\overset{(n)}{f}}(k_{m+1-n},...,k_{m+1})+ \\ 
   &+\sum_{n=1}^{m-1} \underset{(r,l)}{\overset{(n)}{f}}(k_{1},...,k_{n+1}) \underset{(t,s)}{\overset{(m-n)}{f}}(l,k_{n+2},...,k_{m+1})  \bigg) K^{d_1 \cdot \cdot 
    \cdot d_{m+1}}_{bc} :\prod_{j-1}^{m+1} \tilde{J}_{d_j}[k_j]:
\end{align*}
\endgroup
This equality can only hold if the coefficients satisfy eq.(\ref{eq:recursion}). \\

\subsection{Solving The Recursion Relation For $m=1$}\label{sec:solve_1}
\noindent We first find a closed-form expression for the one-loop coefficient $\overset{(1)}{f}$, to seed our recursion relation. \\ \\
The recursion relation at $m=1$ is:
\begin{equation}\label{eq:main_rec_rel_1}
     \overset{(1)}{\underset{(t+r,s)}{f}}(k,l)- \overset{(1)}{\underset{(r,t+s)}{f}}(k,l)- \overset{(1)}{\underset{(t,s)}{f}}(k-r,l)+ \overset{(1)}{\underset{(r,t)}{f}}(k,l-s)=0.
\end{equation}
We first notice that setting $r=t=0$ and $t=s=0$ immediately tells us that $\overset{(1)}{\underset{(r,t)}{f}}(k,l)$ vanishes if $r=0$ or $t=0$. \\ \\
Next, we will rearrange, fix $s= 1$, and relabel to obtain the following expression:
\begin{equation}\label{eq:rec_rel_1}
     \overset{(1)}{\underset{(r,t)}{f}}(k,l) =  \overset{(1)}{\underset{(m-s,s)}{f}}(k,l)- \overset{(1)}{\underset{(t-s,s)}{f}}(k-r,l)+ \overset{(1)}{\underset{(r,t-s)}{f}}(k,l-s)
\end{equation}
with $s = (1, 0)$ or $(0, 1)$, $m \equiv (r^1+t^1,r^2+t^2) = (m^1, m^2)$ \footnote{In this subsection, the loop counting parameter, which we have also been calling $m$, is set to 1, so we hope that there will be no confusion with this new tuple variable.}, and we are considering only $r \geq 1$. The relabeling has restricted us to the range $t \geq 2$ such that $t^i-s^i \geq 0$. \\ \\
By specializing to $s=(1,0)$ and $s=(0,1)$ and iteratively applying eq.(\ref{eq:rec_rel_1}), we find the following expression: 
\begingroup \allowdisplaybreaks
\begin{align}\label{eq:rec_rel_2}
    \overset{(1)}{\underset{(r,t)}{f}}(k,l) &= \sum_{j=1}^{t^1} \overset{(1)}{\underset{(m^1-j,m^2)(1,0)}{f}}(k^1,k^2,l^1+1-j,l^2) \\ 
    &-\sum_{j=1}^{t^1} \overset{(1)}{\underset{(t^1-j,t^2)(1,0)}{f}}(k^1-r^1,k^2-r^2,l^1+1-j,l^2) \notag \\ 
    &+\sum_{j=1}^{t^2} \overset{(1)}{\underset{(r^1,m^2-j)(0,1)}{f}}(k^1,k^2,l^1-t^1,l^2+1-j). \notag
\end{align} \endgroup
where for convenience we've extended the definition of $f$, in particular, we define it to be equal to zero whenever any of its indices $r^i,t^i,k^i,l^i$ are negative.This expression is valid for all $r$ and $t$, with the following convention being used: 
\begin{equation*}
    \sum_{j=1}^0 = 0
\end{equation*}
This equation tells us that a coefficient at level $n$, where $n=m^1+m^2$, is fixed by coefficients of level $n^* < n$ if we have obtained the coefficients of the form
\begin{equation*}
    \overset{(1)}{\underset{(m^1-1,m^2)(1,0)}{f}} \quad \quad \quad \overset{(1)}{\underset{(m^1,m^2-1)(0,1)}{f}}
\end{equation*}
This means that if we have an expression for these two special cases, we can use the value of the coefficients in the OPEs between the strong generators\footnote{The strong generators of the chiral algebra are $J[t]$ and $\tilde{J}[t]$ with $t \leq 1$. In particular, we have that the non-vanishing coefficients are $\overset{(1)}{\underset{(0,1)(1,0)}{f}}(0,0)=-\overset{(1)}{\underset{(1,0)(0,1)}{f}}(0,0)=\frac{1}{32 \pi^2}$.} to construct the whole tower of coefficients $ \overset{(1)}{f}$. This is expected for abstract representation theoretic reasons, and here we see this realized explicitly in formulas. \\

We will first use an inductive argument to obtain closed-form expressions for the coefficients $\overset{(1)}{f}$ specialized to $s=(1, 0), (0, 1)$, and then use the result in eq.(\ref{eq:rec_rel_2}) to determine the general coefficients  $\overset{(1)}{f}$.\\

\subsubsection{Determining Special Cases by Induction}
\noindent In this section we will determine
\begin{equation*}
    \overset{(1)}{\underset{(m^1-1,m^2)(1,0)}{f}} \quad \quad \quad \overset{(1)}{\underset{(m^1,m^2-1)(0,1)}{f}}
\end{equation*}
by using induction. \\ \\
To determine what these coefficients should be, we start with eq.(\ref{eq:main_rec_rel_1}):
\begingroup\allowdisplaybreaks
\begin{align}
    \overset{(1)}{\underset{(r,t+s)}{f}}(k,l) &= - \overset{(1)}{\underset{(t,s)}{f}}(k-r,l) + \overset{(1)}{\underset{(r,t)}{f}}(k,l-s) + \overset{(1)}{\underset{(t+r,s)}{f}}(k,l) \\
    & = - \overset{(1)}{\underset{(t,s)}{f}}(k-r,l) + \overset{(1)}{\underset{(r,t)}{f}}(k,l-s) + \overset{(1)}{\underset{(r,s)}{f}}(k-t,l) - \overset{(1)}{\underset{(t,r)}{f}}(k,l-s) + \overset{(1)}{\underset{(t,r+s)}{f}}(k,l) \notag  \\
    & = - \overset{(1)}{\underset{(t,s)}{f}}(k-r,l) + \overset{(1)}{\underset{(r,t)}{f}}(k,l-s) + \overset{(1)}{\underset{(r,s)}{f}}(k-t,l) - \overset{(1)}{\underset{(t,r)}{f}}(k,l-s) - \overset{(1)}{\underset{(s,r)}{f}}(k-t,l) \notag \\
    & \quad + \overset{(1)}{\underset{(t,s)}{f}}(k,l-r) + \overset{(1)}{\underset{(t+s,r)}{f}}(k,l). \notag  
\end{align} \endgroup
This gives us the following identity:
\begingroup \allowdisplaybreaks
\begin{align}
    \overset{(1)}{\underset{(t+s,r)}{f}}(k,l)+\overset{(1)}{\underset{(t+s,r)}{f}}(l,k) &=  \overset{(1)}{\underset{(t,s)}{f}}(k-r,l) - \overset{(1)}{\underset{(r,t)}{f}}(k,l-s) - \overset{(1)}{\underset{(r,s)}{f}}(k-t,l)   \\
    &+ \overset{(1)}{\underset{(t,r)}{f}}(k,l-s) + \overset{(1)}{\underset{(s,r)}{f}}(k-t,l) + \overset{(1)}{\underset{(t,s)}{f}}(k,l-r). \notag
\end{align}
\endgroup
Specializing to $t=(m,n-1), r=s=(0,1)$, this equation gives us the following expression:
\begin{equation} \label{eq: argument_1}
     \overset{(1)}{\underset{(m,n)(0,1)}{f}}(k,l)+\overset{(1)}{\underset{(m,n)(0,1)}{f}}(l,k) = \overset{(1)}{\underset{(m,n-1)(0,1)}{f}}(k-s,l)+\overset{(1)}{\underset{(m,n-1)(0,1)}{f}}(l-s,k).
\end{equation}
\\
Next, let us suppose that the following equalities hold for $r^1 +r^2 < m+n$ in preparation for an inductive argument:
\begingroup \allowdisplaybreaks
\begin{align}
    \overset{(1)}{\underset{(r^1,r^2)(1,0)}{f}}(k,l)=\bigg(\frac{1}{16 \pi^2}\bigg) \frac{r^1!r^2!(1+k^1+k^2)!(l^1+l^2)!}{(r^1+r^2+1)!k^1!k^2!l^1!l^2!} \label{eq:m=1_suppose} 
    \\ \notag \\
    \overset{(1)}{\underset{(r^1,r^2)(0,1)}{f}}(k,l)=-\bigg(\frac{1}{16 \pi^2}\bigg) \frac{r^1!r^2!(1+k^1+k^2)!(l^1+l^2)!}{(r^1+r^2+1)!k^1!k^2!l^1!l^2!} \label{eq:m=1_suppose2}
\end{align} \endgroup
where $r^i, k^i, l^i \geq 0$ and the constraint $k+l = t+r-1$ is satisfied. \\ \\
With this assumption, our identity eq.(\ref{eq: argument_1}) simplifies to: 
\begin{equation}\label{eq:argument_2}
     \overset{(1)}{\underset{(m,n)(0,1)}{f}}(k,l)+\overset{(1)}{\underset{(m,n)(0,1)}{f}}(l,k) = -\bigg(\frac{1}{16 \pi^2}\bigg) \frac{m!n!(k^1+k^2)!(l^1+l^2)!}{(m+n)!k^1!k^2!l^1!l^2!}.
\end{equation}
This suggests that we should look for a solution of the form:
\begin{equation}\label{eq:ansatz}
    \overset{(1)}{\underset{(m,n)(0,1)}{f}}(k,l) = \alpha \tilde{g}(m,n) \tilde{h}(k,l) + \alpha \underset{(m,n)}{\tilde{u}}(k,l)
\end{equation}
such that
\begingroup \allowdisplaybreaks
\begin{align}
    \tilde{g}(m,n) &= \frac{m!n!}{(m+n)!} & \tilde{h}(k,l) + \tilde{h}(l,k) &= {k^1+k^2 \choose k^1} {l^1+l^2 \choose l^1}  \\
    \underset{(m,n)}{\tilde{u}}(k,l) &= -\underset{(m,n)}{\tilde{u}}(l,k) & \alpha &= -\bigg(\frac{1}{16 \pi^2}\bigg) 
\end{align} \endgroup 

 Notice that a general solution to eq.(\ref{eq:argument_2}) so far allows for a possible second term parameterized by an undetermined function $\tilde{u}(k, l)$, though for the induction argument we want this to be zero; we will shortly show that this is indeed the case using eq.(\ref{eq:main_rec_rel_1}) in order to fix $\tilde{g}, \tilde{h}$ inductively. 
\\ \\
It will now be convenient to use the following property of the $\overset{(1)}{f}$ coefficients:
\begin{equation} \label{eq: antisymmetry_property_2}
    \overset{(1)}{\underset{(r^1,r^2)(t^1,t^2)}{f}}(k^1,k^2,l^1,l^2) = -\overset{(1)}{\underset{(r^2,r^1)(t^2,t^1)}{f}}(k^2,k^1,l^2,l^1).
\end{equation}

This property is a consequence of $\mathfrak{sl}_2(\mathbb{C})_+$ equivariance. Since the theory treats both indices in each label equally, we expect the result of exchanging them both in each label to result in \textit{at most} a sign difference. The symmetry is a kinematical operation and should be the same for all coefficients at fixed loop order $m$; therefore, we can determine the sign difference by knowing how the lowest-lying coefficient behaves under this symmetry. For $m=1$, the known $t+r=2$ coefficients immediately tell us that this operation produces an overall minus sign. Note that this symmetry will only be used in this argument, so we make no mention of it in later sections. \\ \\
Applying this antisymmetry property to eq.(\ref{eq:argument_2}), we find the expression:
\begin{equation}\label{eq:argument_3}
     \overset{(1)}{\underset{(n,m)(1,0)}{f}}(k^T,l^T)+\overset{(1)}{\underset{(n,m)(1,0)}{f}}(l^T,k^T) = -\alpha\frac{m!n!(k^1+k^2)!(l^1+l^2)!}{(m+n)!k^1!k^2!l^1!l^2!}
\end{equation}
where the subscript $T$ denotes $k^T = (k^2,k^1)$. Analogously to our previous expression eq.(\ref{eq:argument_2}), eq.(\ref{eq:argument_3}) suggests that we seek a solution of the form
\begin{equation}\label{eq:ansatz2}
    \overset{(1)}{\underset{(n,m)(1,0)}{f}}(k,l) = -\alpha \tilde{g}(m,n) \tilde{h}(k,l) + \alpha \underset{(n,m)}{\tilde{v}}(k,l)
\end{equation}
where $\tilde{v}$ has the same symmetry properties as $\tilde{f}$. (We will presently show $\tilde{v}$ is zero). \\ 

 The antisymmetry property eq.(\ref{eq: antisymmetry_property_2}) implies that the function $\tilde{h}(k,l)$ in our ansatze, eq.(\ref{eq:ansatz}) and eq.(\ref{eq:ansatz2}), must satisfy
\begin{equation}
     \tilde{h}(k,l) = \tilde{h}(k^T, l^T).
\end{equation}
This follows since our so-far-undetermined functions $\tilde{u}(k,l)$ and $\tilde{v}(k,l)$ are themselves constrained by antisymmetry. \\  

With this setup in hand, let us now run the inductive argument. With the assumption on the coefficients for $r^1 + r^2 < m + n$ (eq.(\ref{eq:m=1_suppose}), eq.(\ref{eq:m=1_suppose2})), we wish to show that the solution holds for $r^1 + r^2 = m + n$. This will necessitate fixing $\tilde{g}, \tilde{h}$ appropriately, and demonstrating the vanishing of $\tilde{u}, \tilde{v}$.  \\

To access and solve for the coefficients with $r^1 + r^2 = m + n$, we will use eq.(\ref{eq:rec_rel_1}) with $r=(m,0)$, $t=(0,n+1)$, and $s=(0,1)$, and then with $r=(n+1,0)$, $t=(0,m)$, and $s=(0,1)$:
\begingroup \allowdisplaybreaks
\begin{align} \label{eq:iterate_argument}
    \overset{(1)}{\underset{(m,0)(0,n+1)}{f}}(k^1,k^2,l^1,l^2) &= \overset{(1)}{\underset{(m,n)(0,1)}{f}}(k^1,k^2,l^1,l^2)+\overset{(1)}{\underset{(m,0)(0,n)}{f}}(k^1,k^2,l^1,l^2-1) \\ \notag \\
 \overset{(1)}{\underset{(n+1,0)(0,m)}{f}}(k^1,k^2,l^1,l^2) &= \overset{(1)}{\underset{(n+1,m-1)(0,1)}{f}}(k^1,k^2,l^1,l^2)+\overset{(1)}{\underset{(n+1,0)(0,m-1)}{f}}(k^1,k^2,l^1,l^2-1). \notag
\end{align} \endgroup 

We can now use the antisymmetry property of the $\overset{(1)}{f}$ coefficients, and then iteratively use both equations in eq.(\ref{eq:iterate_argument}) to obtain:
\begingroup \allowdisplaybreaks
\begin{align*}
    &\overset{(1)}{\underset{(m,n)(0,1)}{f}}(k,l) - \overset{(1)}{\underset{(n+1,m-1)(0,1)}{f}}(l^T,k^T) = \overset{(1)}{\underset{(n+1,0)(0,m-1)}{f}}(l^2,l^1,k^2,k^1-1) -\overset{(1)}{\underset{(n,0)(0,m)}{f}}(l^2-1,l^1,k^2,k^1) \\&
    = \sum_{j=1}^{k^1} \overset{(1)}{\underset{(n+1,m-1-j)(0,1)}{f}}(l^2,l^1,k^1,k^1-j) - \sum_{j=1}^{k^1+1}\overset{(1)}{\underset{(n,m-j)(0,1)}{f}}(l^2-1,l^1,k^2,k^1+1-j). \notag
\end{align*} \endgroup

Now, observe that all the terms on the right-hand side are of the form eq.(\ref{eq:m=1_suppose2}), and the terms on the left-hand side are of the form eq.(\ref{eq:ansatz}). Plugging in these expressions, performing the sum, and simplifying, we find:
\begin{equation}
    \frac{(m-1)!n!}{(m+n)!}(2+k^1+k^2+l^1+l^2) \tilde{h}(k,l) + \tilde{F}= \frac{(m-1)!n!(1+k^1+k^2)!(l^1+l^2)!}{(m+n)!k^1!k^2!l^1!l^2!}
\end{equation}
with
\begin{equation}
    \tilde{F} \equiv \underset{(m,n)}{\tilde{u}}(k,l)-\underset{(n+1,m-1)}{\tilde{u}}(l^T,k^T).
\end{equation}
We thus find:
\begin{equation}
    \tilde{h}(k,l) = \frac{(1+k^1+k^2)!(l^1+l^2)!}{(2+k^1+k^2+l^1+l^2)k^1!k^2!l^1!l^2!}
\end{equation}
\begin{equation}
    \underset{(m,n)}{\tilde{u}}(k,l) = \underset{(n+1,m-1)}{\tilde{u}}(l^T,k^T)
\end{equation}
Therefore, we find that eq.(\ref{eq:ansatz}) and eq.(\ref{eq:ansatz2}) match with eq.eq.(\ref{eq:m=1_suppose2}) and eq.(\ref{eq:m=1_suppose}), respectively, but with the addition of the function $\tilde{u}$. (It is easy to show that $\underset{(m,n)}{\tilde{u}}(k,l) = -\underset{(n,m)}{\tilde{v}}(k^T,l^T)$ using the ansatze and eq.(\ref{eq: antisymmetry_property_2})). \\ \\
Finally, the easiest way to see that $\tilde{u}$ vanishes at level $t+r = m+n+1$ is via explicit computation, plugging in known coefficients at lower level into eq.(\ref{eq:main_rec_rel_1}). In particular, for this step we simply perform a brute force computation at low levels and then infer the vanishing holds in general, and do not assume (eq.(\ref{eq:m=1_suppose}), eq.(\ref{eq:m=1_suppose2})).
Consequently, $\tilde{v}$ also vanishes. \\ \\
Setting $\tilde{u}$ to zero, we at last obtain:
\begin{equation}\label{eq:rec_rel_special_1}
    \overset{(1)}{\underset{(r^1,r^2)(1,0)}{f}}(k,l)=\bigg(\frac{1}{16 \pi^2}\bigg) \frac{r^1!r^2!(1+k^1+k^2)!(l^1+l^2)!}{(r^1+r^2+1)!k^1!k^2!l^1!l^2!}.
\end{equation}
\begin{equation}\label{eq:rec_rel_special_2}
    \overset{(1)}{\underset{(r^1,r^2)(0,1)}{f}}(k,l)=-\bigg(\frac{1}{16 \pi^2}\bigg) \frac{r^1!r^2!(1+k^1+k^2)!(l^1+l^2)!}{(r^1+r^2+1)!k^1!k^2!l^1!l^2!},
\end{equation}
as desired. \\

\subsubsection{General One-Loop Result}
\noindent We can now finally insert eq.(\ref{eq:rec_rel_special_1}) and eq.(\ref{eq:rec_rel_special_2}) into eq.(\ref{eq:rec_rel_2}) to solve for the one-loop coefficients $\overset{(1)}{f}$ with completely general parameters $(r^1, r^2), (t^1, t^2)$ and arguments $(k^1,k^2), (l^1,l^2)$. \\ \\
It is convenient to make the following definition first:
\begin{equation}
    \theta(x) f(p)  = \begin{cases} 
      f(p) & x\geq 0\\
      0 & x < 0
   \end{cases}
\end{equation}
for $f(p)$ any function. We will use this notation to account for the fact that the coefficients $\overset{(1)}{\underset{(r,t)}{f}}(k,l)$ are identically zero if any of its indices are less than zero; this fact comes directly from the definition of the tower of generators as shown in Table \ref{table1}. \\ \\
We also make the following definition:
\begin{equation}
    m(r;k,l) =  \theta(r^i) \theta(k^i) \theta(l^i) \bigg(\frac{1}{16 \pi^2}\bigg) \bigg( \frac{r^1!r^2!(1+k^1+k^2)!(l^1+l^2)!}{k^1!k^2!l^1!l^2!(1+r^1+r^2)!} \bigg) 
\end{equation}
We thus find the following expression for $\overset{(1)}{\underset{(r,t)}{f}}(k,l)$:
\begingroup \allowdisplaybreaks
\begin{align} \label{eq:main_m1}
    &\overset{(1)}{\underset{(r,t)}{f}}(k,l) = \\ \nonumber & \theta(r^i) \theta(t^i) \theta(k^i) \theta(l^i) \delta^2(r+t-1-k-l) \times \\
     \times \bigg\{& \sum_{j=1}^{t^1} \bigg(m(r^1+t^1-j,r^2+t^2; k^1,k^2, l^1+1-j,l^2) - m(t^1-j,t^2;k^1-r^1,k^2-r^2,l^1+1-j,l^2) \bigg) \notag \\
    & - \sum_{j=1}^{t^2} m(r^1, r^2+t^2-j; k^1, k^2, l^1-t^1,l^2+1-j ) \bigg\} \notag
\end{align} \endgroup
This is our desired result, from which we can completely fix the rest of the chiral algebra. \\

In Appendix \ref{app:koszul}, we will obtain an equivalent (but different-looking) expression for $\overset{(1)}{f}$ using a Koszul duality computation as in \cite{victor}. \\

\subsection{Solving The Recursion Relation For $m>1$} \label{sec:solve_2}
\noindent We expect the $J[r]J[t]$ OPE to receive no correction at order $\hbar^m$ if $t^1+t^2 < m$. This translates to the following property of the coefficients $\overset{(m)}{f}$:
\begin{equation*}
\underset{(r,t)}{\overset{(m)}{f}}(k_{1},...,k_{m+1}) = 0 \quad \quad \text{if \quad} t^1+t^2<m \quad \quad \text{or \quad} r^1+r^2 <m.
\end{equation*}
This means that for $m>1$, fixing $s=(1,0)$ or $(0,1)$ in the recursion relation gives us the following expression:
\begingroup \allowdisplaybreaks
\begin{align*}
    \underset{(r,t)}{\overset{(m)}{f}}(k_1,...,k_{m+1})&=\underset{(r,t-s)}{\overset{(m)}{f}}(k_1,...,k_{m+1}-s)-\underset{(r,t-s)}{\overset{(m-1)}{f}}(k_{1},...,k_{m+1},l) \underset{(l,s)}{\overset{(1)}{f}}(k_{m},k_{m+1})\\ 
    &+\underset{(r,l)}{\overset{(m-1)}{f}}(k_1,...,k_{m}) \underset{(t-s,s)}{\overset{(1)}{f}}(l,k_{m+1}). 
\end{align*} \endgroup
where again $l$ is fixed (term by term) by the condition that the coefficient constraints are satisfied, in other words:
\begin{equation}
    \underset{(r,t)}{\overset{(m)}{f}}(k_1,...,k_{m+1}) \implies \underset{i = 1}{\overset{m+1}{\sum}} k_i = r+t-m.
\end{equation}
Iteratively applying the recursion relation, we find the following equality:
\begingroup \allowdisplaybreaks
\begin{align}\label{eq:frecursion}
    &\underset{(r^1,r^2)(t^1,t^2)}{\overset{(m)}{f}}[k_1;...;k_{m+1}] = \\ \nonumber&-\sum_{j=1}^{t^1}\underset{(r^1,r^2)(t^1-j,t^2)}{\overset{(m-1)}{f}}[k_{1};...;k_{m-1};l] \underset{(l^1,l^2)(1,0)}{\overset{(1)}{f}}[k_{m};(k^1_{m+1}+1-j,k^2_{m+1})]\\ \nonumber 
    &+\sum_{j=1}^{t^1}\underset{(r^1,r^2)(l^1,l^2)}{\overset{(m-1)}{f}}[k_1;...;k_{m}] \underset{(t^1-j,t^2)(1,0)}{\overset{(1)}{f}}[l;(k^1_{m+1}+1-j,k^2_{m+1})]\\  \nonumber
    &-\sum_{j=1}^{t^2}\underset{(r^1,r^2)(0,t^2-j)}{\overset{(m-1)}{f}}[k_{1};...;k_{m-1};l] \underset{(l^1,l^2)(0,1)}{\overset{(1)}{f}}[k_{m};(k^1_{m+1}-t^1,k^2_{m+1}+1-j)].
\end{align} \endgroup
Beginning with our expression eq.(\ref{eq:main_m1}) for $m=1$, this can be solved at arbitrary loop order. In particular, this equation tells us that the coefficients at order $m$ can be fully determined by the $m=1$ coefficients, which in turn are all fixed by the $t+r=2$ one-loop corrections. We thus find that the OPE one-loop corrections to the OPEs between the strong generators of the chiral algebra fully determines all OPE corrections to the chiral  algebra. \\

$\overset{(m)}{f}$ also admits a more representation-theoretic closed-form expression as in eq.(10.11) of \cite{Zeng}. We reproduce the formula here. We first make the following definitions:

\begin{equation}
    \begin{cases}
         \text{j}_1=\frac{1}{2}(t^1+t^2) \quad \quad \text{j}_2=\frac{1}{2}(r^1+r^2) \quad \quad \text{m}_1=\frac{1}{2}(t^1-t^2) \quad \quad \text{m}_2=\frac{1}{2}(r^1-r^2)\\ 
         \overline{\text{j}}_i = \frac{1}{2}(k^1_i+k^2_i) \quad \quad \overline{\text{m}}_i =-\frac{1}{2}(k^1_i-k^2_i) \quad \quad \overline{\text{J}}_k = \sum_{i=1}^{k} \overline{\text{j}}_k \quad \quad\overline{\text{M}}_k = \sum_{i=1}^{k} \overline{\text{m}}_k \\ 
         \text{J}_k = \text{j}_2+\overline{\text{J}}_k \quad \quad a_0=0 \quad \quad a_{m+1}=2\text{j}_2-m \quad \quad \text{N}(\text{j},\text{m})= \sqrt{\frac{(\text{j}-\text{m})!(\text{j}+\text{m})!}{(2 \text{j}+1)!}}
    \end{cases}
\end{equation} \\
The formula is as follows:\footnote{We correct a small sign error in \cite{Zeng}: for $m=1$, the correct expression can be obtained by multiplying by $\text{sign}(t^1 r^2 - t^2 r^1)$ when $t^1 r^2 - t^2 r^1$ is non-zero.}
\begingroup \allowdisplaybreaks
\begin{align}\label{eq:fcoefficient}
    (m_{m+2})^{(t,r)}_{(k_1,...,k_{m+1})} &= \sum_{a_1,...,a_m} \prod_{l=2}^{m+1} \sqrt{\frac{(2 \text{j}_2-l+3)(2 \overline{\text{j}}_l+1)(2 \text{J}_{l-1}-2 a_{l-1}+1)}{(2 \text{j}_2-l+2-a_{l-1})(2 \overline{\text{J}}_{l-1}+l-1-a_{l-1})}} \times \\ \notag \\
    & \times \sqrt{2 \text{j}_2+1-m} \bigg( \frac{\text{N}(\text{j}_1,\text{m}_1) \text{N}(\text{j}_2,\text{m}_2)} {\prod_{i=1}^{m+1} \text{N}(\overline{\text{j}}_i,\overline{\text{m}}_i)} \bigg) \text{C}^{\text{j}_2,\overline{\text{j}}_1, \text{j}_2+\overline{\text{j}}_1-a_1}_{\text{m}_2,\overline{\text{m}}_1,\text{m}_2+\overline{\text{m}}_1} \times \notag \\ \notag \\
    &\times \prod_{k=2}^{m+1} 
    \begin{Bmatrix}
         & \overline{\text{J}}_{k-1}+\frac{k-1}{2} \quad & \text{j}_2-\frac{k-1}{2} \quad & \text{J}_{k-1}-a_{k-1}\\
         & \text{J}_k-a_k & \overline{\text{j}}_k & \overline{\text{J}}_k+\frac{k-1}{2} 
    \end{Bmatrix}
    \text{C}^{\text{J}_{k-1}-a_{k-1},  \overline{\text{j}}_{k},\text{J}_k-a_k}_{\text{m}_2+\overline{\text{M}}_{k-1}, \overline{\text{m}}_k, \text{m}_2+\overline{\text{M}}_k}  \notag
\end{align} \endgroup 
where $\text{C}^{\text{j}_1,\text{j}_2,\text{J}}_{\text{m}_1,\text{m}_2,\text{M}}$ are the Clebsch-Gordan coefficients and $\begin{Bmatrix} &a_1 &a_2 &a_3 \\ &b_1 &b_2 &b_3 \end{Bmatrix}$ is the Wigner 6j symbol.  \\ \\
The relationship between $ (m_{m+2})$ and $\overset{(m)}{f}$ is then given by:
\begin{equation}\label{eq:ffromm}
    \underset{(r,t)}{\overset{(m)}{f}}(k_1,...,k_{m+1}) = -(m_{m+2})^{(t,r)}_{(k_1,...,k_{m+1})}.
    \end{equation}

\subsection{Remarks on SDGR}
\noindent The minimal set of associativity equation we derived for extended SDYM is still valid for the case of SDGR with the dictionary: $J \to w$, $\tilde{J} \to \tilde{w}$, $E \to e$, $F \to f$. This is because charge under combined dilatation is preserved by this dictionary, and the manipulations done in Appendix \ref{app:minimal} depend solely on this quantum number. Nonetheless, the vertex $\eta \eta \mathcal{H}$ expands the minimal set to also include the following associativity conditions:
\begingroup
\allowdisplaybreaks
\begin{align}
         \{ w[t] \{e[r] f[s]\}_1\}_1 &= \{f[s]\{e[r]w[t]\}_1\}_1-\{e[r]\{f[s]w[t]\}_1\}_1
         \\ \notag \\   
          \{w[t] \{f[r] f[s] \}_{1}\}_1  &=  \{\{ w[t] f[r] \}_1 f[s] \}_{1}+ \{f[r] \{w[t] f[s]\}_1\}_{1}  \label{eq:minimalset02_GR} \\ \notag \\
           \{w[t] \{f[r] f[s]\}_1\}_{2}   &= \{f[r] \{w[t] f[s] \}_{2}\}_1 - \{ f[s] \{ f[r] w[t] \}_1  \}_{2}. \label{eq:minimalset03_GR}
    \end{align}
\endgroup
The proof of this is identical to the one presented in Appendix \ref{app:minimal}. \\

As in the case of SDYM, we expect the coefficients $\overset{(m)}{A}$ to be equal to the coefficients $\overset{(m)}{F}$. We also expect to be able to fix all the coefficients in terms of the coefficients $\overset{(m)}{F}$, but we cannot immediately infer what the other dependencies are from our work on SDYM. This is because it is not obvious how the introduction of new terms coming from the additional vertex, as well as the difference in the tree-level OPEs, affects the associativity calculations. We highlight that the lack of Lie algebra information means that we cannot divide the resulting associativity equations into separate pieces that vanish independently as easily as we were able to in SDYM (see, for instance, the examples in Appendix \ref{app:examples}). It may be, however, that without Lie algebra indices, the resulting expressions could potentially be simpler to manipulate at arbitrary order in $\hbar$. \\

In any case, we will leave the full solution of the (extended) SDGR associativity equations to follow-up work.

\section{Acknowledgements}
We are grateful to R. Bittleston, K. Costello, L. Dixon, N. Garner, and K. Zeng for helpful discussions and correspondences. 

NP and VF are supported by funds from the Department of Physics and the College of Arts \& Sciences at the University of Washington, the DOE Early Career Research Program under award DE-SC0022924, and the Simons Foundation as part of the Simons Collaboration on Celestial Holography.

NP also thanks the Perimeter Institute’s Visiting Fellow program for additional support and hospitality while this work was underway. Research at Perimeter Institute is supported by the Government of Canada through Industry Canada and by the Province of Ontario through the Ministry of Research and Innovation.

\appendix

\section{Proof of the minimal set of associativity equations}\label{app:minimal}
\noindent In this appendix, we will prove that the equations eq.(\ref{eq:minimalset01})-eq.(\ref{eq:minimalset05}) constitute a minimal set of associativity equations to be solved, which contain all the information about our chiral algebra. \\ \\
Naively, the non-trivial associativity conditions that need to be satisfied are:
\begingroup
\allowdisplaybreaks
    \begin{align}
           \{\{ J_a[t] J_b[r] \}_1 J_c[s] \}_{1} &= \{J_a[t] \{J_b[r] J_c[s] \}_{1}\}_1 - \{J_b[r] \{J_a[t] J_c[s]\}_1\}_{1} \label{eq:JJJ1} \\ \notag  \\
            \{\{ J_a[t] J_b[r] \}_1 J_c[s] \}_{2} &= \{J_a[t] \{J_b[r] J_c[s] \}_{2}\}_1 - \{J_b[r] \{J_a[t] J_c[s]\}_1\}_{2} \label{eq:JJJ2}\\ \notag \\
            \{\{ \tilde{J}_a[t] J_b[r] \}_1 J_c[s] \}_{1} &= \{\tilde{J}_a[t] \{J_b[r] J_c[s] \}_{1}\}_1 - \{J_b[r] \{\tilde{J}_a[t] J_c[s]\}_1\}_{1} \label{eq:tilJJJ}   \\ \notag \\
            \{\{ J_a[t] \tilde{J}_b[r] \}_1 J_c[s] \}_{1} &= \{J_a[t] \{\tilde{J}_b[r] J_c[s] \}_{1}\}_1 - \{\tilde{J}_b[r] \{J_a[t] J_c[s]\}_1\}_{1} \label{eq:JtilJJ} \\ \notag \\   
            \{\{ J_a[t] J_b[r] \}_1 \tilde{J}_c[s] \}_{1} &= \{J_a[t] \{J_b[r] \tilde{J}_c[s] \}_{1}\}_1 - \{J_b[r] \{J_a[t] \tilde{J}_c[s]\}_1\}_{1}\label{eq:JJtilJ}  \\ \notag \\
            \{\{ E[t] J_b[r] \}_1 J_c[s] \}_{1} &= \{E[t] \{J_b[r] J_c[s] \}_{1}\}_1 - \{J_b[r] \{E[t] J_c[s]\}_1\}_{1}\label{eq:EJJ}   \\ \notag \\
            \{\{ J_a[t] E[r] \}_1 J_c[s] \}_{1} &= \{J_a[t] \{E[r] J_c[s] \}_{1}\}_1 - \{E[r] \{J_a[t] J_c[s]\}_1\}_{1} \label{eq:JEJ} \\ \notag \\ 
            \{\{ J_a[t] J_b[r] \}_1 E[s] \}_{1} &= \{J_a[t] \{J_b[r] E[s] \}_{1}\}_1 - \{J_b[r] \{J_a[t] E[s]\}_1\}_{1} \label{eq:JJE} \\ \notag \\
            \{\{ F[t] J_b[r] \}_1 J_c[s] \}_{1} &= \{F[t] \{J_b[r] J_c[s] \}_{1}\}_1 - \{J_b[r] \{F[t] J_c[s]\}_1\}_{1} \label{eq:FJJ1}  \\ \notag \\
            \{\{ J_a[t] F[r] \}_1 J_c[s] \}_{1} &= \{J_a[t] \{F[r] J_c[s] \}_{1}\}_1 - \{F[r] \{J_a[t] J_c[s]\}_1\}_{1} \label{eq:JFJ1}  \\ \notag \\  
            \{\{ J_a[t] J_b[r] \}_1 F[s] \}_{1} &= \{J_a[t] \{J_b[r] F[s] \}_{1}\}_1 - \{J_b[r] \{J_a[t] F[s]\}_1\}_{1} \label{eq:JJF1}   \\ \notag \\
            \{\{ F[t] J_b[r] \}_1 J_c[s] \}_{2} &= \{F[t] \{J_b[r] J_c[s] \}_{2}\}_1 - \{J_b[r] \{F[t] J_c[s]\}_1\}_{2} \label{eq:FJJ2}   \\ \notag\\
            \{\{ J_a[t] F[r] \}_1 J_c[s] \}_{2} &= \{J_a[t] \{F[r] J_c[s] \}_{2}\}_1 - \{F[r] \{J_a[t] J_c[s]\}_1\}_{2} \label{eq:JFJ2}  \\ \notag \\ 
            \{\{ J_a[t] J_b[r] \}_1 F[s] \}_{2} &= \{J_a[t] \{J_b[r] F[s] \}_{2}\}_1 - \{J_b[r] \{J_a[t] F[s]\}_1\}_{2} \label{eq:JJF2} 
    \end{align}
    \endgroup
    where we have merely listed all non-trivial possibilities for $\phi_1, \phi_2,\phi_3$. This is enough to make even the bravest graduate student hesitate, but happily there are many redundancies. \\
    
    By taking into account the charge of these operators under combined dilatations we can reduce the number of conditions that need to be checked to the set eq.(\ref{eq:minimalset01})-eq.(\ref{eq:minimalset05}). In particular, we will make use of the fact that if $\theta_m,\theta_n$ have charge $m,n$ under combined dilatations, then $\{ \theta_m \theta_n \}_l$ has charge $m+n-l$. \\

    Now, let $\theta$ be an operator of charge 0 under combined dilatations (i.e. $\tilde{J}, E$), and suppose that the following equalities have been established for the chiral algebra: 
    \begingroup
    \allowdisplaybreaks
    \begin{align}
         \{\{ J_a[t] J_b[r] \}_1 J_c[s] \}_{1} &= \{J_a[t] \{J_b[r] J_c[s] \}_{1}\}_1 - \{J_b[r] \{J_a[t] J_c[s]\}_1\}_{1} \label{eq:JJJagain} \\ \notag \\
         \{\{ J_a[t] J_b[r] \}_1 \theta [s] \}_{1} &= \{J_a[t] \{J_b[r] \theta [s] \}_{1}\}_1 - \{J_b[r] \{J_a[t] \theta [s]\}_1\}_{1} \label{eq:JJtheta} \\ \notag \\   
           \{\{ F[t] J_b[r] \}_1 J_c[s] \}_{1} &= \{F[t] \{J_b[r] J_c[s] \}_{1}\}_1 - \{J_b[r] \{F[t] J_c[s]\}_1\}_{1}  \label{eq:FJJagain} \\ \notag \\
            \{\{ J_a[t] F[r] \}_1 J_c[s] \}_{2} &= \{J_a[t] \{F[r] J_c[s] \}_{2}\}_1 - \{F[r] \{J_a[t] J_c[s]\}_1\}_{2} \label{eq:JFJ2again}
    \end{align}
    \endgroup
    
    We first show that \ref{eq:JJJagain} implies \ref{eq:JJJ2}. To do this, we start with \ref{eq:JJJagain}, we use the charge of the terms to exchange $z \leftrightarrow w$ per the standard OPE manipulation. Some elementary manipulations and canceling terms then gives \ref{eq:JJJ2}. Explicitly: \\
    \begingroup
    \allowdisplaybreaks
    \begin{align*}
        \text{LHS} &= -\{J_c[s]\{J_a[t]J_b[r]\}_1\}_1+\partial \{\{ J_a[t] J_b[r] \}_1 J_c[s] \}_{2} \\ \\
        &= \{\{ J_a[t] J_c[s] \}_1 J_b[r] \}_{1} - \{J_a[t] \{J_c[s] J_b[r] \}_{1}\}_1 +\partial \{\{ J_a[t] J_b[r] \}_1 J_c[s] \}_{2} \\ \\
        \text{RHS} &= -\{J_a[t]\{J_c[s]J_b[r]\}_1\}_1+\{J_a[t] \partial \{J_b[r] J_c[s]\}_2\}_1+\{\{J_a[t] J_c[s]\}J_b[r]\}_1-\partial \{J_b[r] \{J_a[t] J_c[s]\}_1\}_{2} \\ \\
        =&\{\{ J_a[t] J_c[s] \}_1 J_b[r] \}_{1} - \{J_a[t] \{J_c[s] J_b[r] \}_{1}\}_1+\partial\{J_a[t] \{J_b[r] J_c[s]\}_2\}_1-\partial \{J_b[r] \{J_a[t] J_c[s]\}_1\}_{2}
        \end{align*}
        \endgroup
        \\ 
        This gives us the equality
        \begin{equation*}
            \partial \{\{ J_a[t] J_b[r] \}_1 J_c[s] \}_{2} = \partial \{J_a[t] \{J_b[r] J_c[s]\}_2\}_1-\partial \{J_b[r] \{J_a[t] J_c[s]\}_1\}_{2}.
        \end{equation*}
        The overall derivative can only act on the resulting defect operators and annihilates none of them, so we can simply drop it and recover \ref{eq:JJJ2}. \\
        
       \noindent We now show that \ref{eq:JJtheta} implies  \ref{eq:tilJJJ} (taking $\theta = \tilde{J}$) and \ref{eq:EJJ} (taking $\theta = E$) and similarly implies \ref{eq:JtilJJ}, \ref{eq:JEJ}. To do this, we first show that \ref{eq:JJtheta} implies \ref{eq:tilJJJ}, \ref{eq:EJJ}, and then we show that \ref{eq:tilJJJ}, \ref{eq:EJJ} imply \ref{eq:JtilJJ}, \ref{eq:JEJ}. To wit: \\
       
       1. \ref{eq:JJtheta} $\implies$ \ref{eq:tilJJJ} , \ref{eq:EJJ}: 
        \begin{align*}
            \{\{ \theta [t] J_b[r] \}_1 J_c[s] \}_{1} &= - \{ J_c[s] \{ \theta[t] J_b[r] \}_1 \}_{1} \\ \\
            &= \{ J_c[s] \{ J_b[r] \theta[t]\}_1 \}_{1} \\ \\
            &= \{ \{ J_c[s] J_b[r] \}_1 \theta[t] \}_{1} + \{ J_b[r] \{ J_c[s]  \theta[t] \}_1\}_{1} \\ \\
             &= \{ \theta[t] \{ J_b[r] J_c[s] \}_1 \}_{1} - \{ J_b[r] \{  \theta[t] J_c[s] \}_1\}_{1}.
        \end{align*}

        2. \ref{eq:JJtheta} $\implies$ \ref{eq:JtilJJ} , \ref{eq:JEJ}: 
        \begin{align*}
            \{\{ J_a [t] \theta [r] \}_1 J_c[s] \}_{1} &= - \{\{  \theta [r] J_a [t] \}_1 J_c[s] \}_{1} \\ \\
            &= \{J_a[t] \{\theta [r] J_c[s] \}_1\}_{1} -\{\theta [r] \{J_a[t] J_c[s] \}_{1}\}_1.
        \end{align*}
        \\
        To conclude the proof, we will next show that \ref{eq:FJJagain} implies \ref{eq:JFJ1} and \ref{eq:FJJ2}; \ref{eq:JFJ2again} implies \ref{eq:JJF2}; and \ref{eq:FJJagain}, together with \ref{eq:JFJ2again}, imply the remaining equation \ref{eq:JJF1}. \\ \\
        1. \ref{eq:FJJagain} $\implies$ \ref{eq:JFJ1}: 
        \begin{align*}
       \{J_a[t] \{F[r] J_c[s]\}_1\}_{1} -\{F[r] \{J_a[t] J_c[s] \}_{1}\}_1  &= -\{\{ F[r] J_a[t] \}_1 J_c[s] \}_{1} \\ \\
       &= \{\{ J_a[t] F[r]\}_1J_c[s]\}_1-\{\partial \{J_a[t]F[r]\}J_c[s]\}_1 \\ \\
       &= \{\{ J_a[t] F[r]\}_1J_c[s]\}_1.
        \end{align*}
         2. \ref{eq:FJJagain} $\implies$ \ref{eq:FJJ2}: \\ \\
         Start with \ref{eq:FJJagain}. \\
        \begin{align*}
            \text{LHS} &= -\{J_c[s]\{F[t]J_b[r]\}_1\}_1+\partial\{\{F[t]J_b[r]\}_1J_c[s]\}_2 \\ \\
            \text{RHS} &= \{F[t]\{J_b[r]J_c[s]\}_1\}_1+\{\{F[t]J_c[s]\}_1 J_b[r]\}_1-\partial\{\{F[t]J_c[s]\}_1 J_b[r]\}_2 \\ \\
            &=-\{F[t]\{J_c[s] J_b[r]\}_1\}_1+\{F[t] \partial\{J_b[r]J_c[s]\}_2\}_1+\{\{F[t]J_c[s]\}_1 J_b[r]\}_1-\partial\{\{F[t]J_c[s]\}_1 J_b[r]\}_2
        \end{align*}
        This gives us the equality 
        \begin{equation*}
             \partial \{\{ F[t] J_b[r]\}_1 J_c[s] \}_{2} = \partial \{F[t] \{J_b[r] J_c[s] \}_{2}\}_1 - \{J_b[r] \{F[t] J_c[s]\}_1\}_{2}. 
        \end{equation*} 
        Dropping the overall derivative, we obtain \ref{eq:FJJ2}.\\
        
        3. \ref{eq:JFJ2again} $\implies$ \ref{eq:JJF2}: 
          \begin{align*}
 \{\{ J_a[t] J_b[r] \}_1 F[s] \}_{2} &=  \{ F[s] \{ J_a[t] J_b[r] \}_1 \}_{2} \\ \\
 &=  \{J_a[t] \{F[s] J_b[r] \}_{2}\}_1-\{\{ J_a[t] F[s] \}_1 J_b[r] \}_{2} \\ \\
 &=  \{J_a[t] \{ J_b[r]F[s]  \}_{2}\}_1-\{ J_b[r] \{ J_a[t] F[s] \}_1 \}_{2}
          \end{align*}
         4. \ref{eq:FJJagain} and \ref{eq:JFJ2again} $\implies$ \ref{eq:JJF1}:\\ 
         \\ We will expand the left-hand-side and the right-hand-side of \ref{eq:JJF1} and, using the two parent identities, show that they agree. \\
          \begin{align*}
             \text{LHS} &= -\{F[t]\{J_b[r]J_c[s]\}_1\}_1+\partial\{F[t]\{J_b[r]J_c[s]\}_1\}_2 \\ \\
              \text{RHS} &= -\{J_b[r]\{F[t]J_c[s]\}_1\}_1+\{J_c[s]\{F[t]J_b[r]\}_1\}_1+\{J_b[r]\partial \{F[t] J_c[s]\}_2\}_1-\{J_c[s]\partial \{F[t]J_b[r]\}_2\}_1 \\ \\
              &= -\{J_b[r]\{F[t]J_c[s]\}_1\}_1+\{J_c[s]\{F[t]J_b[r]\}_1\}_1+\partial \{J_b[r] \{F[t] J_c[s]\}_2\}_1-\partial\{J_c[s] \{F[t]J_b[r]\}_2\}_1
          \end{align*}
         \\
          Rearranging, \\
        \begin{align*}
               \{J_b[r] \{F[t] J_c[s]\}_1\}_1-\{F[t]\{J_b[r]J_c[s]\}_1\}_1-\{J_c[s]\{F[t]J_b[r]\}_1\}_1 &\stackrel{?}{=} \\ \\
              \stackrel{?}{=} \partial \{J_b[r]\{F[t] J_c[s]\}_2\}_1-\partial\{J_c[s] \{F[t]&J_b[r]\}_2\}_1-\partial\{F[t]\{J_b[r]J_c[s]\}_1\}_2
        \end{align*}
        \\
        Expanding out both sides, \\
        \begingroup
         \allowdisplaybreaks
        \begin{align*}
            \text{LHS} &= -\{\{F[t] J_b[r]\}_1 J_c[s]\}_1-\{J_c[s]\{F[t]J_b[r]\}_1\}_1 = -\partial \{\{F[t] J_b[r]\}_1 J_c[s]\}_2
            \\ \\
            \text{RHS} &= \partial \{\{J_b[r] F[t]\}_1 J_c[s]\}_2-\partial\{J_c[s] \{F[t]J_b[r]\}_2\}_1 \\ \\
            &= -\partial \{\{ F[t] J_b[r]\}_1 J_c[s]\}_2+\partial \{ \partial \{J_b[r] F[t]\}_2 J_c[s]\}_2-\partial\{J_c[s] \{F[t]J_b[r]\}_2\}_1 \\ \\
            &= -\partial \{\{ F[t] J_b[r]\}_1 J_c[s]\}_2-\partial \{  \{ F[t] J_b[r]\}_2 J_c[s]\}_1-\partial\{J_c[s] \{F[t]J_b[r]\}_2\}_1  \\ \\
            &= -\partial \{\{ F[t] J_b[r]\}_1 J_c[s]\}_2+\partial \{   J_c[s] \{ F[t] J_b[r] \}_2\}_1-\partial\{J_c[s] \{F[t]J_b[r]\}_2\}_1
            \\ \\
            &= -\partial \{\{ F[t] J_b[r]\}_1 J_c[s]\}_2
        \end{align*}
        \endgroup
        \\
        We thus find that both sides agree, meaning that \ref{eq:JJF1} holds. \\ \\
Putting this together, the remaining independent equalities are precisely those in eq.(\ref{eq:minimalset01})-eq.(\ref{eq:minimalset05}):
            \begin{align*}
         \{\{ J_a[t] J_b[r] \}_1 J_c[s] \}_{1} &= \{J_a[t] \{J_b[r] J_c[s] \}_{1}\}_1 - \{J_b[r] \{J_a[t] J_c[s]\}_1\}_{1}  \\ \notag \\
         \{\{ J_a[t] J_b[r] \}_1 \tilde{J}_c [s] \}_{1} &= \{J_a[t] \{J_b[r] \tilde{J}_c[s] \}_{1}\}_1 - \{J_b[r] \{J_a[t] \tilde{J}_c[s]\}_1\}_{1}  \\ \notag \\
         \{\{ J_a[t] J_b[r] \}_1 E[s] \}_{1} &= \{J_a[t] \{J_b[r] E[s] \}_{1}\}_1 - \{J_b[r] \{J_a[t] E[s]\}_1\}_{1} 
         \\ \notag \\   
           \{\{ F[t] J_b[r] \}_1 J_c[s] \}_{1} &= \{F[t] \{J_b[r] J_c[s] \}_{1}\}_1 - \{J_b[r] \{F[t] J_c[s]\}_1\}_{1}   \\ \notag \\
            \{\{ J_a[t] F[r] \}_1 J_c[s] \}_{2} &= \{J_a[t] \{F[r] J_c[s] \}_{2}\}_1 - \{F[r] \{J_a[t] J_c[s]\}_1\}_{2} 
    \end{align*}
    This completes the proof.

\section{Sample Associativity Computations}\label{app:examples}
\noindent We present in this appendix several sample derivations of the OPE coefficients, in terms of $f$, that we presented in \S \ref{sec:coeffs}. Since there is much tedious repetition in these computations, we restrict ourselves to presenting several illuminating examples; we present them in decreasing order of straightforwardness, with the most intricate example displayed last. 

\subsection{Determination of $l$}
\noindent We will first determine the coefficient $\overset{(m)}{l}$ which governs the following single-pole term in the $FJ$ OPE:
\begin{equation*}
F[t](z)J_b[r](0)  \sim  \overset{\sum_{j=1}^{m+1} k_j = t+r-m}{\sum_{ m \geq 1 \quad k_j^i \geq 0}} \hat{\lambda}_{\mathfrak{g}, R} \hbar^{m+\frac{1}{2}} \bigg( \frac{1}{z}\overset{(m)}{\underset{(t,r)}{l}}[k_1,...,k_{m+1}]^{d_1 \cdot \cdot \cdot d_{m+1}}_{b} \hat{\partial_1} \bigg):\prod_{j=1}^{m+1} \tilde{J}_{d_j}[k_j]:.
\end{equation*}
Consider eq.(\ref{eq:minimalset05}) with $s=0$. We expand both sides of the equation.
\begin{align}
    \textit{lhs} &= \overset{\sum_{j=1}^{m+1} k_j = t+r-m}{\sum_{ m \geq 1 \quad k_j^i \geq 0}}   \hat{\lambda}_{\mathfrak{g}, R} \hbar^{m+\frac{1}{2}}\overset{(m)}{\underset{(t,r)}{l}}[k_1,...,k_{m+1}]^{e_1 d_2 \cdot \cdot \cdot d_{m+1}}_b f^{d_1}_{c e_1} :\prod_{j=1}^{m+1} \tilde{J}_{d_j}[k_j]: \\ \notag \\
    \textit{rhs} &= -\overset{\sum_{j=1}^{m+1} k_j = t+r-m}{\sum_{ m \geq 1 \quad k_j^i \geq 0}}   \hat{\lambda}_{\mathfrak{g}, R} \hbar^{m+\frac{1}{2}} \overset{(m)}{\underset{(t,r)}{f}}[k_1,...,k_{m+1}]^{d_1 \cdot \cdot \cdot d_{m+1}}_{cb} :\prod_{j=1}^{m+1} \tilde{J}_{d_j}[k_j]:
\end{align}
Comparing both sides, we immediately obtain:
\begin{equation}
    \overset{(m)}{\underset{(t,r)}{l}}[k_1,...,k_{m+1}]^{d_1 \cdot \cdot \cdot d_{m+1}}_b = - \overset{(m)}{\underset{(t,r)}{f}}(k_1,...,k_{m+1}) K^{d_1 e_1} K^{d_2 \cdot \cdot \cdot d_{m+1}}_{e_1 b}.
\end{equation}

\subsection{Determination of $a$}
\noindent The coefficient $\overset{(m)}{a}$ appears in the following term of the $JJ$ OPE:
\begin{equation*}
J_a[t](z)J_b[r](0)  \sim  \quad \frac{1}{z} \overset{\sum_{j=1}^{m+1} k_j = t+r-m}{\sum_{ m \geq 1 \quad k_j^i \geq 0}}  \hbar^m \overset{(m)}{\underset{(t,r)}{a}}[k_1,...,k_{m+1}]^{d_1 \cdot \cdot \cdot d_{m+1}}_{ab}:J_{d_1}[k_1] \prod_{j=2}^{m+1} \tilde{J}_{d_j}[k_j]:.
\end{equation*}
We first show that the following equality holds for $m \geq 1$, where $K^{i_1 \ldots i_{m+1}}_{ab}$ is defined in eq.(\ref{eq:Kdef}):
\begin{equation}
 K^{d_1 \cdot \cdot \cdot d_{m+1}}_{e_1 a} f^{e_1}_{cb} = K^{e_1 d_2 \cdot \cdot \cdot d_{m+1}}_{ba} f^{d_1}_{c e_1} +...+K^{d_1 \cdot \cdot \cdot d_{m} e_1}_{ba} f^{d_{m+1}}_{c e_1}+K^{d_1 \cdot \cdot \cdot d_{m+1}}_{b e_1} f^{e_1}_{ac}. \label{lemma.b.1}
\end{equation}
To do this, we use induction on $m$. Note that for the algebras of interest, the Killing form $K_{ab}$ can always be brought to a form $\propto \delta_{ab}$. Let $\kappa$ denote the proportionality constant. This will allow us to use the Jacobi identity. \\ \\
\textit{Base Case}
\begingroup \allowdisplaybreaks
\begin{align}
    K^{d_1 d_2}_{e_0 a} f^{e_0}_{cb} &= K^{e_1 e_2} f^{d_1}_{e_0 e_1} f^{d_2}_{a e_2} f^{e_0}_{cb} =  K^{e_1 e_2} f^{d_2}_{a e_2} f^{d_1}_{c e_0} f^{e_0}_{b e_1}+ K^{e_1 e_2} f^{d_2}_{a e_2} f^{d_1}_{b e_0} f^{e_0}_{e_1 c} \notag \\ 
    &= K^{e_0 d_2}_{ba} f^{d_1}_{c e_0} -\kappa f^{d_1}_{b e_0} f^{d_2}_{c e_1} f^{e_1}_{e_0 a}-\kappa f^{d_1}_{b e_0}f^{d_2}_{e_0 e_1} f^{e_1}_{ac} \notag \\
     &= K^{e_0 d_2}_{ba} f^{d_1}_{c e_0} + K^{d_1 e_0}_{ba} f^{d_2}_{c e_0}+K^{d_1 d_2}_{b e_0} f^{e_0}_{ac}.
\end{align} \endgroup
Therefore, the statement holds for $m=1$. \\ \\
\textit{Inductive Step}
\begingroup \allowdisplaybreaks
\begin{align}
     K^{d_1 \cdot \cdot \cdot d_{m+1}}_{e_0 a} f^{e_0}_{cb} &= -K^{d_1 \cdot \cdot \cdot d_m}_{e_0 e_1} f^{e_0}_{cb} K^{e_1 e_2} f^{d_{m+1}}_{a e_2} \\ &=-(K^{e_0 d_2 \cdot \cdot \cdot d_m}_{b e_1} f^{d_1}_{c e_0}  +...+K^{d_1 \cdot \cdot \cdot d_{m-1} e_0}_{b e_1} f^{d_m}_{c e_0} + K^{d_1 \cdot \cdot \cdot d_m}_{b e_0} f^{e_0}_{e_1 c}) K^{e_1 e_2} f^{d_{m+1}}_{a e_2}\notag \\ &= K^{e_0 d_2 \cdot \cdot \cdot d_{m+1}}_{ba} f^{d_1}_{c e_0}+...+K^{d_2 \cdot \cdot \cdot d_{m-1} e_0 d_{m+1}}_{ba} f^{d_m}_{c e_0}-\kappa K^{d_1 \cdot \cdot \cdot d_m}_{b e_0} f^{e_0}_{e_1 c} f^{d_{m+1}}_{a e_1} \notag \\
    &= K^{e_0 d_2 \cdot \cdot \cdot d_{m+1}}_{ba} f^{d_1}_{c e_0}+...+K^{ d_1 \cdot \cdot \cdot d_{m} e_0}_{ba} f^{d_{m+1}}_{c e_0}+K^{d_1 \cdot \cdot \cdot d_{m+1}}_{b e_0} f^{e_0}_{ac}. \notag
\end{align} \endgroup
Thus, the statement holds for $m \geq 1$. \\ 

We now consider associativity condition eq.(\ref{eq:minimalset01}) with $s=0$. We expand both sides of the equation.
\begingroup \allowdisplaybreaks
\begin{align}
    \textit{lhs} =& \overset{\sum_{j=1}^{m+1} k_j = t+r-m}{\sum_{ m \geq 1 \quad k_j^i \geq 0}}  \hbar^m \overset{(m)}{\underset{(t,r)}{a}}[k_1,...,k_{m+1}]^{e_1 d_2 \cdot \cdot \cdot d_{m+1}}_{ab} f^{d_1}_{c e_1} :\prod_{j=1}^{m+1} \tilde{J}_{d_j}[k_j]: \\ \notag \\
    \textit{rhs} = &-\overset{\sum_{j=1}^{m+1} k_j = t+r-m}{\sum_{ m \geq 1 \quad k_j^i \geq 0}}  \hbar^m \overset{(m)}{\underset{(r,t)}{f}}[k_1,...,k_{m+1}] K^{d_1 \cdot \cdot \cdot d_{m+1}}_{e_1 a} f^{e_1}_{c b} :\prod_{j=1}^{m+1} \tilde{J}_{d_j}[k_j]: \\
    &+\overset{\sum_{j=1}^{m+1} k_j = t+r-m}{\sum_{ m \geq 1 \quad k_j^i \geq 0}}  \hbar^m \overset{(m)}{\underset{(t,r)}{f}}[k_1,...,k_{m+1}] K^{d_1 \cdot \cdot \cdot d_{m+1}}_{e_1 b} f^{e_1}_{c a} :\prod_{j=1}^{m+1} \tilde{J}_{d_j}[k_j]: \notag \\
    =&-\overset{\sum_{j=1}^{m+1} k_j = t+r-m}{\sum_{ m \geq 1 \quad k_j^i \geq 0}}  \hbar^m (\overset{(m)}{\underset{(r,t)}{f}}[k_1,...,k_{m+1}]+(-1)^{m-1}\overset{(m)}{\underset{(t,r)}{f}}[k_{m+1},...,k_{1}]) K^{d_1 \cdot \cdot \cdot d_{m+1}}_{e_1 a} f^{e_1}_{c b} :\prod_{j=1}^{m+1} \tilde{J}_{d_j}[k_j]: \notag \\
    +&\overset{\sum_{j=1}^{m+1} k_j = t+r-m}{\sum_{ m \geq 1 \quad k_j^i \geq 0}}  \hbar^m (\overset{(m)}{\underset{(t,r)}{f}}[k_1,...,k_{m+1}]^{e_1 d_2 \cdot \cdot \cdot d_{m+1}}_{ab} +...+\overset{(m)}{\underset{(t,r)}{f}}[k_2,...,k_{m+1},k_1]^{d_2 \cdot \cdot \cdot d_{m+1} e_1}_{ab} ) f^{d_{1}}_{c e_1}:\prod_{j=1}^{m+1} \tilde{J}_{d_j}[k_j]:\notag  
\end{align} \endgroup
where we have used eq.(\ref{lemma.b.1}) to simplify the right-hand-side. \\

Comparing both sides and matching the coefficients of the terms with structure constant $f^{i_1}_{cj}$ we finally obtain:
\begin{equation*}
    \overset{(m)}{\underset{(t,r)}{a}}[k_1,...,k_{m+1}]^{d_1 \cdot \cdot \cdot d_{m+1}}_{ab} =  \overset{(m)}{\underset{(t,r)}{f}}[k_1,...,k_{m+1}]^{d_1 \cdot \cdot \cdot d_{m+1}}_{ab}+...+ \overset{(m)}{\underset{(t,r)}{f}}[k_2,...,k_{m+1},k_1]^{d_2 \cdot \cdot \cdot d_{m+1} d_1}_{ab}
\end{equation*}
We also find, by examining the coefficients of the terms with $f^j_{cb}$, that $\overset{(m)}{f}$ satisfy the following relation:
\begin{equation}
    \overset{(m)}{\underset{(t,r)}{f}}(k_1,...,k_{m+1}) =  (-1)^{m}\overset{(m)}{\underset{(r,t)}{f}}(k_{m+1},...,k_{1}).
\end{equation}

\subsection{Determination of $j$}
\noindent The coefficient $\overset{(m)}{j}$ that we will next determine appears in the OPE:
\begin{equation*}
    E[t](z)J_b[r](0)  \sim \quad \frac{1}{z}  \overset{\sum_{j=1}^{m+1} k_j = t+r-m-1}{\sum_{ m \geq 1 \quad k_j^i \geq 0}}  \hat{\lambda}_{\mathfrak{g}, R} \hbar^{m+\frac{1}{2}}  \overset{(m)}{\underset{(t,r)}{j}}[k_1,...,k_{m+1}]^{d_1 \cdot \cdot \cdot d_{m+1}}_{b}: \prod_{j=1}^{m+1} \tilde{J}_{d_j}[k_j]:.
\end{equation*}

The form of the axion interaction, $\eta \mathcal{A} \partial \mathcal{A}$, tells us that the vertex factor is proportional to that of $\mathcal{B} \mathcal{A}^2$ (after stripping away the Lie algebra information), with the proportionality constant some function of the momentum variables. Since the external legs of the diagrams in Figure \ref{figure_1} do not contribute propagators, this means that we should expect the contribution of the diagram with external legs $\eta$ and $J$ to be proportional to the contribution of the diagram with external legs $\mathcal{B}$ and $\mathcal{A}$ upon replacing $E[t] \to \tilde{J}[t-1]$, where we have matched the quantum numbers spin and scaling dimension in an $\mathfrak{sl}_2(\mathbb{C})_+$-invariant way. Explicitly, 
\begin{equation}
    \overset{(m)}{\underset{(t,r)}{j}}(k_1,...,k_{m+1}) = \overset{(m)}{q}(t,r;k_1,...,k_{m+1}]) \overset{(m)}{\underset{(t-1,r)}{f}}(k_1,...,k_{m+1}),
\end{equation} for some expression $\overset{(m)}{q}$ we have yet to determine. \\

It turns out that we can fix $\overset{(m)}{q}$ directly from the tree-level OPEs, since the tree-level diagrams encode the vertex factor proportionality constant we mentioned above:
\begin{equation}
    \overset{(m)}{q}(t,r;k_1,...,k_{m+1}) = -\frac{\alpha(t,k_1)}{\beta(t)}
\end{equation}
using the definitions eq.(\ref{eq:alphabeta}), and we thus obtain:
\begin{equation}
     \underset{(t,r)}{\overset{(m)}{j}}[k_1,...,k_{m+1}]_{b}^{d_1 \cdot \cdot \cdot d_{m+1}} = - \bigg(\frac{\alpha(t,k_1)}{\beta(t)}\bigg) \underset{(t-1,r)}{\overset{(m)}{f}}(k_1,...,k_{m+1}) K^{d_1 e_1} K^{d_2 \cdot \cdot \cdot d_{m+1}}_{e_1 b}. 
\end{equation}

We can plug this solution into eq.(\ref{eq:minimalset03}) order by order in $m$ to verify that the associativity condition holds. To illustrate, we work this out explicitly for $m=1$ and $m=2$. Note that this process can be continued to higher order in $m$, but the structure of the diagrams do not change, meaning that going beyond $m>2$ yields no new dynamics but the difficulty of showing that the solutions hold increases significantly due to the proliferation of terms and Lie algebra indices. Similar comments apply for other coefficients. \\ \\
For notation convenience let $p_m = t+r+s-1-m$. The methods for determining $\overset{(m)}{k}$ are identical to the ones used in this section, so we omit those details. 

\subsubsection{Solving For $j$ at 1 Loop}
\noindent The $m=1$ equation is
\begingroup \allowdisplaybreaks
\begin{align*}
    \textit{lhs} &= \overset{\sum k_j = p_1}{\sum_{ k_j^i \geq 0}}  \bigg(\underset{(t,r+s)}{\overset{(1)}{j}}[k_1,k_2]^{d_1 d_2}_{e_1} f^{e_1}_{bc} -\bigg(\frac{\alpha(t,k_1)}{\beta(t)}\bigg) \underset{(r,s)}{\overset{(1)}{a}}[k_1+1-t,k_2]^{d_1 d_2}_{bc}\bigg) :\prod_{j=1}^{2} \tilde{J}_{d_j}[k_j]:  \\
    \textit{rhs} &= \overset{\sum k_j = p_1}{\sum_{ k_j^i \geq 0}} :\prod_{j=1}^{2} \tilde{J}_{d_j}[k_j]: \bigg( \underset{(t,s)}{\overset{(1)}{j}}[k_1-r,k_2]^{e_1 d_2}_c f^{d_1}_{b e_1}+\underset{(t,s)}{\overset{(1)}{j}}[k_1,k_2-r]^{d_1 e_1}_c f^{d_2}_{b e_1} -\underset{(t,r)}{\overset{(1)}{j}}[k_1-s,k_2]^{e_1 d_2}_b f^{d_1}_{c e_1} \\ 
       -&\underset{(t,r)}{\overset{(1)}{j}}[k_1,k_2-s]^{d_1 e_1}_b f^{d_2}_{c e_1}-\bigg(\frac{\alpha(s,t-1)}{\beta(t)}\bigg) \underset{(t+s-1,r)}{\overset{(1)}{f}}[k_1,k_2]^{d_1 d_2}_{cb}
      +\bigg(\frac{\alpha(r,t-1)}{\beta(t)}\bigg) \underset{(t+r-1,s)}{\overset{(1)}{f}}[k_1,k_2]^{d_1 d_2}_{bc} \bigg)  
\end{align*} \endgroup \\
Manipulating indices, this simplifies to
\begingroup \allowdisplaybreaks
\begin{align*}
0 &= \overset{\sum k_j = p_1}{\sum_{ k_j^i \geq 0}}  \bigg( \underset{(t,r+s)}{\overset{(1)}{j}}(k_2,k_1)-\underset{(t,r+s)}{\overset{(1)}{j}}(k_1,k_2) +\underset{(t,s)}{\overset{(1)}{j}}(k_1-r,k_2) +\underset{(t,r)}{\overset{(1)}{j}}(k_1,k_2-s)-\underset{(t,s)}{\overset{(1)}{j}}(k_2,k_1-r) \\
& -\underset{(t,r)}{\overset{(1)}{j}}(k_2-s,k_1) + \bigg( \frac{\alpha((r,t-1)}{\beta(t)}\bigg) \underset{(t+r-1,s)}{\overset{(1)}{f}}(k_1,k_2)  -\bigg( \frac{\alpha(s,t-1)}{\beta(t)}\bigg) \underset{(t+s-1,r)}{\overset{(1)}{f}}(k_2,k_1) \\
& + \bigg( \frac{\alpha(t,k_1)}{\beta(t)}\bigg) \underset{(r,s)}{\overset{(1)}{f}}(k_1+1-t,k_2) - \bigg( \frac{\alpha(t,k_2)}{\beta(t)}\bigg) \underset{(r,s)}{\overset{(1)}{f}}(k_1,k_2+1-t) \bigg) K^{d_1 d_2}_{bc} :\prod_{j=1}^{2} \tilde{J}_{d_j}[k_j]: 
\end{align*} \endgroup
Using eq.(\ref{eq:recursion}), it is easy to see this equation is satisfied. \\

\subsubsection{Solving For $j$ at 2 Loops}
\noindent The $m=2$ equation is
\begingroup \allowdisplaybreaks
\begin{align*}
    \textit{lhs} &= \overset{\sum k_j = p_2}{\sum_{ k_j^i \geq 0}}  \bigg( \underset{(t,r+s)}{\overset{(2)}{j}}[k_1,k_2,k_3]^{d_1 d_2 d_3}_{e_1} f^{e_1}_{bc}+\underset{(r,s)}{\overset{(1)}{a}}[l,k_3]^{e_1 d_3}_{bc} \underset{(t,l)}{\overset{(1)}{j}}[k_1,k_2]^{d_1 d_2}_{e_1}  \\
    & -\bigg(\frac{\alpha(t,k_1)}{\beta(t)}\bigg) \underset{(r,s)}{\overset{(2)}{a}}[k_1+1-t,k_2,k_3]^{d_1 d_2 d_3}_{bc} \bigg) :\prod_{j=1}^{3} \tilde{J}_{d_j}[k_j]: \\ \\
    \textit{rhs} &= \overset{\sum k_j = p_2}{\sum_{ k_j^i \geq 0}} \bigg( \underset{(t,s)}{\overset{(2)}{j}}[k_1-r,k_2,k_3]^{e_1 d_2 d_3}_c f^{d_1}_{b e_1}+\underset{(t,s)}{\overset{(2)}{j}}[k_1,k_2-r,k_3]^{d_1 e_1 d_3}_c f^{d_2}_{b e_1} +\underset{(t,s)}{\overset{(2)}{j}}[k_1,k_2,k_3-r]^{d_1 d_2 e_1}_c f^{d_3}_{b e_1} \\ 
    & - \underset{(t,s)}{\overset{(1)}{j}}[l,k_3]^{e_1 d_3}_c \underset{(l,r)}{\overset{(1)}{f}}[k_1,k_2]^{d_1 d_2}_{e_1 b} - \underset{(t,s)}{\overset{(1)}{j}}[k_3,l]^{d_3 e_1}_c \underset{(l,r)}{\overset{(1)}{f}}[k_1,k_2]^{d_1 d_2}_{e_1 b} - \bigg( \frac{\alpha(s,t-1)}{\beta(t)} \bigg) \underset{(t+s-1,r)}{\overset{(2)}{f}}[k_1,k_2,k_3]^{d_1 d_2 d_3}_{cb} \\
    &   -\underset{(t,r)}{\overset{(2)}{j}}[k_1-s,k_2,k_3]^{e_1 d_2 d_3}_b f^{d_1}_{c e_1} -\underset{(t,r)}{\overset{(2)}{j}}[k_1,k_2-s,k_3]^{d_1 e_1 d_3}_b f^{d_2}_{c e_1}  -\underset{(t,r)}{\overset{(2)}{j}}[k_1,k_2,k_3-s]^{d_1 d_2 e_1}_b f^{d_3}_{c e_1} \\
    &  + \underset{(t,r)}{\overset{(1)}{j}}[l,k_3]^{e_1 d_3}_b \underset{(l,s)}{\overset{(1)}{f}}[k_1,k_2]^{d_1 d_2}_{e_1 c}  + \underset{(t,r)}{\overset{(1)}{j}}[k_3,l]^{d_3 e_1}_b \underset{(l,s)}{\overset{(1)}{f}}[k_1,k_2]^{d_1 d_2}_{e_1 c} \\
    & + \bigg( \frac{\alpha(r,t-1)}{\beta(t)} \bigg) \underset{(t+r-1,s)}{\overset{(2)}{f}}[k_1,k_2,k_3]^{d_1 d_2 d_3}_{bc} \bigg) :\prod_{j=1}^{3} \tilde{J}_{d_j}[k_j]: 
\end{align*} \endgroup
Manipulating the indices through repeated use of the Jacobi identity
\begingroup \allowdisplaybreaks
\begin{align*}
    0 &= \overset{\sum k_j = p_2}{\sum_{ k_j^i \geq 0}}  \bigg( \underset{(t,r+s)}{\overset{(2)}{j}}(k_1,k_2,k_3)+\underset{(t,r+s)}{\overset{(2)}{j}}(k_3,k_2,k_1) - \underset{(t,r+s)}{\overset{(2)}{j}}(k_2,k_3,k_1) -\underset{(t,r+s)}{\overset{(2)}{j}}(k_2,k_1,k_3) \\
     & - \underset{(t,s)}{\overset{(2)}{j}}(k_1-r,k_2,k_3) + \underset{(t,s)}{\overset{(2)}{j}}(k_2,k_1-r,k_3) +\underset{(t,s)}{\overset{(2)}{j}}(k_2,k_3,k_1-r) -\underset{(t,s)}{\overset{(2)}{j}}(k_3,k_2,k_1-r) +\underset{(t,r)}{\overset{(2)}{j}}(k_2,k_3-s,k_1) \\
     &   -\underset{(t,r)}{\overset{(2)}{j}}(k_3-s,k_2,k_1)-\underset{(t,r)}{\overset{(2)}{j}}(k_1,k_2,k_3-s) +\underset{(t,r)}{\overset{(2)}{j}}(k_2,k_1,k_3-s) - \underset{(t,s)}{\overset{(1)}{j}}(l,k_3) \underset{(l,r)}{\overset{(1)}{f}}(k_2,k_1) \\
     &  +\underset{(t,s)}{\overset{(1)}{j}}(k_3,l) \underset{(l,r)}{\overset{(1)}{f}}(k_2,k_1) -\underset{(t,r)}{\overset{(1)}{j}}(l,k_1) \underset{(l,s)}{\overset{(1)}{f}}(k_2,k_3) +\underset{(t,r)}{\overset{(1)}{j}}(k_1,l) \underset{(l,s)}{\overset{(1)}{f}}(k_2,k_3)-\underset{(t,l)}{\overset{(1)}{j}}(k_2,k_3) \underset{(r,s)}{\overset{(1)}{f}}(k_1,l)  \\
      & +\underset{(t,l)}{\overset{(1)}{j}}(k_3,k_2) \underset{(r,s)}{\overset{(1)}{f}}(k_1,l) -\underset{(t,l)}{\overset{(1)}{j}}(k_1,k_2) \underset{(r,s)}{\overset{(1)}{f}}(l,k_3) + \underset{(t,l)}{\overset{(1)}{j}}(k_2,k_1) \underset{(r,s)}{\overset{(1)}{f}}(l,k_3)+\bigg( \frac{\alpha(s,t-1)}{\beta(t)}\bigg) \underset{(t+s-1,r)}{\overset{(2)}{f}}(k_3,k_2,k_1) \\
       &  +\bigg( \frac{\alpha(r,t-1)}{\beta(t)}\bigg) \underset{(t+r-1,s)}{\overset{(2)}{f}} (k_1,k_2,k_3) +\bigg( \frac{\alpha(t,k_1)}{\beta(t)}\bigg) \underset{(r,s)}{\overset{(2)}{f}}(k_1+1-t,k_2,k_3) \\
       & +\bigg( \frac{\alpha(t,k_2)}{t^1+t^2}\bigg) \underset{(r,s)}{\overset{(2)}{f}}(k_1,k_2+1-t,k_3) +\bigg( \frac{\alpha(t,k_3)}{t^1+t^2}\bigg) \underset{(r,s)}{\overset{(2)}{f}}(k_1,k_2,k_3+1-t)\bigg) K^{d_1 d_2 d_3}_{bc} :\prod_{j=1}^3 \tilde{J}_{d_j}[k_j]: 
\end{align*} \endgroup
Fix $k_1,k_2,k_3$ such that $\sum k_j = t+r+s-3$, and plug in the expression for $\overset{(m)}{j}$. Grouping terms, we find:
\begingroup \allowdisplaybreaks
\begin{align*}
    0 &= \alpha(t,k_1)\bigg( -\underset{(t-1,r+s)}{\overset{(2)}{f}}(k_1,k_2,k_3)+\underset{(t-1,s)}{\overset{(2)}{f}}(k_1-r,k_2,k_3)+\underset{(t-1,r)}{\overset{(2)}{f}}(k_1,k_2,k_3-s) \\
    &-\underset{(r,s)}{\overset{(2)}{f}}(k_1+1-t,k_2,k_3)-\underset{(t-1,r)}{\overset{(1)}{f}}(k_1,l) \underset{(l,s)}{\overset{(1)}{f}}(k_2,k_3)+\underset{(t,l)}{\overset{(1)}{f}}(k_1,k_2) \underset{(r,s)}{\overset{(1)}{f}}(l,k_3)\bigg)  \\
    &+\alpha(t,k_2) \bigg( \underset{(t-1,r+s)}{\overset{(2)}{f}}(k_2,k_3,k_1)+\underset{(t-1,r+s)}{\overset{(2)}{f}}(k_2,k_1,k_3)-\underset{(t-1,s)}{\overset{(2)}{f}}(k_2,k_1-r,k_3)  \\
    &- \underset{(t-1,s)}{\overset{(2)}{f}}(k_2,k_3,k_1-r)- \underset{(t-1,r)}{\overset{(2)}{f}}(k_2,k_3-s,k_1)- \underset{(t-1,s)}{\overset{(2)}{f}}(k_2,k_3,k_1-r)-\underset{(t-1,r)}{\overset{(2)}{f}}(k_2,k_1,k_3-s)  \\
    &- \underset{(r,s)}{\overset{(2)}{f}}(k_1,k_2+1-t,k_3)+\underset{(t-1,l)}{\overset{(1)}{f}}(k_2,k_3) \underset{(r,s)}{\overset{(1)}{f}}(k_1,l)-\underset{(t,l)}{\overset{(1)}{f}}(k_2,k_1) \underset{(r,s)}{\overset{(1)}{f}}(l,k_3) \bigg)  \\
    &+\alpha(t,k_3) \bigg( - \underset{(t-1,r+s)}{\overset{(2)}{f}}(k_3,k_2,k_1)+\underset{(t-1,r)}{\overset{(2)}{f}}(k_3-s,k_2,k_1)+\underset{(t-1,s)}{\overset{(2)}{f}}(k_3,k_2,k_1-r)  \\
    &-\underset{(s,r)}{\overset{(2)}{f}}(k_3+1-t,k_2,k_1)-\underset{(t-1,s)}{\overset{(1)}{f}}(k_3,l) \underset{(l,r)}{\overset{(1)}{f}}(k_2,k_1)+\underset{(t,l)}{\overset{(1)}{f}}(k_3,k_2) \underset{(s,r)}{\overset{(1)}{f}}(l,k_1) \bigg)  \\
    &+\alpha(t,l) \underset{(t-1,s)}{\overset{(1)}{f}}(l,k_3) \underset{(l,r)}{\overset{(1)}{f}}(k_2,k_1)+\alpha(t,l) \underset{(t-1,r)}{\overset{(1)}{f}}(l,k_1) \underset{(l,s)}{\overset{(1)}{f}}(k_2,k_3)  +\alpha(t,r-1)\underset{(t+r-1,s)}{\overset{(2)}{f}}(k_1,k_2,k_3) \\
    &- \alpha(t,r-1) \underset{(t-1,s)}{\overset{(2)}{f}}(k_1-r,k_2,k_3) +\alpha(t,s-1) \underset{(t+s-1,r)}{\overset{(2)}{f}}(k_3,k_2,k_1) - \alpha(t,s-1)\underset{(t-1,r)}{\overset{(2)}{f}}(k_3-s,k_2,k_1)  
\end{align*} \endgroup
Using eq.(\ref{eq:recursion}), we can show that the right-hand-side vanishes, thus the equation is satisfied. 

\subsection{Determination of $i$}
\noindent Let us next determine $\overset{(m)}{i}$, which comes from the following term in the $JJ$ OPE:
\begin{equation*}
J_a[t](z)J_b[r](0) \sim  \overset{\sum_{j=1}^{m+1} k_j = t+r-m-1}{\sum_{ m \geq 1 \quad k_j^i \geq 0}}  \hat{\lambda}_{\mathfrak{g}, R} \hbar^{m+\frac{1}{2}} \frac{1}{z}\overset{(m)}{\underset{(t,r)}{i}}[k_1,...,k_{m+1}]^{d_2 \dotsm d_{m+1}}_{ab}:F[k_1]\prod_{j=2}^{m+1} \tilde{J}_{d_j}[k_j]:.
\end{equation*}

Most of the logic from the previous section goes through unchanged, but there are two key differences: the placement of the $\eta-$vertex is not fixed, and now there is an $\eta-$propagator. For the purposes of this section it is sufficient to know that the $\eta$-propagator is proportional to the derivative of the gauge propagator, $P^\eta_{12} \sim \partial_1 P_{12}$; see \cite{BSS} for more details and, e.g., \cite{victor} for the explicit definition of $P_{12}$. As a result, we pick up functions of momentum variables coming both from the $\eta-$vertex and the $\eta-$propagator. With the replacement $F[k_1] \to J[k_1+1]$, we expect the contribution from the relevant diagram to be proportional to the diagram with external legs $\mathcal{A}$ and $\mathcal{A}$ and defect operator content $J$ and $\tilde{J}$, with the placement of $F$ in the relevant diagram matching that of $J$ in the other. In summary, we are looking a solution of the form
\begin{align}
     \overset{(m)}{\underset{(t,r)}{i}}(k_1,...,k_{m+1}) =& \underset{(1)}{\overset{(m)}{q}}(t,r;k_1,...,k_{m+1}) \overset{(m)}{\underset{(t,r)}{f}}(k_1,...,k_{m+1})+... \\ &...+\underset{(m+1)}{\overset{(m)}{q}}(t,r;k_1,...,k_{m+1}) \overset{(m)}{\underset{(t,r)}{f}}(k_2,...,k_{m+1},k_1). \notag
\end{align}

\begin{figure}[t]
    \centering
    \includegraphics[scale=0.4]{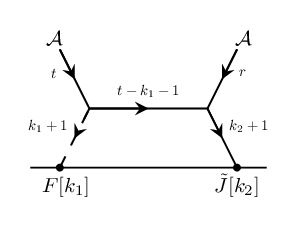}
     \caption{ Illustration of the flow of momentum. }
     \label{figure_appendix_b_01}
\end{figure}
We can again determine an expression for $\overset{(m)}{q}$ directly from the tree-level OPEs. Note that in extracting the $\eta$-vertex factor from tree-level OPE, we must be more careful than in the previous section; in particular, we must account for the different possible locations of the $\eta-$vertex in the diagram. In the context of these diagrams, the avatar of momentum conservation is $t+r=\sum_{j=1}^{m+1} (k_j+1)$. We interpret this as $t+r$ momentum flowing in from the external legs, $t$ from the left-most external leg and $r$ from the right-most external leg, and $\sum_{j=1}^{m+1} (k_j+1)$ momentum flowing out in the direction of the defect operators, $k_j+1$ of momentum per defect operator. We illustrate this in Figure \ref{figure_appendix_b_01}. \\

With this interpretation of momentum flow, we see that the tree-level OPE gives us $\underset{(j)}{\overset{(m)}{q}}(t,r;k_1,...,k_{m+1})$ is given by:
\begin{equation}
    \underset{(j)}{\overset{(m)}{q}}(t,r;k_1,...,k_{m+1}) = \alpha( p_j,k_1)
\end{equation}
where we are reading the flow of momentum from left to right, as in Figure \ref{figure_appendix_b_01}, and $p_j$ is interpreted as the incoming momentum, which is given by the simple expression:
\begin{equation}
    p_{j} = t-\sum_{i=2}^{j}(k_j+1).
\end{equation}
This finally gives us the equality
\begin{equation*}
    \overset{(m)}{\underset{(t,r)}{i}}(k_1,...,k_{m+1}) = \alpha(t,k_1) \overset{(m)}{\underset{(t,r)}{f}}(k_1,...,k_{m+1})+...+\alpha(p_{m+1},k_1) \overset{(m)}{\underset{(t,r)}{f}}(k_2,...,k_{m+1},k_1). 
\end{equation*}
Let us explicitly demonstrate that the associativity condition eq.(\ref{eq:minimalset02}) is satisfied with this solution for $m=1$ and $m=2$. For notation convenience let $p_m = t+r+s-1-m$.

\subsubsection{Solving For $i$ at 1 Loop}
\noindent The $m=1$ equation is
\begingroup \allowdisplaybreaks \begin{align*}
 0 = \overset{\sum k_j = p_1}{\sum_{ k_j^i \geq 0}} \bigg( &-\overset{(1)}{\underset{(t,r+s)}{i}}(k_1,k_2) f^{d_2}_{a e_1} f^{e_1}_{bc}-\overset{(1)}{\underset{(r,t+s)}{i}}(k_1,k_2) f^{d_2}_{b e_1} f^{e_1}_{ac}+\overset{(1)}{\underset{(s,t+r)}{i}}(k_1, k_2) f^{d_2}_{c e_1} f^{e_1}_{ab} 
 \\ 
 & +\overset{(1)}{\underset{(r,s)}{i}}(k_1,k_2-t) f^{e_1}_{b c} f^{d_2}_{a e_1}-\overset{(1)}{\underset{(t,s)}{i}}(k_1,k_2-r) f^{e_1}_{a c} f^{d_2}_{b e_1}+\overset{(1)}{\underset{(t,r)}{i}}(k_1,k_2-s) f^{e_1}_{a b} f^{d_2}_{c e_1} 
 \\
 & - \alpha(t,k_1) \overset{(1)}{\underset{(r,s)}{f}}(k_1+1-t,k_2) f^{e_1}_{a b} f^{d_2}_{c e_1} -\alpha(r,k_1) \overset{(1)}{\underset{(t,s)}{f}}(k_1+1-r,k_2) f^{e_1}_{a b} f^{d_2}_{c e_1} 
 \\
 & +\alpha(s,k_1) \overset{(1)}{\underset{(t,r)}{f}}(k_1+1-s,k_2) f^{e_1}_{a c} f^{d_2}_{b e_1}-\alpha(t,k_1) \overset{(1)}{\underset{(r,s)}{f}}(k_2,k_1+1-t) f^{e_1}_{a c} f^{d_2}_{b e_1}
 \\
 & +\alpha(r,k_1) \overset{(1)}{\underset{(t,s)}{f}}(k_2,k_1+1-r) f^{e_1}_{b c} f^{d_2}_{a e_1}+\alpha(s,k_1) \overset{(1)}{\underset{(t,r)}{f}}(k_2,k_1+1-s) f^{e_1}_{b c} f^{d_2}_{a e_1}\bigg) :F[k_1]\tilde{J}_{d_2}[k_2]:
\end{align*} \endgroup
Plugging in our expression for $i$ and using eq.(\ref{eq:recursion}), it is easy to see the right-hand-side vanishes as expected. 

\subsubsection{Solving For $i$ at 2 Loops}
\noindent The $m=2$ equation is
\begingroup \allowdisplaybreaks \begin{align*}
    0 &= \overset{\sum k_j = p_2}{\sum_{ k_j^i \geq 0}} \bigg( \overset{(2)}{\underset{(t,r+s)}{i}}(k_1,k_2,k_3)^{d_2 d_3}_{a e_1} f^{e_1}_{bc} +\overset{(2)}{\underset{(r,s)}{i}}(k_1,k_2-t,k_3)^{e_1 d_3}_{bc} f^{d_2}_{a e_1}  
    \\ 
    & \quad \quad \quad \quad +\overset{(2)}{\underset{(r,s)}{i}}(k_1,k_2,k_3-t)^{d_2 e_1}_{bc} f^{d_3}_{a e_1}-\overset{(2)}{\underset{(r,t+s)}{i}}(k_1,k_2,k_3)^{d_2 d_3}_{b e_1} f^{e_1}_{a c} -\overset{(2)}{\underset{(t,s)}{i}}(k_1,k_2-r,k_3)^{e_1 d_3}_{ac} f^{d_2}_{b e_1}  \\
    &\quad \quad \quad \quad - \overset{(2)}{\underset{(t,s)}{i}}(k_1,k_2,k_3-r)^{d_2 e_1}_{ac} f^{d_3}_{b e_1}+\overset{(2)}{\underset{(s,t+r)}{i}}(k_1,k_2,k_3)^{d_2 d_3}_{c e_1} f^{e_1}_{ab}+\overset{(2)}{\underset{(t,r)}{i}}(k_1,k_2-s,k_3)^{e_1 d_3}_{ab} f^{d_2}_{c e_1}  \\
    & \quad \quad \quad \quad +\overset{(2)}{\underset{(t,r)}{i}}(k_1,k_2,k_3-s)^{d_2 e_1}_{ab} f^{d_3}_{c e_1}-\overset{(1)}{\underset{(r,s)}{i}}(k_1,l)^{e_1}_{bc} \overset{(1)}{\underset{(l,t)}{f}}(k_2,k_3)^{d_2 d_3}_{e_1 a}+\overset{(1)}{\underset{(t,l)}{i}}(k_1,k_2)^{d_2}_{a e_1} \overset{(1)}{\underset{(r,s)}{a}}(l,k_3)^{e_1 d_3}_{bc}  \\
    & \quad \quad \quad \quad +\overset{(1)}{\underset{(t,s)}{i}}(k_1,l)^{e_1}_{ac} \overset{(1)}{\underset{(l,r)}{f}}(k_2,k_3)^{d_2 d_3}_{e_1 b}-\overset{(1)}{\underset{(r,l)}{i}}(k_1,k_2)^{d_2}_{b e_1} \overset{(1)}{\underset{(t,s)}{a}}(l,k_3)^{e_1 d_3}_{ac}-\overset{(1)}{\underset{(t,r)}{i}}(k_1,l)^{e_1}_{ab} \overset{(1)}{\underset{(l,s)}{f}}(k_2,k_3)^{d_2 d_3}_{e_1 c}  \\
    &\quad \quad \quad \quad +\overset{(1)}{\underset{(s,l)}{i}}(k_1,k_2)^{d_2}_{c e_1} \overset{(1)}{\underset{(t,r)}{a}}(l,k_3)^{e_1 d_3}_{a b}+\alpha(t,k_1) K_{a e_1} \overset{(2)}{\underset{(r,s)}{a}}(k_1-t+1,k_2,k_3)^{e_1 d_2 d_3}_{bc}  \\
    &\quad \quad \quad \quad +\alpha(t,k_1) K_{a e_1} \overset{(2)}{\underset{(r,s)}{a}}(k_1-t+1,k_2,k_3)^{e_1 d_2 d_3}_{bc}-\alpha(r,k_1) K_{b e_1} \overset{(2)}{\underset{(t,s)}{a}}(k_1-r+1,k_2,k_3)^{e_1 d_2 d_3}_{ac}  \\
    & \quad \quad \quad \quad +\alpha(s,k_1) K_{c e_1} \overset{(2)}{\underset{(t,r)}{a}}(k_1-s+1,k_2,k_3)^{e_1 d_2 d_3}_{ab}\bigg) :F[k_1] \prod_{j=2}^{3} \tilde{J}_{d_j}[k_j]: \notag
\end{align*} \endgroup
Fix $k_1,k_2,k_3$ such that their sum equals $p_2$. Manipulating the indices, we can separate the expression into three pieces that must vanish independently:
\begingroup \allowdisplaybreaks \begin{align*}
    0 &= \bigg( -\overset{(2)}{\underset{(t,r+s)}{i}}(k_1,k_2,k_3)+\overset{(2)}{\underset{(s,t+r)}{i}}(k_1,k_2,k_3)+\overset{(2)}{\underset{(t,s)}{i}}(k_1,k_2,k_3-r) \\
    &\quad \quad +\overset{(2)}{\underset{(t,s)}{i}}(k_1,k_3-r,k_2)+\overset{(2)}{\underset{(r,s)}{i}}(k_1,k_3,k_2-t)+\overset{(2)}{\underset{(t,r)}{i}}(k_1,k_2-s,k_3)-\overset{(1)}{\underset{(r,s)}{i}}(k_1,l) \overset{(1)}{\underset{(l,t)}{f}}(k_3,k_2) \\
    &\quad \quad + \overset{(1)}{\underset{(t,r)}{i}}(k_1,l) \overset{(1)}{\underset{(l,s)}{f}}(k_3,k_2) -\overset{(1)}{\underset{(r,l)}{i}}(k_1,k_3) \overset{(1)}{\underset{(t,s)}{f}}(k_2,l) -\overset{(1)}{\underset{(t,l)}{i}}(k_1,k_2) \overset{(1)}{\underset{(r,s)}{f}}(k_3,l) - \overset{(1)}{\underset{(r,l)}{i}}(k_1,k_3) \overset{(1)}{\underset{(t,s)}{f}}(l,k_2)    \\
    & \quad \quad +\overset{(1)}{\underset{(s,l)}{i}}(k_1,k_2) \overset{(1)}{\underset{(r,t)}{f}}(k_3,l) +\alpha(r,k_1) \overset{(2)}{\underset{(t,s)}{f}}(k_2,k_3,k_1+1-r) +\alpha(s,k_1) \overset{(2)}{\underset{(t,r)}{f}}(k_2,k_3,k_1+1-s)  \\
    & \quad \quad -\alpha(t,k_1) \overset{(2)}{\underset{(r,s)}{f}}(k_1+1-t,k_3,k_2)-\alpha(r,k_1) \overset{(2)}{\underset{(t,s)}{f}}(k_1+1-r,k_3,k_2)+\alpha(s,k_1) \overset{(2)}{\underset{(t,r)}{f}}(k_2,k_1+1-s,k_3)  \\
    & \quad \quad -\alpha(t,k_1) \overset{(2)}{\underset{(r,s)}{f}}(k_3,k_1+1-t,k_2)\bigg)K^{d_2 d_3}_{a e_1} f^{e_1}_{bc}  \\
    &+\bigg( -\overset{(2)}{\underset{(t,r+s)}{i}}(k_1,k_3,k_2)+\overset{(2)}{\underset{(t,r)}{i}}(k_1,k_3,k_2-s)-\overset{(2)}{\underset{(r,s)}{i}}(k_1,k_3-t,k_2)-\overset{(2)}{\underset{(s,t+r)}{i}}(k_1,k_2,k_3)  \\
    & \quad \quad -\overset{(1)}{\underset{(r,s)}{i}}(k_1,l) \overset{(1)}{\underset{(l,t)}{f}}(k_2,k_3) + \overset{(1)}{\underset{(t,l)}{i}}(k_1,k_3) \overset{(1)}{\underset{(r,s)}{f}}(l,k_2)-\overset{(1)}{\underset{(t,r)}{i}}(k_1,l) \overset{(1)}{\underset{(l,s)}{f}}(k_3,k_2)-\overset{(1)}{\underset{(s,l)}{i}}(k_1,k_2) \overset{(1)}{\underset{(t,r)}{f}}(k_3,l)  \\
    & \quad \quad +\alpha(r,k_1) \overset{(2)}{\underset{(t,s)}{f}}(k_3,k_2,k_1+1-r)+\alpha(s,k_1) \overset{(2)}{\underset{(t,r)}{f}}(k_3,k_2,k_1+1-s)+\alpha(r,k_1) \overset{(2)}{\underset{(t,s)}{f}}(k_3,k_1+1-r,k_2)  \\
    &\quad \quad +\alpha(r,k_1) \overset{(2)}{\underset{(t,s)}{f}}(k_1+1-r,k_3,k_2)+\alpha(t,k_1) \overset{(2)}{\underset{(r,s)}{f}}(k_1+1-t,k_3,k_2)\bigg) K^{d_3 d_2}_{a e_1}f^{e_1}_{bc}  \\
    & + \bigg( \overset{(2)}{\underset{(s,t+r)}{i}}(k_1,k_2,k_3)+\overset{(2)}{\underset{(s,t+r)}{i}}(k_1,k_3,k_2)-\overset{(2)}{\underset{(t,r)}{i}}(k_1,k_2,k_3-s)+\overset{(2)}{\underset{(r,s)}{i}}(k_1,k_2-t,k_3)  \\
    & \quad \quad +\overset{(2)}{\underset{(t,s)}{i}}(k_1,k_2,k_3-r)+\overset{(2)}{\underset{(r,s)}{i}}(k_1,k_3,k_2-t)+\overset{(2)}{\underset{(t,s)}{i}}(k_1,k_3-r,k_2)+\overset{(2)}{\underset{(t,r)}{i}}(k_1,k_2-s,k_3)  \\
    & \quad \quad - \overset{(1)}{\underset{(t,l)}{i}}(k_1,k_2) \overset{(1)}{\underset{(r,s)}{f}}(l,k_3)+\overset{(1)}{\underset{(t,r)}{i}}(k_1,l) \overset{(1)}{\underset{(l,s)}{f}}(k_3,k_2)-\overset{(1)}{\underset{(r,l)}{i}}(k_1,k_3) \overset{(1)}{\underset{(t,s)}{f}}(k_2,l)+\overset{(1)}{\underset{(s,l)}{i}}(k_1,k_3) \overset{(1)}{\underset{(t,r)}{f}}(k_2,l)  \\
    & \quad \quad -\overset{(1)}{\underset{(t,l)}{i}}(k_1,k_2) \overset{(1)}{\underset{(r,s)}{f}}(k_3,l)-\overset{(1)}{\underset{(r,l)}{i}}(k_1,k_3) \overset{(1)}{\underset{(t,s)}{f}}(l,k_2)+\overset{(1)}{\underset{(s,l)}{i}}(k_1,k_2) \overset{(1)}{\underset{(r,t)}{f}}(k_3,l)+\overset{(1)}{\underset{(t,r)}{i}}(k_1,l) \overset{(1)}{\underset{(l,s)}{f}}(k_2,k_3)  \\
    &\quad \quad -\alpha(r,k_1) \overset{(2)}{\underset{(t,s)}{f}}(k_2,k_1+1-r,k_3)-\alpha(r,k_1) \overset{(2)}{\underset{(t,s)}{f}}(k_1+1-r,k_3,k_2)-\alpha(r,k_1) \overset{(2)}{\underset{(t,s)}{f}}(k_1+1-r,k_2,k_3)  \\
    & \quad \quad -\alpha(t,k_1) \overset{(2)}{\underset{(r,s)}{f}}(k_1+1-t,k_3,k_2)-\alpha(t,k_1) \overset{(2)}{\underset{(r,s)}{f}}(k_3,k_1+1-t,k_2)-\alpha(t,k_1) \overset{(2)}{\underset{(r,s)}{f}}(k_1+1-t,k_2,k_3)  \\
    & \quad \quad +\alpha(s,k_1) \overset{(2)}{\underset{(t,r)}{f}}(k_2,k_1+1-s,k_3) \bigg) K^{d_2 e_1}_{ab} f^{d_3}_{c e_1} -(a,t) \leftrightarrow (b,r). 
\end{align*} \endgroup
Using eq.(\ref{eq:recursion}), we can show that all three equations vanish, and so the associativity condition is satisfied. \\ 

The methods for determining $\overset{(m)}{g}$ are identical to the ones used in this section, so we omit those details. 

\subsection{Determination of $a_*$ and $d_*$}
\noindent The determination of the coefficients in the matter-only sector have been labeled with a subscript $*$ to facilitate structural comparisons to their gauge/axion counterparts. The derivations are essentially identical, upon extending the Lie algebra to $\mathfrak{g}_R$ as described in the main text. 
The only new technical tool is to use a generalized Jacobi identity for all the $g, f$ structure constants of $\mathfrak{g}_R$ (in addition to the usual Jacobi identity constraining the $f$s). The relations can be easily derived using the Jacobi identity for Lie superalgebras:
\begin{equation}
    (-1)^{F_x F_z}[x, [y, z]] + (-1)^{F_x F_y}[y, [z, x]] + (-1)^{F_z F_y}[z, [x, y]]= 0
\end{equation}
and specializing to our superalgebra $\mathfrak{g}_R$ become
\begin{align}
0&= f^e_{ad}f^d_{bc} + f^e_{bd}f^d_{ca} + f^e_{cd}f^d_{ab}\\
0&= g^{k}_{ja}g^{j}_{ib}- g^k_{jb}g^j_{ia} + g^k_{id}f^d_{ab} \\
0& = g^l_{kd} g^d_{ij} + g^l_{id}g^d_{jk} + g^l_{jd}g^d_{ki}.
\end{align}
For example, eq.(\ref{eq:matter1}), when expanded, depends on $a, a_*, d_*$. We have solved for the former already in this appendix, and by analogous steps find
\begingroup \allowdisplaybreaks\begin{align*} 
    \overset{(m)}{\underset{(t,r)}{a_*}}[k_1,...,k_{m+1}]^{d_3 \cdot \cdot \cdot d_{m+1},ij}_{ab} &= \overset{(m)}{\underset{(t,r)}{f}}(k_1,...,k_{m+1}) G^{ij d_3 \cdot \cdot \cdot d_{m+1}}_{ab}+...+\overset{(m)}{\underset{(t,r)}{f}}(k_1,...,k_{m+1},k_2) G^{i d_3 \cdot \cdot \cdot d_{m+1} j}_{ab}  \\
    &+ \overset{(m)}{\underset{(t,r)}{f}}(k_2,k_1...,k_{m+1}) G^{j i d_3 \cdot \cdot \cdot d_{m+1}}_{ab}+...+\overset{(m)}{\underset{(t,r)}{f}}(k_3,k_1,...,k_{m+1},k_2) G^{ d_3 i \cdot \cdot \cdot d_{m+1} j}_{ab}+...  \\
    &+ \overset{(m)}{\underset{(t,r)}{f}}(k_2,...,k_{m+1},k_1) G^{j d_3 \cdot \cdot \cdot d_{m+1} i}_{ab}+...+\overset{(m)}{\underset{(t,r)}{f}}(k_3,...,k_{m+1},k_2,k_1) G^{d_3 \cdot \cdot \cdot d_{m+1} j i}_{ab}  \\  \\
    \overset{(m)}{\underset{(t,r)}{d_*}}[k_1,...,k_{m+1}]^{d_2 \cdot \cdot \cdot d_{m+1},j}_{b,i} &= \overset{(m)}{\underset{(t,r)}{f}}(k_1,...,k_{m+1}) G^{j d_2 \cdot \cdot \cdot d_{m+1}}_{ib}+...+\overset{(m)}{\underset{(t,r)}{f}}(k_2,...,k_{m+1},k_1) G^{d_2 \cdot \cdot \cdot d_{m+1} j}_{ib}
\end{align*} \endgroup

\subsubsection{Solving For $a_*$ and $d_*$ at 2 Loops}
\noindent We expand both sides of eq.\eqref{eq:matter1} and focus on the coefficients of the operators 
\begin{equation}
    :M_{i_1}[k_1]\overset{3}{\underset{j=2}{\prod}}\tilde{J}_{d_j}[k_2]:
\end{equation} 
Let $p = t+r-1$. The coefficients on the two sides of the equation at order $\hbar^2$ are: 
\begin{equation}
    \textit{lhs} = \overset{\sum k_j = p}{\sum_{ k_j^i \geq 0}} \left(\overset{(2)}{\underset{(t,r)}{a}}[k_1, k_2, k_3]^{d_1 d_2 d_3}_{ab} g^{i_1}_{i d_1} - \overset{(2)}{\underset{(t,r)}{a_*}}[k_1, k_2, k_3]^{d_3; i_1 i_2}_{ab} g^{d_2}_{i_2 i} \right)
\end{equation}
\begin{equation}
    \textit{rhs} =  \overset{\sum k_j = p}{\sum_{ k_j^i \geq 0}} \left( \overset{(2)}{\underset{(t,r)}{d_*}}[k_1, k_2, k_3]^{d_2 d_3; i_1}_{b; j_1} g^{j_1}_{i a} - \overset{(2)}{\underset{(r, t)}{d_*}}[k_1, k_2, k_3]^{d_2 d_3; i_1}_{a; j_1} g^{j_1}_{i b} \right)
\end{equation}
To each side we plug in the proposed solutions: 
\begingroup \allowdisplaybreaks \begin{align*}
 \overset{(2)}{\underset{(t,r)}{a}}[k_1, k_2, k_3]^{d_1 d_2 d_3}_{ab} & = \overset{(2)}{\underset{(t,r)}{f}}(k_1, k_2, k_3)K^{d_1 d_2 d_3}_{ab} +  \overset{(2)}{\underset{(t,r)}{f}}(k_2, k_1, k_3)K^{d_2 d_1 d_3}_{ab} + \overset{(2)}{\underset{(t,r)}{f}}(k_2, k_3, k_1)K^{d_2 d_3 d_1}_{ab} 
 \\ \\
 \overset{(2)}{\underset{(t,r)}{a_*}}[k_1, k_2, k_3]^{d_3; i_1 i_2}_{ab} & = \overset{(2)}{\underset{(t,r)}{f}}(k_1, k_2, k_3)G_{ab}^{i_1 i_2 d_3} + \overset{(2)}{\underset{(t,r)}{f}}(k_1, k_3, k_2)G_{ab}^{i_1 d_3 i_2} + \overset{(2)}{\underset{(t,r)}{f}}(k_2, k_1, k_3) G_{ab}^{i_2 i_1 d_3} 
 \\ 
 & + \overset{(2)}{\underset{(t,r)}{f}}(k_3, k_1, k_2)G_{ab}^{d_3 i_1 i_2} + \overset{(2)}{\underset{(t,r)}{f}}(k_2, k_3, k_1)G_{ab}^{i_2 d_3 i_1}+ \overset{(2)}{\underset{(t,r)}{f}}(k_3, k_2, k_1) G_{a b}^{d_3 i_2 i_1}
 \\ \\
 \overset{(2)}{\underset{(t,r)}{d_*}}[k_1, k_2, k_3]^{d_2 d_3; i_1}_{b; j_1} &= \overset{(2)}{\underset{(t,r)}{f}}(k_1, k_2, k_3)G_{j_1 b}^{i_1 d_2 d_3} +  \overset{(2)}{\underset{(t,r)}{f}}(k_2, k_1, k_3)g^{d_2}_{j_1 j_2}G_{j_1 b}^{d_2 i_1 d_3} + \overset{(2)}{\underset{(t,r)}{f}}(k_2, k_3, k_1)G_{j_1 b}^{d_2 d_3 i_1}.
\end{align*} \endgroup
The left-hand side, after grouping terms and using generalized Jacobi, becomes (suppressing the overall sum for tidiness and the normalizing factors of $K$ and $G$)
\begingroup \allowdisplaybreaks \begin{align*}
    \textit{lhs}&= \overset{(2)}{\underset{(t,r)}{f}}(k_1, k_2, k_3)K^{d_1 d_2 d_3}_{ab}g^{i_1}_{i d_1} +  \overset{(2)}{\underset{(t,r)}{f}}(k_2, k_1, k_3)K^{d_2 d_1 d_3}_{ab} g^{i_1}_{i d_1} + \overset{(2)}{\underset{(t,r)}{f}}(k_2, k_3, k_1)K^{d_2 d_3 d_1}_{ab}g^{i_1}_{i d_1} \\ 
    & - \overset{(2)}{\underset{(t,r)}{f}}(k_1, k_2, k_3)g^{i_1}_{aj_1}g^{i_2}_{j_1 e_1}f^{d_3}_{b e_1}g^{d_2}_{i_2 i} - \overset{(2)}{\underset{(t,r)}{f}}(k_1, k_3, k_2)g^{i_1}_{aj_1}g^{d_3}_{j_1 j_2}g^{i_2}_{b j_2}g^{d_2}_{i_2 i} - \overset{(2)}{\underset{(t,r)}{f}}(k_2, k_1, k_3)g^{i_1}_{j_1e_2}f^{d_3}_{b e_2}g^{d_2}_{j_1 i_2}g^{i_2}_{i a} \\
    &- \overset{(2)}{\underset{(t,r)}{f}}(k_3, k_1, k_2)f^{d_3}_{a e_1}g^{i_1}_{e_1 j_1}g^{i_2}_{b j_1}g^{d_2}_{i_2 i} - \overset{(2)}{\underset{(t,r)}{f}}(k_2, k_3, k_1)g^{i_2}_{a j_1}g^{d_3}_{j_1 j_2}g^{i_1}_{bj_2}g^{d_2}_{i_2 i} - \overset{(2)}{\underset{(t,r)}{f}}(k_3, k_2, k_1)f^{d_3}_{a e_1}g^{i_2}_{e j_1}g^{i_1}_{bj_1}g^{d_2}_{i_2 i} \\
    &-\overset{(2)}{\underset{(t,r)}{f}}(k_2, k_1, k_3)f^{d_2}_{a e_1}f^{d_3}_{b e_2}(g^{i_1}_{i d_1}f^{d_1}_{e_1 e_2} + g^{i_1}_{j_1 e_1}g^{j_1}_{i e_2}).
\end{align*} \endgroup
Several terms in the above immediately cancel, so that the left-hand side collapses to
\begingroup \allowdisplaybreaks \begin{align*}
    \textit{lhs}&= \overset{(2)}{\underset{(t,r)}{f}}(k_1, k_2, k_3)K^{d_1 d_2 d_3}_{ab}g^{i_1}_{i d_1} +  \overset{(2)}{\underset{(t,r)}{f}}(k_2, k_3, k_1)K^{d_2 d_3 d_1}_{ab}g^{i_1}_{i d_1} \\ 
    & -\overset{(2)}{\underset{(t,r)}{f}}(k_1, k_2, k_3)f^{d_3}_{b e_2}g^{i_1}_{a j_1}f^{d_2}_{e_1 e_2}g^{j_1}_{i e_1} + \overset{(2)}{\underset{(t,r)}{f}}(k_1, k_2, k_3) g^{i_1}_{a j_1}g^{d_2}_{j_1 j_2}g^{i_2}_{i b}g^{d_3}_{i_2 j_2}\\
    & -\overset{(2)}{\underset{(t,r)}{f}}(k_2, k_1, k_3)g^{i_1}_{j_1 e_2}f^{d_3}_{b e_2}g^{d_2}_{j_1 i_2}g^{i_2}_{ia}  -\overset{(2)}{\underset{(t,r)}{f}}(k_3, k_1, k_2)f^{d_3}_{a e_1}g^{i_1}_{e_1 j_1}g^{i_2}_{b j_1}g^{d_2}_{i_2 i} \\
    & -\overset{(2)}{\underset{(t,r)}{f}}(k_2, k_3, k_1)g^{d_3}_{j_1 j_2}g^{i_1}_{bj_2}g^{i_2}_{ia}g^{d_2}_{j_1 i_2}-\overset{(2)}{\underset{(t,r)}{f}}(k_2, k_1, k_3)f^{d_2}_{a e_1}f^{d_3}_{b e_2}g^{i_1}_{j_1 e_1}g^{j_1}_{i e_2} + \overset{(2)}{\underset{(t,r)}{f}}(k_3, k_2, k_1)f^{d_3}_{a e_2}g^{i_1}_{b j_1}f^{d_2}_{e_1 e_2}g^{j_1}_{i e_1}
\end{align*} \endgroup
Similarly, plugging in to the right hand side gives
\begingroup \allowdisplaybreaks \begin{align*}
    \textit{lhs}&= \overset{(2)}{\underset{(t,r)}{f}}(k_1, k_2, k_3)g^{i_1}_{j_1 e_1}f^{d_2}_{e_1 e_2}f^{d_3}_{b e_2}g^{j_1}_{ia} +  \overset{(2)}{\underset{(t,r)}{f}}(k_2, k_1, k_3)g^{d_2}_{j_1 j_2}g^{i_1}_{j_2 e_1}f^{d_3}_{b e_1}g^{j_1}_{ia} +  \overset{(2)}{\underset{(t,r)}{f}}(k_2, k_3, k_1)g^{d_2}_{j_1 j_2}g^{d_3}_{j_2 j_3}g^{i_1}_{b j_3}g^{j_1}_{ia} \\
    & - \overset{(2)}{\underset{(t,r)}{f}}(k_1, k_2, k_3)g^{i_1}_{j_1 e_1}f^{d_2}_{e_1 e_2}f^{d_3}_{a e_2}g^{j_1}_{ib} - \overset{(2)}{\underset{(t,r)}{f}}(k_2, k_1, k_3)g^{d_2}_{j_1 j_2}g^{i_1}_{j2 e_1}f^{d_3}_{a e_1}g^{j_1}_{ib}- \overset{(2)}{\underset{(t,r)}{f}}(k_2, k_3, k_1)g^{d_2}_{j_1 j_2}g^{d_3}_{j_2 j_3}g^{i_1}_{a j_3}g^{j_1}_{ib}
\end{align*} \endgroup

Grouping together like terms and again using the Jacobi identities for the structure constants of $\mathfrak{g}_R$ gives an expression identical to the left-hand side, verifying that the associativity equation is satisfied at two-loops.

\subsection{Determination of $c$ and $b$}
\noindent The final coefficients we will determine,  $c_{\lambda}$ and $b_{\lambda}$, appear in the following OPE:
\begin{equation*}
    J_a[t](z)J_b[r](0)  \sim  \quad \overset{\sum_{j=1}^{m} k_j = t+r-m}{\sum_{ m \geq 1 \quad k_j^i \geq 0}} \hbar^m \bigg( \frac{1}{z^2}\overset{(m)}{\underset{(t,r)}{b_{\lambda}}}[k_1,...,k_{m}]^{d_1 \cdot \cdot \cdot d_m}_{ab}+\frac{1}{z}\overset{(m)}{\underset{(t,r)}{c_{\lambda}}}[k_1,...,k_{m}]^{d_1 \cdot \cdot \cdot d_m}_{ab} \hat{\partial_1} \bigg):\prod_{j=1}^{m} \tilde{J}_{d_j}[k_j]:
\end{equation*}
For notational convenience, let $p_m = t+r-m$. These coefficients are related (see eq.\eqref{eq:b}), and it is helpful to approach their solutions in parallel. \\

We begin by working through some illustrative low order examples, which will help set us up for the general solution.

\subsubsection{Solving For $c$ and $b$ at 1 Loop}\label{sec:c_1}
\noindent Let us begin by deriving the solution for $c$ at one loop, which will then enable us to solve for $b$ at one loop as well. \\ \\
We first consider eq.(\ref{eq:minimalset02}) with $r = 0$.
\begin{equation}
    0 = \overset{(1)}{\underset{(t,s)}{c_{\lambda}}}[t+s-1]_{e_1 c}^{d_1} f_{a b}^{e_1} + \overset{(1)}{\underset{(t,s)}{c_{\lambda}}}[t+s-1]_{a c}^{e_1} f_{b e_1}^{d_1} - \overset{(1)}{\underset{(t,s)}{c_{\lambda}}}[t+s-1]_{a e_1}^{d_1} f_{b c}^{e_1}
\end{equation}
Comparing this to the Jacobi identity
\begin{equation}
    0 = f_{e_1 c}^{d_1} f_{a b}^{e_1} + f_{a c}^{e_1} f_{b e_1}^{d_1} - f_{a e_1}^{d_1} f_{b c}^{e_1},
\end{equation}
we see that we should first look for a solution of the form
\begin{equation} \label{eq: c^1 form}
    \overset{(1)}{\underset{(t,s)}{c_{\lambda}}}[t+s-1]_{ab}^{d_1} = \overset{(1)}{\underset{(t,s)}{c_{\lambda}}} f_{ab}^{d_1}. 
\end{equation}
Next, we consider eq.(\ref{eq:minimalset02}) with $s = 0$. We will explicitly diagonalize $K^{d_1 d_2} = \kappa \delta^{d_1 d_2}$ and $R^{i_1 i_2} = \kappa \delta^{i_1 i_2}$, as in the main text.
\begingroup \allowdisplaybreaks
\begin{align}
    -\overset{(1)}{\underset{(t,r)}{c_{\lambda}}}f_{a e_1 d_1} f_{b c e_1} + \overset{(1)}{\underset{(r,t)}{c_{\lambda}}} f_{b e_1 d_1} f_{a c e_1} &= \hat{\lambda}_{\mathfrak{g}, R}^2 \kappa^{-1}(t^2 r^1 - t^1 r^2) ( \delta_{ab} \delta_{c d} + \delta_{ac} \delta_{b d} + \delta_{bc} \delta_{a d}) \notag \\
     &+ \overset{\sum l_j = p_1}{\sum_{l_j^i \geq 0}} \overset{(1)}{\underset{(t,r)}{f}}(l_1, l_2) \kappa \text{Tr}_{\mathfrak{g}_R}(t_at_b\{t_c,t_{d_1}\})
\end{align} \endgroup
We plug in the form (\ref{eq: c^1 form}) to the preceding equation, and compare the number of pairs of structure constants between the $f$ term and the  $c_\lambda$ terms. \\ \\
Using the result, and employing the Lie algebra identities
\begin{align}
K^{cd}f^a_{cd_1}f^{d_1}_{d b} &= 2h^{\vee}\delta^a_b \\ \notag \\
R^{ij}g^k_{bi}g^a_{kj} &= R^{ij}g_{bik}g_{akj} =-T(R) \delta^a_b
\end{align}\label{eq: algebraIdentity h}
 we can deduce that $\overset{(1)}{c_\lambda}$ can be expressed as
\begin{equation}
    \overset{(1)}{\underset{(t,r)}{c_{\lambda}}}[p_1]_{ab}^{c} = (2 h^\vee - T(r)) \overset{(1)}{\underset{(t,r)}{c}} f_{ab}^c.
\end{equation}
Notice that $c(p_1)$ has no algebra dependence (recall that $f(k_1, k_2)$ is algebra-independent). This means that we can specialize to any anomaly-cancelling choice of $\mathfrak{g}_R$ (and the corresponding $\lambda_{\mathfrak{g}, R}^2$) and determine the numerical value of the undetermined coefficient. \\ 

To this end, we specialize to, for simplicity, the case  $\mathfrak{g} = \mathfrak{su}(2)$, coupled only to the axion-like field, where, in our conventions, $f_{a b c} \to \epsilon_{abc}$, $h^\vee = 2$, $K^{ab} = -2 \delta^{ab}$ and $\lambda_{\mathfrak{g}}^2 = 8$. (It is easily verified that other choices will give the same result). We will use this choice of data throughout this appendix as a convenient intermediate step; the general solutions for an arbitrary choice of anomaly-cancelling data is presented in the main text.  \\

To isolate the desired coefficient, we fix gauge indices so that $ a = c$ and $a \neq b = d$.
Using the $\mathfrak{su}(2)$ identity
\begin{equation}
    \text{Tr}_{\text{adj}}(t_at_b\{t_c,t_d\}) = \delta_{ad}\delta_{bc}+\delta_{ac}\delta_{bd}+2\delta_{ab}\delta_{cd}
\end{equation}
we find the following expression for $\overset{(1)}{c}$:
\begin{equation}
    \overset{(1)}{\underset{(t,r)}{c}} = \frac{1}{2} \overset{\sum l_j = p_1}{\sum_{l_j^i \geq 0}} \overset{(1)}{\underset{(t,r)}{f}}(l_1, l_2) -\bigg(\frac{1}{48 \pi^2}\bigg) (t^2 r^1 - t^1 r^2) 
\end{equation} 
To check the solution, we consider all other nontrivial, nonvanishing combinations of gauge indices and verify that each equation is satisfied. \\ \\
For example, fix $a = b \neq c = d$ 
\begingroup \allowdisplaybreaks
\begin{align*}
    - \overset{(1)}{\underset{(t,r)}{c}} + \overset{(1)}{\underset{(r,t)}{c}} = \overset{\sum l_j = p_1}{\sum_{l_j^i \geq 0}} \overset{(1)}{\underset{(t,r)}{f}}(l_1, l_2) - \frac{1}{48 \pi^2}(t^2 r^1 - t^1 r^2)
    \implies \overset{\sum l_j = p_1}{\sum_{l_j^i \geq 0}} \overset{(1)}{\underset{(t,r)}{f}}(l_1, l_2)  = \frac{1}{32 \pi^2}(t^2r^1-t^1r^2)
\end{align*} \endgroup
which can be proven to hold by using induction on the recursion relation eq.(\ref{eq:rec_rel_1}). \\ \\
Finally, consider $t = (0,1)$ and $r = (1,0)$:
\begin{equation}
    \overset{(1)}{\underset{(0,1)(1,0)}{c_{\lambda}}}[0]_{ab}^{d} = \bigg(-\frac{1}{24 \pi^2} + \frac{1}{32 \pi^2} \bigg) h^\vee f_{ab}^d = -\bigg(\frac{h^\vee}{96 \pi^2}\bigg) f_{ab}^d.
\end{equation}
Similarly, we find:
\begin{equation}
    \overset{(1)}{\underset{(1,0)(0,1)}{c_{\lambda}}}[0]_{ab}^{d} = \bigg(\frac{1}{24 \pi^2} - \frac{1}{32 \pi^2} \bigg) h^\vee f_{ab}^d = \bigg(\frac{h^\vee}{96 \pi^2}\bigg) f_{ab}^d.
\end{equation}
These specializations agree precisely with the coefficients found independently using Koszul duality in \cite{victor} (as well as the rather simpler associativity computation of \cite{CPassoc}).

From this, we can also determine $\overset{(1)}{b}$ by plugging $\overset{(1)}{c}$ into (\eqref{eq:b}). For example, the specialization to the $\mathfrak{g} = \mathfrak{su}(2)$ axion-only case is
\begingroup \allowdisplaybreaks
\begin{align}
    \overset{(1)}{\underset{(t,r)}{c}}[p_1]_{a e_1}^d f_{b c}^{e_1} - \overset{(1)}{\underset{(r,t)}{c}}[p_1]_{b e_1}^d f_{a c}^{e_1}  &=
    - \hat{\lambda}_{\mathfrak{g}}^2 (t^2 r^1 - t^1 r^2)(K_{a b} \delta_c^d + K_{b c} \delta_a^d + K_{a c} \delta_b^d) \notag \\
    & +\underset{l_j^i \geq 0}{\overset{\sum l_j=p_1}{\sum}} \overset{(1)}{\underset{(t,r)}{a}}[l_1,l_2]_{a b}^{e_1 e_2} f_{c e_1}^{e_3} f_{e_3 e_2}^d
\end{align} \endgroup
In the above, it is to be understood that we have plugged in $\lambda_{\mathfrak{g}}^2=8$ (equivalently, $\hat{\lambda}_{\mathfrak{g}}^2 = {1 \over (2 \pi i )^2}\frac{8}{12}$).
This reduces to
\begingroup \allowdisplaybreaks
\begin{align}
    \overset{(1)}{\underset{(t,r)}{b}}(p_1) = \bigg( \frac{1}{96 \pi^2} \bigg) (t^1 r^2 - t^2 r^1).
\end{align} \endgroup

We can easily generalize the Lie algebraic data to the case of any anomaly-cancelling $\mathfrak{g}_R$ (indeed, at any loop order)\footnote{Notice that if the anomaly is cancelled with ordinary matter only, one sets the $i, j$ terms to zero. The matter contribution to the associativity equation, as illustrated explicitly in the next subsection, arises in the form of $a_*$ terms which combine with the $a$'s to simply shift $2 h^{\vee} \mapsto 2h^{\vee} - T(R)$.}. Moreover, we can also rewrite the numerical coefficient suggestively, in anticipation of the $m$-loop generalization. These improvements yield:
\begin{equation}
    \overset{(1)}{\underset{(t,r)}{b}}[p_1]_{ab}^c = (2 h^\vee - T(R)) \bigg( \bigg(\frac{1}{48 \pi^2}\bigg) \overset{(0)}{\underset{(t,r)}{i}} +
    \underset{l_j^i \geq 0}{\overset{\sum l_j=p_1}{\sum}} \overset{(1)}{\underset{(t,r)}{f}}(l_1,l_2)
    \bigg)
\end{equation}

\subsubsection{Solving For $c$ and $b$ at 2 Loops}\label{sec:c_2}
\noindent We now derive the solution at two loops. We first consider (\ref{eq:minimalset02}) with $r = 0$.
\begin{equation}
    0 = -\overset{(2)}{\underset{(t,s)}{c_{\lambda}}}[k_1, k_2]_{e_1 c}^{d_1 d_2} f_{a b}^{e_1} + \overset{(2)}{\underset{(t,s)}{c_{\lambda}}}[k_1, k_2]_{a e_1}^{d_1 d_2} f_{b c}^{e_1} - \overset{(2)}{\underset{(t,s)}{c_{\lambda}}}[k_1, k_2]_{a c}^{e_1 d_2} f_{b e_1}^{d_1} - \overset{(2)}{\underset{(t,s)}{c_{\lambda}}}[k_1, k_2]_{a c}^{d_1 e_1} f_{b e_1}^{d_2}
\end{equation}
The most general compatible expression which solves this equation is 
\begingroup \allowdisplaybreaks
\begin{align}
    \overset{(2)}{\underset{(t,r)}{c_{\lambda}}}[k_1, k_2]_{ab}^{d_1 d_2} &= (2 h^\vee - T(R)) \overset{(2)}{\underset{(t,r)}{c_1}}(k_1, k_2) K_{ab}^{d_1 d_2} + (2 h^\vee - T(R)) \overset{(2)}{\underset{(t,r)}{c_2}}(k_1, k_2) K_{ab}^{d_2 d_1} \\
    & \quad + \hat{{\lambda}}_{\mathfrak{g}, R}^2 \overset{(2)}{\underset{(t,r)}{e_1}}(k_1, k_2) \delta_a^{d_1} \delta_b^{d_2} + \hat{{\lambda}}_{\mathfrak{g}, R}^2 \overset{(2)}{\underset{(t,r)}{e_2}}(k_1, k_2) \delta_a^{d_2} \delta_b^{d_1}\notag \label{eq:cdecomp}
\end{align}
However, we find by explicitly solving the associativity equations (below) that the $\overset{(2)}{e}$ terms are zero, so we will anticipate this and drop them in what follows (this also will enable us to simplify our notation somewhat: $b_{\lambda}, c_{\lambda} \mapsto b, c$). The $e$-coefficients come from diagrams where a BF propagator in a loop is replaced by an axion propagator, and the latter contributes to the holomorphic integrand with additional derivatives; consequently, we expect that one can argue more generally that their contribution should vanish using arguments analogous to those in Appendix B of \cite{FPW}.  \\ \\
We now consider eq.(\ref{eq:minimalset02}) with $s = 0$. Again, we will explicitly use $K^{d_1 d_2} = \kappa \delta^{d_1 d_2}$ and $R^{i_1 i_2} = \kappa \delta^{i_1 i_2}$.
\begingroup \allowdisplaybreaks
\begin{align}
    \textit{lhs} &= (2 h^\vee - T(R)) \kappa (\overset{(2)}{\underset{(t,r)}{c_1}}(k_1,k_2) - \overset{(2)}{\underset{(t,r)}{c_2}}(k_2,k_1)) f_{a d_2 d} f_{b d_2 d_1} f_{c d_1 e} \\
    &+ (2 h^\vee - T(R)) \kappa (\overset{(2)}{\underset{(t,r)}{c_2}}(k_1,k_2) - \overset{(2)}{\underset{(t,r)}{c_1}}(k_2,k_1)) f_{a d_2 d_1} f_{b d_2 d} f_{c d_1 e} \notag \\
    &+ (2 h^\vee - T(R)) \kappa \overset{(2)}{\underset{(t,r)}{c_1}}(k_1,k_2) f_{a d_2 d} f_{d_1 d_2 e} f_{b c d_1} + (2 h^\vee - T(R)) \kappa \overset{(2)}{\underset{(t,r)}{c_2}}(k_1,k_2) f_{a d_2 e} f_{d_1 d_2 d} f_{b c d_1} \notag \\
    &- (2 h^\vee - T(R)) \kappa \overset{(2)}{\underset{(r,t)}{c_1}}(k_1,k_2) f_{b d_2 d} f_{d_1 d_2 e} f_{a c d_1} - (2 h^\vee - T(R)) \kappa \overset{(2)}{\underset{(r,t)}{c_2}}(k_1,k_2) f_{b d_2 e} f_{d_1 d_2 d} f_{a c d_1} \notag
\end{align} \endgroup
\begingroup \allowdisplaybreaks
\begin{align}
    \textit{rhs}  &= \hat{\lambda}_{\mathfrak{g}, R}^2 \overset{(1)}{\underset{(t,r)}{i}}(k_1, k_2) f_{abe} \delta_{cd} + \hat{\lambda}_{\mathfrak{g}, R}^2 (t^1+t^2)\overset{(1)}{\underset{(t,r)}{j}}(k_1, k_2) f_{d b e} \delta_{ac} - \hat{\lambda}_{\mathfrak{g}, R}^2 (r^1+r^2)\overset{(1)}{\underset{(r,t)}{j}}(k_1, k_2) f_{d a e} \delta_{bc} \notag \\
    & + \kappa^2 \overset{\sum l_j = p_1}{\sum_{l_j^i \geq 0}} \overset{(1)}{\underset{(t,r)}{f}}(l_1,l_2) \overset{(1)}{\underset{(l_2,l_1)}{f}}(k_1,k_2) \text{Tr}_{\mathfrak{g}_R}(t_at_bt_dt_et_c) - \kappa^2 \overset{\sum l_j = p_1}{\sum_{l_j^i \geq 0}} \overset{(1)}{\underset{(t,r)}{f}}(l_2,l_1) \overset{(1)}{\underset{(l_2,l_1)}{f}}(k_1,k_2) \text{Tr}_{\mathfrak{g}_R}(t_at_bt_ct_et_d) \notag \\
    & + \kappa^2 \overset{\sum l_j = k_1}{\sum_{l_j^i \geq 0}} \overset{(2)}{\underset{(t,r)}{f}}(l_1,l_2, k_2) \text{Tr}_{\mathfrak{g}_R}(t_a[t_b,t_e]\{t_c,t_d\}) - \kappa^2 \overset{\sum l_j = k_1}{\sum_{l_j^i \geq 0}} \overset{(2)}{\underset{(t,r)}{f}}(l_1,k_2, l_2) \text{Tr}_{\mathfrak{g}_R}(t_at_et_b\{t_c,t_d\}) \notag \\
    & + \kappa^2 \overset{\sum l_j = k_1}{\sum_{l_j^i \geq 0}} \overset{(2)}{\underset{(t,r)}{f}}(k_2,l_1, l_2) \text{Tr}_{\mathfrak{g}_R}(t_a\{t_b,\{t_c,t_d\}\} t_e) + (k_1,d) \leftrightarrow (k_2,e)
\end{align}
One can also use various adjoint-trace identities to rewrite some of the coefficients in more convenient forms, such as $\text{Tr}_{\mathfrak{g}_R}(t_a\{t_b,\{t_c,t_d\}\} t_e) = \text{Tr}_{\mathfrak{g}_R}([t_a, t_e] t_b \{t_c, t_d\})$, $\text{Tr}_{\mathfrak{g}_R}(t_a [t_b, t_e] t_c t_d) = f_{bex}\text{Tr}_{\mathfrak{g}_R}(t_a t_x t_c t_d)$, etc. Expressions of the latter type can also appear in the anomaly-cancelling quartic identity after multiplying the identity on both sides by an additional structure constant and summing over the repeated index although, in contrast to the one-loop case \cite{CPassoc}, we did not find this the most convenient method for simplifying the two-loop equation. \\

We again specialize to the case where $\mathfrak{g}=\mathfrak{su}(2)$ and the anomaly is cancelled only by the axion to solve for $\overset{(2)}{\underset{(t,r)}{c_1}}$ and $\overset{(2)}{\underset{(t,r)}{c_2}}$. \\ \\
Fixing $a=c=e \neq b \neq d$ with $a \neq d$
\begingroup \allowdisplaybreaks
\begin{align}
    \overset{(2)}{\underset{(t,r)}{c_1}}(k_1,k_2) &= \bigg( \frac{1}{48 \pi^2}\bigg)\overset{(1)}{\underset{(t,r)}{i}}(k_2,k_1) - \bigg( \frac{1}{48 \pi^2}\bigg) (t^1+t^2) \overset{(1)}{\underset{(t,r)}{j}}(k_1,k_2) + \bigg( \frac{1}{48 \pi^2}\bigg) (t^1+t^2) \overset{(1)}{\underset{(t,r)}{j}}(k_2,k_1) \notag \\
    &+\frac{1}{2} \overset{\sum l_j = p_1}{\sum_{l_j^i \geq 0}} \overset{(1)}{\underset{(t,r)}{f}}(l_1,l_2) \overset{(1)}{\underset{(l_2,l_1)}{f}}(k_1,k_2) - \frac{1}{2}\overset{\sum l_j = k_1}{\sum_{l_j^i \geq 0}} \overset{(2)}{\underset{(t,r)}{f}}(l_1,l_2, k_2) \notag \\
    & - \frac{1}{2}\overset{\sum l_j = k_1}{\sum_{l_j^i \geq 0}} \overset{(2)}{\underset{(t,r)}{f}}(l_1,k_2, l_2) + 2 \overset{\sum l_j = k_2}{\sum_{l_j^i \geq 0}} \overset{(2)}{\underset{(t,r)}{f}}(l_1,l_2, k_1) + \overset{\sum l_j = k_2}{\sum_{l_j^i \geq 0}} \overset{(2)}{\underset{(t,r)}{f}}(l_1,k_1,l_2) \notag \\
    & + \overset{\sum l_j = k_2}{\sum_{l_j^i \geq 0}} \overset{(2)}{\underset{(t,r)}{f}}(k_1,l_1,l_2) 
\end{align} \endgroup
Combining the equations which result from the specializations $a \neq b = d = e \neq c, a \neq c$, and $a = b = e \neq c \neq d, a \neq d$ yields
\begingroup \allowdisplaybreaks
\begin{align}
    \overset{(2)}{\underset{(t,r)}{c_2}}(k_1,k_2) &= \bigg( \frac{1}{96 \pi^2}\bigg)\overset{(1)}{\underset{(t,r)}{i}}(k_1,k_2) - \bigg( \frac{1}{96 \pi^2}\bigg)\overset{(1)}{\underset{(r,t)}{i}}(k_2,k_1) \\
    &- \bigg( \frac{1}{96 \pi^2}\bigg) (t^1+t^2) \overset{(1)}{\underset{(t,r)}{j}}(k_2,k_1) + \bigg( \frac{1}{96 \pi^2}\bigg) (t^1+t^2) \overset{(1)}{\underset{(t,r)}{j}}(k_1,k_2) \notag \\
    &+ \bigg( \frac{1}{96 \pi^2}\bigg) (r^1+r^2) \overset{(1)}{\underset{(r,t)}{j}}(k_1,k_2) - \bigg( \frac{1}{96 \pi^2}\bigg) (r^1+r^2) \overset{(1)}{\underset{(r,t)}{j}}(k_2,k_1) \notag \\ 
    &+\frac{1}{2} \overset{\sum_{i=1}^2 l_i = t+r-1}{\sum} \overset{(1)}{\underset{(t,r)}{f}}(l_1,l_2) \overset{(1)}{\underset{(l_2,l_1)}{f}}(k_1,k_2)  \notag \\
    &+\overset{\sum_{i=1}^2 l_i = k_1}{\sum} \overset{(2)}{\underset{(t,r)}{f}}(l_1,l_2, k_2) +\frac{1}{2}\overset{\sum_{i=1}^2 l_i = k_1}{\sum} \overset{(2)}{\underset{(t,r)}{f}}(l_1,k_2, l_2) +\frac{1}{2}\overset{\sum_{i=1}^2 l_i = k_1}{\sum} \overset{(2)}{\underset{(t,r)}{f}}(k_2,l_1, l_2) \notag \\
    & - \overset{\sum_{i=1}^2 l_i = k_2}{\sum} \overset{(2)}{\underset{(t,r)}{f}}(l_1,l_2, k_1) - \overset{\sum_{i=1}^2 l_i = k_2}{\sum} \overset{(2)}{\underset{(t,r)}{f}}(l_1,k_1,l_2) - \overset{\sum_{i=1}^2 l_i = k_2}{\sum} \overset{(2)}{\underset{(t,r)}{f}}(k_1,l_1,l_2) \notag
\end{align} \endgroup
For completeness, we list the other specializations of the gauge indices that give nontrivial equations:
\begin{equation*}
    a = d = e \neq b \neq c \quad a \neq c \quad \quad \quad a \neq b \neq c = d = e \quad a \neq c \quad \quad \quad a \neq b = c = d \neq e \quad a \neq e
\end{equation*}
\begin{equation*}
    a = b = d \neq c \neq e \quad a \neq e \quad \quad \quad a = b = c \neq d \neq e \quad a \neq e
\end{equation*}
We can plug our solutions for $c$ into the resulting equations and find that they are satisfied, though we omit this here. \\

Our solution for $c$ is again sufficient to determine $b$ at the same loop order. The equations of course contain increasingly more terms as we increase the loop order, so to avoid clutter we will illustrate how this works for the case where the anomaly is cancelled by the axion only. In that case, the left-hand-side of the $m=2$ equation is
\begin{equation}
     \overset{(2)}{\underset{(t,r)}{c}}[k_1,k_2]_{a b}^{d_1 e_1} f_{c e_1}^{d_2} -
     \overset{(2)}{\underset{(t,r)}{c}}[k_2,k_1]_{a b}^{e_1 d_e} f_{c e_1}^{d_2} +
     \overset{(2)}{\underset{(t,r)}{c}}[k_1,k_2]_{a e_1}^{d_1 d_2} f_{b c}^{d_1} -
     \overset{(2)}{\underset{(r,t)}{c}}[k_1,k_2]_{b e_1}^{d_1 d_2} f_{a c}^{d_1},
\end{equation} \\
while the right-hand-side is 
\begingroup \allowdisplaybreaks
\begin{align}
    & -\hat{\lambda}_{\mathfrak{g}}^2 \overset{(1)}{\underset{(t,r)}{i}}[k_1,k_2]_{a b}^{d_1} \delta_c^{d_1} + \hat{\lambda}_{\mathfrak{g}}^2 (t^1+t^2) K_{a c} \overset{(1)}{\underset{(t,r)}{j}}[k_1,k_2]_{b}^{d_1 d_2} - \hat{\lambda}_{\mathfrak{g}}^2 (r^1+r^2) K_{b c} \overset{(1)}{\underset{(r,t)}{j}}[k_1,k_2]_{a}^{d_1 d_2} \notag \\
    &+ \underset{l_j^i \geq 0}{\overset{\sum l_j=p_1}{\sum}} \overset{(1)}{\underset{(t,r)}{a}}[l_1,l_2]_{a b}^{e_1 e_2} \overset{(1)}{\underset{(l_2,l_1)}{f}}[k_1,k_2]_{e_2 e_3}^{d_1 d_2} f_{e_1 c}^{e_3} + \underset{l_j^i \geq 0}{\overset{\sum l_j=k_1}{\sum}} \overset{(2)}{\underset{(t,r)}{a}}[l_1,l_2,k_2]_{a b}^{e_1 e_2 d_2} f_{c e_1}^{e_3} f_{e_3 e_2}^{d_1} \notag \\
    & + \underset{l_j^i \geq 0}{\overset{\sum l_j=k_1}{\sum}} \overset{(2)}{\underset{(t,r)}{a}}[l_1,k_2,l_2]_{a b}^{e_1 d_2 e_2} f_{c e_1}^{e_3} f_{e_3 e_2}^{d_1} + (d_1,d_2;k_1,k_2) \rightarrow (d_2,d_1;k_2,k_1).
\end{align} \endgroup
This reduces to
\begingroup \allowdisplaybreaks
\begin{align}
    \overset{(2)}{\underset{(t,r)}{b}}(k_1,k_2)+\overset{(2)}{\underset{(r,t)}{b}}(k_1,k_2) &= \bigg( \frac{1}{48 \pi^2} \bigg) \overset{(1)}{\underset{(r,t)}{i}}(k_1,k_2)+\bigg(\frac{1}{48 \pi^2} \bigg) \overset{(1)}{\underset{(t,r)}{i}}(k_2,k_1) \notag \\
    & + \underset{l_j^i \geq 0}{\overset{\sum l_j=p_1}{\sum}} \overset{(1)}{\underset{(t,r)}{f}}(l_1,l_2) \overset{(1)}{\underset{(l_2,l_1)}{f}}(k_2,k_1) + \underset{l_j^i \geq 0}{\overset{\sum l_j=p_1}{\sum}} \overset{(1)}{\underset{(r,t)}{f}}(l_1,l_2) \overset{(1)}{\underset{(l_2,l_1)}{f}}(k_2,k_1) \notag \\
    & + \underset{l_j^i \geq 0}{\overset{\sum l_j=k_1}{\sum}} \overset{(2)}{\underset{(t,r)}{f}}(k_2,l_1,l_2) + \underset{l_j^i \geq 0}{\overset{\sum l_j=k_2}{\sum}} \overset{(2)}{\underset{(t,r)}{f}}(l_1,l_2,k_1) \notag \\
    & + \underset{l_j^i \geq 0}{\overset{\sum l_j=k_1}{\sum}} \overset{(2)}{\underset{(r,t)}{f}}(k_2,l_1,l_2) + \underset{l_j^i \geq 0}{\overset{\sum l_j=k_2}{\sum}} \overset{(2)}{\underset{(r,t)}{f}}(l_1,l_2,k_1).
\end{align} \endgroup
The following expression is a solution to this equation
\begingroup \allowdisplaybreaks
\begin{align}
    \overset{(2)}{\underset{(t,r)}{b}}(k_1,k_2) &= \bigg( \frac{1}{96 \pi^2} \bigg) \overset{(1)}{\underset{(t,r)}{i}}(k_2,k_1) + \bigg( \frac{1}{96 \pi^2} \bigg) \overset{(1)}{\underset{(t,r)}{i}}(k_1,k_2)
    \\
    &+ \bigg( \frac{1}{96 \pi^2} \bigg) \bigg( (t^1+t^2)\overset{(1)}{\underset{(t,r)}{j}}(k_2,k_1) - (r^1+r^2)\overset{(1)}{\underset{(r,t)}{j}}(k_2,k_1) - (t^1+t^2)\overset{(1)}{\underset{(t,r)}{j}}(k_1,k_2) \notag \\
    & + (r^1+r^2)\overset{(1)}{\underset{(r,t)}{j}}(k_1,k_2)\bigg) \notag \\
    &+ \bigg(\frac{1}{2}\bigg) \underset{l_j^i \geq 0}{\overset{\sum l_j=p_1}{\sum}} \bigg( \overset{(1)}{\underset{(t,r)}{f}}(l_1,l_2) + \overset{(1)}{\underset{(r,t)}{f}}(l_1,l_2) \bigg)\overset{(1)}{\underset{(l_2,l_1)}{f}}(k_2,k_1) \notag \\
    & + \underset{l_j^i \geq 0}{\overset{\sum l_j=k_2}{\sum}} \overset{(2)}{\underset{(t,r)}{f}}(l_1,l_2,k_1) + \underset{l_j^i \geq 0}{\overset{\sum l_j=k_1}{\sum}} \overset{(2)}{\underset{(t,r)}{f}}(k_2,l_1,l_2) 
\end{align} \endgroup

Compared to the $c$ coefficients, the $b$ coefficients clearly exhibit a great deal of symmetry. In fact, the preceding expression turns out to be consistent with a symmetry we will shortly derive in eq.(\ref{sym:b_m}). From these explicit low-order solutions, we now proceed to derive a general solution for $b$. 

\subsubsection{The general solution for $b$}
\noindent Let us consider again the OPE
\begin{equation*}
    J_a[t](z)J_b[r](0)  \sim  \quad \overset{\sum_{j=1}^{m} k_j = t+r-m}{\sum_{ m \geq 1 \quad k_j^i \geq 0}} \hbar^m \bigg( \frac{1}{z^2}\overset{(m)}{\underset{(t,r)}{b}}[k_1,...,k_{m}]^{d_1 \cdot \cdot \cdot d_m}_{ab}+\frac{1}{z}\overset{(m)}{\underset{(t,r)}{c}}[k_1,...,k_{m}]^{d_1 \cdot \cdot \cdot d_m}_{ab} \hat{\partial_1} \bigg):\prod_{j=1}^{m} \tilde{J}_{d_j}[k_j]:
\end{equation*}
The left-hand-side is invariant under the exchange of $z$ and $w$ followed by renaming $z \leftrightarrow w$ and the relabeling $(t,a) \leftrightarrow (r,b)$. \\

Performing these operations on the right-hand-side and demanding invariance of the double-pole coefficient leads to the following equality
\begin{equation}
    \overset{(m)}{\underset{(t,r)}{b}}(k_1, k_2,\ldots,k_{m}) = (-1)^m \overset{(m)}{\underset{(r,t)}{b}}(k_{m}, k_{m-1}, \ldots, k_2,k_1) \label{sym:b_m}
\end{equation}
Invariance of the single-pole coefficient gives us
\begin{equation} \label{eq:c_m-to-b_m}
    \overset{(m)}{\underset{(t,r)}{b}}(k_2,..., k_j, k_1, k_{j+1},...,k_{m}) = \overset{(m)}{\underset{(t,r)}{c_j}}(k_1, k_2, \ldots, k_m) + (-1)^m \overset{(m)}{\underset{(r,t)}{c_{j^*}}}(k_1,k_m, k_{m-1}, \ldots, k_2)
\end{equation}
where $1 \leq j \leq m$ and we have defined $j^* = m+1-j$. The constituent terms $c_j$ are defined via the obvious $m$-loop generalization of eq.\eqref{eq:cdecomp},
\begin{equation*}
    \overset{(m)}{\underset{(t,r)}{c}}[k_1,,...,k_m]_{a b}^{d_1 \cdot \cdot \cdot d_m} = (2 h^\vee - T(R)) \bigg(
    \overset{(m)}{\underset{(t,r)}{c_1}}[k_1,,...,k_m]_{a b} K_{a b}^{d_1 \cdot \cdot \cdot d_m}
    +...+ \overset{(m)}{\underset{(t,r)}{c_m}}[k_1,,...,k_m]_{a b} K_{a b}^{d_2 \cdot \cdot \cdot d_m d_1}\bigg),
\end{equation*} in which the $d_1$ gauge index moves sequentially from left to right. 

Consider the equation (\ref{eq:minimalset02}) with $s=0$. For ease of presentation, and WLOG, we will again specialize to the case $\mathfrak{g} = \mathfrak{su}(2)$ with an anomaly-cancelling axion only for what follows. \\

Further, let us fix gauge indices indices such that $a,b \neq c= d_j$ for all $j$. For notational convenience, let $p_m = t+r-m$. \\ \\
It turns out that we can explicitly evaluate this equation for general $m$. First, let us determine the contributions that involve only the terms at the highest possible loop order. We will call these contributions the ``maximal terms'':
\begin{equation*}
    \overset{(m)}{b} \quad \quad \overset{(m-1)}{i} \quad \quad \overset{(m-1)}{j} \quad \quad \overset{(m)}{f}
\end{equation*}
Notice that the coefficients coming from axion exchange are at one lower loop order, by the Green-Schwarz mechanism.
The solution also has important contributions from ``sub-maximal'' terms, which we will determine subsequently, involving composite contributions of lower loop order:
\begin{equation*}
    \overset{(n)}{f} \overset{(m-n)}{f}.
\end{equation*}
We illustrate the diagrammatic contributions of such terms to $b$ in Figure \\
\ref{figure_contractions}.
\begin{figure}[t]
    \centering
    \begin{subfigure}[b]{0.4\textwidth}
        \centering
        \includegraphics[scale=0.30]{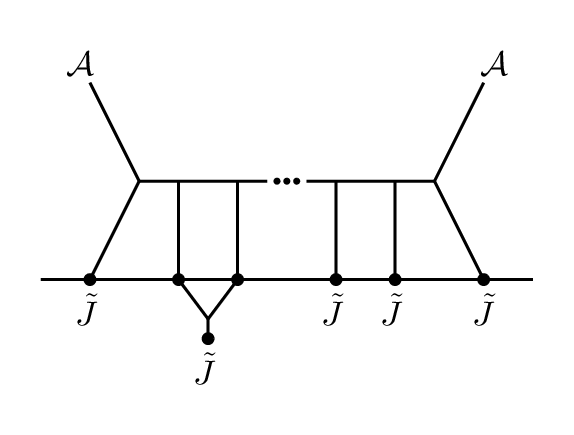}
    \end{subfigure}
    \begin{subfigure}[b]{0.4\textwidth}
        \centering
         \includegraphics[scale=0.30]{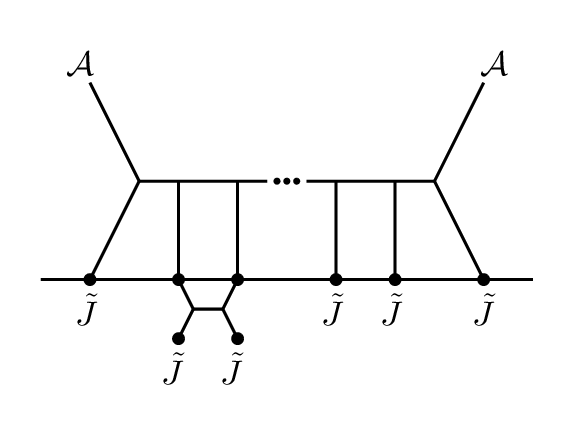}
    \end{subfigure}
     \caption{These diagrams contribute respectively to the maximal (left) and sub-maximal (right) $f$-terms, as defined in the main text. In both cases, only adjacent defect operators on the defect may contract with one another, which will produce contributions proportional to either $\overset{(m)}{f}$ as on the left or $\overset{(n)}{f} \overset{(m-n)}{f}$ as on the right.}
     \label{figure_contractions}
\end{figure}

We first fix $k_1,...,k_m$ such that $\sum k_j = p_m$. Per our specialization in the derivation below, it is understood that $\hat{\lambda}^2_{\mathfrak{g}} = {8 \over 12}{1 \over (2 \pi i)^2}$.
\begingroup \allowdisplaybreaks \begin{align*}
    \underset{\sigma \in S_{m-1}}{\sum} \underset{(t,r)}{\overset{(m)}{c}}[k_1&,k_{\sigma(2)},...,k_{\sigma(m)}]_{a e_1}^{c_1 \cdot \cdot \cdot c_m} f_{b c}^{e_1} - (a,t) \leftrightarrow (b,r) \sim - \hat{\lambda}_{\mathfrak{g}}^2 \underset{\sigma \in S_{m}}{\sum} \underset{(t,r)}{\overset{(m-1)}{i}}[k_{\sigma(1)},...,k_{\sigma(m)}]_{a b}^{c_1 \cdot \cdot \cdot c_{m-1}} 
    \\
    +&\overset{\sum l_j = k_{\sigma(1)}}{\underset{\sigma \in S_{m}}{\sum}} \bigg(
    \underset{(t,r)}{\overset{(m)}{a}}[l_1,l_2,k_{\sigma(2)},...,k_{\sigma(m)}]_{a b}^{e_1 e_2 c_1 \cdot \cdot \cdot c_{m-1}} +
    \underset{(t,r)}{\overset{(m)}{a}}[l_1,k_{\sigma(2)},l_2,...,k_{\sigma(m)}]_{a b}^{e_1 c_1 e_2 \cdot \cdot \cdot c_{m-1}} 
    \\
    & \quad \quad \quad \quad \quad +...+\underset{(t,r)}{\overset{(m)}{a}}[l_1,k_{\sigma(2)},...,k_{\sigma(m)},l_2]_{a b}^{e_1  c_1 \cdot \cdot \cdot c_{m-1} e_2} \bigg) f_{c e_1}^{e_3} f_{e_3 e_2}^c
\end{align*} \endgroup
where $c_j = c$ for all $j$, and $S_n$ is the symmetry group on $n$. \\ \\
When we take all the upper gauge indices to agree, the decomposition for $c$ in terms of its constituents becomes simply
\begin{equation*}
    \underset{(t,r)}{\overset{(m)}{c}}[k_1,...,k_m]_{a b}^{c_1 \cdot \cdot \cdot c_m} =
    2 h^\vee \bigg(\underset{(t,r)}{\overset{(m)}{c_1}}(k_1,...,k_m) + ...
    + \underset{(t,r)}{\overset{(m)}{c_m}}(k_1,...,k_m) \bigg) K_{a b}^{c_1 \cdot \cdot \cdot c_m}
\end{equation*}
We now use the following identities
\begin{equation}
    K_{a e_1}^{c_1 \cdot \cdot \cdot c_m} f_{b c}^{e_1} = \kappa^{-1} K_{ab}^{c_1 \cdot \cdot \cdot c_{m+1}} 
    \quad \quad
    K_{a b}^{c_1 \cdot \cdot \cdot c_{m+1}} = -(-1)^m K_{b a}^{c_{m+1} \cdot \cdot \cdot c_1}
\end{equation}
to simplify the left-hand-side to
\begingroup \allowdisplaybreaks \begin{align*}
    4 \kappa^{-1} \underset{\sigma \in S_{m-1}}{\sum} \bigg(\underset{(t,r)}{\overset{(m)}{c_1}}(k_1&,k_{\sigma(2)},...,k_{\sigma(m)}) + ... + \underset{(t,r)}{\overset{(m)}{c_m}}(k_1,k_{\sigma(2)},...,k_{\sigma(m)}) \\
    &+ (-1)^m \underset{(r,t)}{\overset{(m)}{c_1}}(k_1,k_{\sigma(2)},...,k_{\sigma(m)}) + ... + (-1)^m \underset{(r,t)}{\overset{(m)}{c_m}}(k_1,k_{\sigma(2)},...,k_{\sigma(m)}) \bigg) K_{a b}^{c_1 \cdot \cdot \cdot c_{m+1}}
\end{align*} \endgroup
Using equation (\ref{eq:c_m-to-b_m}) this becomes
\begin{equation*}
    4 \kappa^{-1} \underset{\sigma \in S_{m-1}}{\sum} \bigg(\underset{(t,r)}{\overset{(m)}{b}}(k_1,k_{\sigma(2)},...,k_{\sigma(m)}) + ... + \underset{(t,r)}{\overset{(m)}{b}}(k_{\sigma(2)},...,k_{\sigma(m)},k_1)\bigg) K_{a b}^{c_1 \cdot \cdot \cdot c_{m+1}}
\end{equation*}
which can in turn be expressed as a sum over $S_m$, the group of permutations on $m$ elements:
\begin{equation*}
    4 \kappa^{-1} \underset{\sigma \in S_m}{\sum} \underset{(t,r)}{\overset{(m)}{b}}(k_{\sigma(1)},...,k_{\sigma(m)}) K_{a b}^{c_1 \cdot \cdot \cdot c_{m+1}}
\end{equation*}
To simplify the right-hand-side, we first note that we have the following identity
\begin{equation}
    K_{ab}^{c_1 \cdot \cdot \cdot c_{m-1}} = - \kappa^{-2} K_{ab}^{c_1 \cdot \cdot \cdot c_{m+1}}
\end{equation} \label{id:K^m-to-K^m-2}
To see this, start with the right-hand-side and contract the indices of the last two Levi-Civita symbols:
\begin{equation*}
    - \kappa^{-2} K_{ab}^{c_1 \cdot \cdot \cdot c_{m+1}} = K_{a e_1}^{c_1 \cdot \cdot \cdot c_{m-1}} \epsilon_{e_1 e_2 c} \epsilon_{b e_2 c} =  K_{a b}^{c_1 \cdot \cdot \cdot c_{m-1}}.   
\end{equation*}
This simplifies the $\overset{(m)}{i}$ term (recalling $\kappa = -2, \lambda_{\mathfrak{g}}^2 = 8$)
\begin{equation*}
    -\hat{\lambda}_{\mathfrak{g}}^2 \underset{\sigma \in S_{m}}{\sum} \underset{(t,r)}{\overset{(m-1)}{i}}[k_{\sigma(1)},...,k_{\sigma(m)}]_{a b}^{c_1 \cdot \cdot \cdot c_{m-1}} 
    \to - \bigg( \frac{\kappa^{-2} \lambda_{\mathfrak{g}}^2}{48 \pi^2} \bigg) 
    \underset{\sigma \in S_{m}}{\sum} \underset{(t,r)}{\overset{(m-1)}{i}}(k_{\sigma(1)},...,k_{\sigma(m)}) K_{a b}^{c_1 \cdot \cdot \cdot c_{m+1}}
\end{equation*}
To understand the remaining terms, let us contract the two Levi-Civita symbols in front and expand the expression:
\begingroup \allowdisplaybreaks \begin{align*}
    -2 \overset{\sum l_j = k_{\sigma(1)}}{\underset{\sigma \in S_{m}}{\sum}} \bigg(
    &\underset{(t,r)}{\overset{(m)}{f}}(l_1,l_2,k_{\sigma(2)},...,k_{\sigma(m)})K_{a b}^{d d c_1 \cdot \cdot \cdot c_{m-1}} +
    \underset{(t,r)}{\overset{(m)}{f}}(l_1,k_{\sigma(2)},l_2,...,k_{\sigma(m)})K_{a b}^{d c_1 d \cdot \cdot \cdot c_{m-1}} + 
    \\
    ...+ &\underset{(t,r)}{\overset{(m)}{f}}(k_{\sigma(2)},...,l_1, k_{\sigma(m)},l_2)K_{a b}^{c_1 \cdot \cdot \cdot d c_{m-1} d} + \underset{(t,r)}{\overset{(m)}{f}}(k_{\sigma(2)},...,k_{\sigma(m)},l_1,l_2)K_{a b}^{c_1 \cdot \cdot \cdot c_{m-1} d d}\bigg)(\delta_{d,a}+\delta_{d,\overline{a}})
\end{align*} \endgroup
where $\overline{a} \notin \{a,c\}$. \\ \\
We now argue that only those terms which have $l_1, l_2$ next to each other can be non-vanishing. In other words, 
\begin{equation*}
    K_{a b}^{\cdot \cdot \cdot c d c \cdot \cdot \cdot} = 0 \quad \quad \quad K_{a b}^{d c \cdot \cdot \cdot c d} = 0.
\end{equation*}
for $d \neq c$:
\begin{equation*}
    K_{a b}^{\cdot \cdot \cdot c d c \cdot \cdot \cdot} \sim f_{e_1 g_1}^c f_{g_1 g_2}^d f_{g_2 e_2}^c \to -\epsilon_{g_1 e_1 c} \epsilon_{g_1 g_2 d} \epsilon_{g_2 e_2 c} = \delta_{e_1 d} \delta_{g_2 c} \epsilon_{g_2 e_2 c} = 0
\end{equation*}
\begin{equation*}
    K_{a b}^{d c \cdot \cdot \cdot c d} \sim f_{a g_1}^d f_{g_1 g_2}^c \to -\epsilon_{g_1 a d} \epsilon_{g_1 g_2 c} = 0
\end{equation*}
This leaves us with
\begingroup \allowdisplaybreaks \begin{align*}
    -2 \overset{\sum l_j = k_{\sigma(1)}}{\underset{\sigma \in S_{m}}{\sum}} \bigg(
    &\underset{(t,r)}{\overset{(m)}{f}}(l_1,l_2,k_{\sigma(2)},...,k_{\sigma(m)})K_{a b}^{d d c_1 \cdot \cdot \cdot c_{m-1}} +
    \underset{(t,r)}{\overset{(m)}{f}}(k_{\sigma(2)},l_1,l_2,...,k_{\sigma(m)})K_{a b}^{c_1 d d \cdot \cdot \cdot c_{m-1}} + 
    \\
    ...+ &\underset{(t,r)}{\overset{(m)}{f}}(k_{\sigma(2)},...,l_1, l_2, k_{\sigma(m)})K_{a b}^{c_1 \cdot \cdot \cdot d d c_{m-1}} + \underset{(t,r)}{\overset{(m)}{f}}(k_{\sigma(2)},...,k_{\sigma(m)},l_1,l_2)K_{a b}^{c_1 \cdot \cdot \cdot c_{m-1} d d}\bigg)(\delta_{d,a}+\delta_{d,\overline{a}})
\end{align*} \endgroup
We now prove the following identity
\begin{equation}
    (\delta_{d,a}+\delta_{d,\overline{a}}) K_{ab}^{c_1\cdot \cdot \cdot c_n d d c_{n+1} \cdot \cdot \cdot c_m} = - \kappa^2 K_{a b}^{c_1 \cdot \cdot \cdot c_m} = K_{a b}^{c_1 \cdot \cdot \cdot c_{m+2}}.
\end{equation}
Consider first $n=1$
\begin{equation*}
    K_{ab}^{c_1 d d c_{2} \cdot \cdot \cdot c_m} = \kappa^{4} \epsilon_{a g_1 c} \epsilon_{g_1 g_2 d} \epsilon_{g_2 g_3 d} \epsilon_{g_3 g_4 c_2} K_{g_4 b}^{c_3 \cdot \cdot \cdot c_m} = \kappa^4 \delta_{a d} K_{a b}^{c_1 \cdot \cdot \cdot c_{m-2}} = - \delta_{a d} \kappa^2 K_{a b}^{c_1 \cdot \cdot \cdot c_m}
\end{equation*} \endgroup
Now let $n > 1$
\begingroup \allowdisplaybreaks \begin{align*}
    K_{ab}^{c_1\cdot \cdot \cdot c_n d d c_{n+1} \cdot \cdot \cdot c_m} &=-\kappa K_{a g_{n-1}}^{c_1 \cdot \cdot \cdot c_{n-1}} K_{g_{n-1} b}^{c_n d d \cdot \cdot \cdot c_{m-2}} =-\kappa^{5}K_{a g_{n-1}}^{c_1 \cdot \cdot \cdot c_{n-1}} \delta_{g_{n-1} d} K_{g_{n-1} b}^{c_n \cdot \cdot \cdot c_{m-2}} \\ &= -\kappa^2 K_{a b}^{c_1 \cdot \cdot \cdot c_m}
    \begin{cases}
        \delta_{d a}, & \text{if $n$ odd} \\
        \delta_{d \overline{a}}, & \text{if $n$ even}
    \end{cases}
\end{align*} \endgroup
where we used the following identity
\begin{equation}
    K_{ab}^{c_1 \cdot \cdot \cdot c_m} = 
    \begin{cases}
        \kappa(-\kappa^2)^{\frac{m-2}{2}} \delta_{ab}, & \text{if $m$ even} \\
        -(-\kappa^2)^{\frac{m-1}{2}} \epsilon_{abc}, & \text{if $m$ odd}
    \end{cases}
\end{equation}
which follows from eq.(\ref{id:K^m-to-K^m-2}). \\ \\
We thus find that the maximal $f$ terms simplify to
\begin{equation*}
    -2 \overset{\sum l_j = k_{\sigma(1)}}{\underset{\sigma \in S_{m}}{\sum}} \bigg(
    \underset{(t,r)}{\overset{(m)}{f}}(l_1,l_2,k_{\sigma(2)},...,k_{\sigma(m)}) +...
    + \underset{(t,r)}{\overset{(m)}{f}}(k_{\sigma(2)},...,k_{\sigma(m)},l_1,l_2)\bigg) K_{a b}^{c_1  \cdot \cdot \cdot c_{m+1}}
\end{equation*} \endgroup
Putting it all together,
\begingroup \allowdisplaybreaks \begin{align*}
    \underset{\sigma \in S_m}{\sum} \underset{(t,r)}{\overset{(m)}{b}}(k_{\sigma(1)},...,k_{\sigma(m)}) &\sim - \bigg( \frac{\kappa^{-1} \lambda_{\mathfrak{g}}^2}{4 \cdot 48 \pi^2} \bigg) 
    \underset{\sigma \in S_{m}}{\sum} \underset{(t,r)}{\overset{(m-1)}{i}}(k_{\sigma(1)},...,k_{\sigma(m)}) 
    \\
    &-\frac{1}{2} \kappa \overset{\sum l_j = k_{\sigma(1)}}{\underset{\sigma \in S_{m}}{\sum}} \bigg(
    \underset{(t,r)}{\overset{(m)}{f}}(l_1,l_2,k_{\sigma(2)},...,k_{\sigma(m)}) +...
    + \underset{(t,r)}{\overset{(m)}{f}}(k_{\sigma(2)},...,k_{\sigma(m)},l_1,l_2)\bigg)
\end{align*}
and finally plugging in $\kappa = -2$ and $\lambda_{\mathfrak{g}}^2 = 8$,
\begingroup \allowdisplaybreaks \begin{align*}
    \underset{\sigma \in S_m}{\sum} \underset{(t,r)}{\overset{(m)}{b}}(k_{\sigma(1)},...,k_{\sigma(m)}) &\sim \bigg( \frac{1}{48 \pi^2} \bigg) 
    \underset{\sigma \in S_{m}}{\sum} \underset{(t,r)}{\overset{(m-1)}{i}}(k_{\sigma(1)},...,k_{\sigma(m)}) 
    \\ 
    &+ \overset{\sum l_j = k_{\sigma(1)}}{\underset{\sigma \in S_{m}}{\sum}} \bigg(
    \underset{(t,r)}{\overset{(m)}{f}}(l_1,l_2,k_{\sigma(2)},...,k_{\sigma(m)}) +
    \underset{(t,r)}{\overset{(m)}{f}}(k_{\sigma(2)}, l_1,l_2,k_{\sigma(3)},...,k_{\sigma(m)})
    \\
    & \quad \quad \quad \quad ...+ \underset{(t,r)}{\overset{(m)}{f}}(k_{\sigma(2)},...,k_{\sigma(m)},l_1,l_2)\bigg).
\end{align*}
 This equation is solved by the following expression
 \begingroup \allowdisplaybreaks \begin{align*}
     \underset{(t,r)}{\overset{(m)}{b}}(k_1,...,k_m) &\sim 
     \bigg( \frac{1}{96\pi^2} \bigg) \underset{(t,r)}{\overset{(m-1)}{i}}(k_m,...,k_1) +
     \bigg( \frac{1}{96\pi^2} \bigg) \underset{(t,r)}{\overset{(m-1)}{i}}(k_1,k_m,k_{m-1},...,k_2)
     \\
     & + \underset{l_j^i \geq 0}{\overset{\sum l_j = k_m}{\sum}}  \underset{(t,r)}{\overset{(m)}{f}}(l_1,l_2,k_{m-1},...,k_{1})
     + \underset{l_j^i \geq 0}{\overset{\sum l_j = k_{m-1}}{\sum}}  \underset{(t,r)}{\overset{(m)}{f}}(k_{m},l_1,l_2,k_{m-2}...,k_{1}) + 
     \\
     ...&+ \underset{l_j^i \geq 0}{\overset{\sum l_j = k_1}{\sum}}  \underset{(t,r)}{\overset{(m)}{f}}(k_{m},...,k_{2}, l_1,l_2)
 \end{align*}
 Recall that we have the following symmetries
 \begin{equation*}
     \underset{(t,r)}{\overset{(m-1)}{i}}(k_1,...,k_m) = (-1)^m \underset{(r,t)}{\overset{(m-1)}{i}}(k_1,k_m,k_{m-1},...,k_2) 
     \quad \quad
     \underset{(t,r)}{\overset{(m)}{f}}(k_1,...,k_{m+1}) = (-1)^m \underset{(r,t)}{\overset{(m)}{f}}(k_{m+1},k_{m-1},...,k_1)
 \end{equation*}
 It immediately follows from these that this solution respects the symmetry eq.(\ref{sym:b_m}). Although we have used the specialization of gauge algebra and axion-only matter in the intermediate steps, the final answer for the numerical coefficient just derived is the same for all $\mathfrak{g}_R$ on general grounds. 
 \\ \\ 
 
 We now consider the sub-maximal terms. Let $0 < n < m$. We will determine the terms of the form  
 \begin{equation*}
    \overset{(n)}{f} \overset{(m-n)}{f}
\end{equation*}
Fixing $k_1,...,k_m$ such that $\sum k_j = p_m$,
\begingroup \allowdisplaybreaks \begin{align*} 
    -\overset{\sum l_j = \overline{p}_{\sigma(i)}}{\underset{\sigma \in S_{m}}{\sum}} \bigg(
    &\underset{(t,r)}{\overset{(n)}{a}}[l_1,l_2,k_{\sigma(1)},...,k_{\sigma(n-1)}]_{a b}^{e_1 e_2 c_1 \cdot \cdot \cdot c_{n-1}} +
    \underset{(t,r)}{\overset{(n)}{a}}[l_1,k_{\sigma(1)},l_2,k_{\sigma(2)},...,k_{\sigma(n-1)}]_{a b}^{e_1 c_1 e_2 c_2\cdot \cdot \cdot c_{n-1}} +...
    \\
    + &\underset{(t,r)}{\overset{(n)}{a}}[l_1,k_{\sigma(1)},...,k_{\sigma(n-1)},l_2]_{a b}^{e_1 c_1 \cdot \cdot \cdot c_{n-1} e_2} \bigg)
    \underset{(l_2,l_1)}{\overset{(m-n)}{f}}[k_{\sigma(n)},...,k_{\sigma(m)}]_{e_3 e_2}^{c_n \cdot \cdot \cdot c_{m}} f_{c e_1}^{e_3} + (n \to m-n)
\end{align*} \endgroup
where we defined $\overline{p}_i = \overline{p}_{i_n \cdot \cdot \cdot i_{m+1}} = p_{m-n} + k_{i_n} +...+k_{i_{m+1}}$ and $\sigma(i) = \sigma(i_n) \cdot \cdot \cdot \sigma(i_{m+1})$.
\begingroup \allowdisplaybreaks \begin{align*} 
    -2 \overset{\sum l_j = \overline{p}_{\sigma(i)}}{\underset{\sigma \in S_{m}}{\sum}} \bigg(&\underset{(t,r)}{\overset{(n)}{f}}(l_1,l_2,k_{\sigma(1)},...,k_{\sigma(n-1)}) K_{a b}^{e_1 e_2 c_1 \cdot \cdot \cdot c_{n-1}} +
    \underset{(t,r)}{\overset{(n)}{f}}(l_1,k_{\sigma(1)},l_2,k_{\sigma(2)},...,k_{\sigma(n-1)}) K_{a b}^{e_1 c_1 e_2 c_2\cdot \cdot \cdot c_{n-1}} +
    \\
    ...+ &\underset{(t,r)}{\overset{(n)}{f}}(k_{\sigma(1)},...,k_{\sigma(n-1)},l_1,l_2) K_{a b}^{ c_1 \cdot \cdot \cdot c_{n-1} e_1 e_2} \bigg)
    \underset{(l_2,l_1)}{\overset{(m-n)}{f}}(k_{\sigma(n)},...,k_{\sigma(m)}) \kappa^{-1} K_{e_1 e_2}^{c_n \cdot \cdot \cdot c_{m+1}} + (n \to m-n)
\end{align*} \endgroup
Note that it immediately follows that $e_1, e_2 \neq c$, and so we again find that only those terms which have $l_1, l_2$ next to each other can be non-vanishing. \\ \\
We now prove the following identity
\begin{equation}
     K_{ab}^{c_1\cdot \cdot \cdot c_k e_1 e_2 c_{k+1} \cdot \cdot \cdot c_{n-1}} 
     K_{e_1 e_2}^{c_n \cdot \cdot \cdot c_{m+1}}
     = - \kappa^3 K_{a b}^{c_1 \cdot \cdot \cdot c_{m-1}} 
     = \kappa K_{a b}^{c_1 \cdot \cdot \cdot c_{m+1}}.
\end{equation}
Consider first $k=1$
\begin{equation*}
    K_{ab}^{e_1 e_2 c_1\cdot \cdot \cdot c_{n-1}} 
    K_{e_1 e_2}^{c_n \cdot \cdot \cdot c_{m+1}}
    = -\kappa^4 \epsilon_{a g_1 e_1} \epsilon_{e_1 g_3 c} \epsilon_{g_1 g_2 e_2} \epsilon_{e_2 g_4 c} K_{g_2 b}^{c_1 \cdot \cdot \cdot c_{n-1}} K_{g_3 g_4}^{c_{n+1} \cdot \cdot \cdot c_{m}}
    = - \kappa^3 K_{ab}^{c_{1} \cdot \cdot \cdot c_{m-1}}
\end{equation*} \endgroup
Now let $k > 1$
\begingroup \allowdisplaybreaks \begin{align*}
    K_{ab}^{c_1\cdot \cdot \cdot c_k e_1 e_2 c_{k+1} \cdot \cdot \cdot c_{n-1}} 
    K_{e_1 e_2}^{c_n \cdot \cdot \cdot c_{m+1}} 
    = -\kappa K_{a g_1}^{c_1 \cdot \cdot \cdot c_{k}} K_{g_1 b}^{e_1 e_2 c_{k+1} \cdot \cdot \cdot c_{n-1}} K_{e_1 e_2}^{c_n \cdot \cdot \cdot c_{m+1}}
    = - \kappa^3 K_{ab}^{c_{1} \cdot \cdot \cdot c_{m-1}}
\end{align*} \endgroup
We thus find that these terms simplify to 
\begingroup \allowdisplaybreaks \begin{align*} 
    -2 \overset{\sum l_j = \overline{p}_{\sigma(i)}}{\underset{\sigma \in S_{m}}{\sum}} \bigg(&\underset{(t,r)}{\overset{(n)}{f}}(l_1,l_2,k_{\sigma(1)},...,k_{\sigma(n-1)}) 
    + \underset{(t,r)}{\overset{(n)}{f}}(k_{\sigma(1)},l_1,l_2,k_{\sigma(2)},...,k_{\sigma(n-1)}) +
    \\
    ...+ &\underset{(t,r)}{\overset{(n)}{f}}(k_{\sigma(1)},...,k_{\sigma(n-1)},l_1,l_2) \bigg)
    \underset{(l_2,l_1)}{\overset{(m-n)}{f}}(k_{\sigma(n)},...,k_{\sigma(m)}) K_{a b}^{c_1 \cdot \cdot \cdot c_{m+1}} + (n \to m-n)
\end{align*} \endgroup
This gives us the equation
\begingroup \allowdisplaybreaks \begin{align*}
    \underset{\sigma \in S_m}{\sum} \underset{(t,r)}{\overset{(m)}{b}}(k_{\sigma(1)},...,k_{\sigma(m)}) \sim \sum_{n = 1}^{m-1} \overset{\sum l_j = \overline{p}_{\sigma(i)}}{ \underset{\sigma \in S_{m}}{\sum}} \bigg(&\underset{(t,r)}{\overset{(n)}{f}}(l_1,l_2,k_{\sigma(1)},...,k_{\sigma(n-1)})  + \underset{(t,r)}{\overset{(n)}{f}}(k_{\sigma(1)},l_1,l_2,k_{\sigma(2)},...,k_{\sigma(n-1)}) \\
    ...+ &\underset{(t,r)}{\overset{(n)}{f}}(k_{\sigma(1)},...,k_{\sigma(n-1)},l_1,l_2) \bigg)
    \underset{(l_2,l_1)}{\overset{(m-n)}{f}}(k_{\sigma(n)},...,k_{\sigma(m)})
\end{align*} \endgroup
This equation is solved by the following expression
\begingroup \allowdisplaybreaks \begin{align*}
     \underset{(t,r)}{\overset{(m)}{b}}(k_1,...,k_m) \sim 
     \frac{1}{2} \sum_{n = 1}^{m-1} \bigg(
     \underset{l_j^i \geq 0}{\overset{\sum l_j = q^{m}_{n}}{\sum}}  &\underset{(t,r)}{\overset{(n)}{f}}(l_1,l_2,k_{n-1},...,k_{1}) \underset{(l_2,l_1)}{\overset{(m-n)}{f}}(k_{m},...,k_{n}) + \\
     +\underset{l_j^i \geq 0}{\overset{\sum l_j = q^{m-1}_{n-1}}{\sum}}  &\underset{(t,r)}{\overset{(n)}{f}}(k_m,l_1,l_2,k_{n-2},...,k_{1}) \underset{(l_2,l_1)}{\overset{(m-n)}{f}}(k_{m-1},...,k_{n-1})+...\\
     +\underset{l_j^i \geq 0}{\overset{\sum l_j = q^{m-n+1}_{1}}{\sum}}  &\underset{(t,r)}{\overset{(n)}{f}}(k_m,...,k_{m-n+2},l_1,l_2) \underset{(l_2,l_1)}{\overset{(m-n)}{f}}(k_{m-n+1},...,k_{1}) \bigg) + (-1)^m (t \leftrightarrow r)
 \end{align*} \endgroup
 where we have defined $q^m_n = (m-n)+k_{m}+...+k_{n}$. \\ 
 
 Lastly, we see that the general $m$ equation has no $j$ terms. The following expression is consistent with this, the symmetries of $b$, and our previous 1- and 2-loop solutions:
\begingroup \allowdisplaybreaks \begin{align*}
     \underset{(t,r)}{\overset{(m)}{b}}(k_1,...,k_m) \sim 
     &\bigg( \frac{1}{96\pi^2} \bigg) (t^1+t^2) \underset{(t,r)}{\overset{(m-1)}{j}}(k_m,...,k_1) 
     -(-1)^m\bigg( \frac{1}{96\pi^2} \bigg) (r^1+r^2) \underset{(r,t)}{\overset{(m-1)}{j}}(k_m,...,k_1)
     \\
     -&\bigg( \frac{1}{96\pi^2} \bigg) (t^1+t^2) \underset{(t,r)}{\overset{(m-1)}{j}}(k_1,...,k_m) 
     +(-1)^m\bigg( \frac{1}{96\pi^2} \bigg) (r^1+r^2) \underset{(r,t)}{\overset{(m-1)}{j}}(k_1,...,k_m)
\end{align*} \endgroup
Putting all the pieces together results in the expression for $\overset{(m)}{b}$ presented in \S \ref{sec:coeffs} of the main text.

As a sanity check, we will verify that our general solution solves the associativity equation at three and four loops. Again, for cleanliness, we will focus on the specialization of $\mathfrak{g}=\mathfrak{su}(2)$ plus the axion. 
\subsubsection{A check of $b$ at 3 loops}
\noindent The specialized $m=3$ equation is
\begingroup \allowdisplaybreaks \begin{align*}
    \underset{\sigma \in S_3}{\sum} \underset{(t,r)}{\overset{(3)}{b}}(k_{\sigma(1)},&k_{\sigma(2)},k_{\sigma(3)}) = 
    \bigg( \frac{1}{48 \pi^2} \bigg) \underset{\sigma \in S_3}{\sum} \underset{(t,r)}{\overset{(2)}{i}}(k_{\sigma(1)},k_{\sigma(2)},k_{\sigma(3)}) 
    \\
    &+ \overset{\sum l_j = k_{\sigma(1)}}{\underset{\sigma \in S_{3}}{\sum}} \bigg(
    \underset{(t,r)}{\overset{(3)}{f}}(l_1,l_2,k_{\sigma(2)},k_{\sigma(3)})
    + \underset{(t,r)}{\overset{(3)}{f}}(k_{\sigma(2)}, l_1,l_2,k_{\sigma(3)})
    + \underset{(t,r)}{\overset{(3)}{f}}(k_{\sigma(2)},k_{\sigma(3)},l_1,l_2)\bigg)
    \\
    &+\overset{\sum l_j = \overline{p}_{\sigma(i)}}{ \underset{\sigma \in S_{3}}{\sum}} \underset{(t,r)}{\overset{(1)}{f}}(l_1,l_2) \underset{(l_2,l_1)}{\overset{(2)}{f}}(k_{\sigma(1)},k_{\sigma(2)},k_{\sigma(3)})
    \\
    &+\overset{\sum l_j = \overline{p}_{\sigma(i)}}{ \underset{\sigma \in S_{3}}{\sum}} \bigg(\underset{(t,r)}{\overset{(2)}{f}}(l_1,l_2,k_{\sigma(1)})  + \underset{(t,r)}{\overset{(2)}{f}}(k_{\sigma(1)},l_1,l_2) \bigg)
    \underset{(l_2,l_1)}{\overset{(1)}{f}}(k_{\sigma(2)},k_{\sigma(3)})
\end{align*} \endgroup
Plugging in our expression for $b$ the left-hand-side becomes
\begingroup \allowdisplaybreaks \begin{align*}
    \underset{\sigma \in S_3}{\sum} \underset{(t,r)}{\overset{(3)}{b}}(k_{\sigma(1)},&k_{\sigma(2)},k_{\sigma(3)}) = 
    \bigg( \frac{1}{96\pi^2} \bigg) \underset{\sigma \in S_3}{\sum} \bigg( \underset{(t,r)}{\overset{(2)}{i}}(k_{\sigma(3)},k_{\sigma(2)},k_{\sigma(1)}) +
    \underset{(t,r)}{\overset{(2)}{i}}(k_{\sigma(1)},k_{\sigma(3)},k_{\sigma(2)}) \bigg)
    \\
    & + \underset{\sigma \in S_3}{\sum} \underset{l_j^i \geq 0}{\overset{\sum l_j = k_{\sigma(3)}}{\sum}} \underset{(t,r)}{\overset{(3)}{f}}(l_1,l_2,k_{\sigma(2)},k_{\sigma(1)})
    + \underset{\sigma \in S_3}{\sum} \underset{l_j^i \geq 0}{\overset{\sum l_j = k_{\sigma(2)}}{\sum}} \underset{(t,r)}{\overset{(3)}{f}}(k_{\sigma(3)},l_1,l_2,k_{\sigma(1)}) 
    \\
    &+ \underset{\sigma \in S_3}{\sum} \underset{l_j^i \geq 0}{\overset{\sum l_j = k_{\sigma(1)}}{\sum}} \underset{(t,r)}{\overset{(3)}{f}}(k_{\sigma(3)},k_{\sigma(2)},l_1,l_2)
    \\
    &+ \frac{1}{2} \underset{\sigma \in S_3}{\sum} \underset{l_j^i \geq 0}{\overset{\sum l_j = \overline{p}_{\sigma(i)}}{\sum}} 
    \bigg( \underset{(t,r)}{\overset{(1)}{f}}(l_1,l_2) - \underset{(r,t)}{\overset{(1)}{f}}(l_1,l_2) \bigg) \underset{(l_2,l_1)}{\overset{(2)}{f}}(k_{\sigma(3)},k_{\sigma(2)},k_{\sigma(1)})
    \\
    &+\frac{1}{2} \underset{\sigma \in S_3}{\sum} 
    \underset{l_j^i \geq 0}{\overset{\sum l_j = \overline{p}_{\sigma(i)}}{\sum}} \bigg( \underset{(t,r)}{\overset{(2)}{f}}(l_1,l_2,k_{\sigma(1)}) - \underset{(r,t)}{\overset{(2)}{f}}(l_1,l_2,k_{\sigma(1)}) \bigg)\underset{(l_2,l_1)}{\overset{(1)}{f}}(k_{\sigma(3)},k_{\sigma(2)}) 
    \\
    &+\frac{1}{2} \underset{\sigma \in S_3}{\sum} 
    \underset{l_j^i \geq 0}{\overset{\sum l_j = \overline{p}_{\sigma(i)}}{\sum}} \bigg( \underset{(t,r)}{\overset{(2)}{f}}(k_{\sigma(3)},l_1,l_2) - \underset{(r,t)}{\overset{(2)}{f}}(k_{\sigma(3)},l_1,l_2) \bigg)\underset{(l_2,l_1)}{\overset{(1)}{f}}(k_{\sigma(2)},k_{\sigma(1)}) 
    \\ \\
    &= \bigg( \frac{1}{48\pi^2} \bigg) \underset{\sigma \in S_3}{\sum} \underset{(t,r)}{\overset{(2)}{i}}(k_{\sigma(3)},k_{\sigma(2)},k_{\sigma(1)}) 
    \\
    &+\underset{\sigma \in S_3}{\overset{\sum l_j = k_{\sigma(1)}}{\sum}} \bigg( \underset{(t,r)}{\overset{(3)}{f}}(l_1,l_2,k_{\sigma(2)},k_{\sigma(3)})
    + \underset{(t,r)}{\overset{(3)}{f}}(k_{\sigma(2)},l_1,l_2,k_{\sigma(3)}) + 
    \underset{(t,r)}{\overset{(3)}{f}}(k_{\sigma(2)},k_{\sigma(3)},l_1,l_2) \bigg)
    \\
    &+\underset{\sigma \in S_3}{\overset{\sum l_j = \overline{p}_{\sigma(i)}}{\sum}} 
    \underset{(t,r)}{\overset{(1)}{f}}(l_1,l_2) \underset{(l_2,l_1)}{\overset{(2)}{f}}(k_{\sigma(3)},k_{\sigma(2)},k_{\sigma(1)})
    \\
    &+\underset{\sigma \in S_3}{\overset{\sum l_j = \overline{p}_{\sigma(i)}}{\sum}} \bigg( \underset{(t,r)}{\overset{(2)}{f}}(l_1,l_2,k_{\sigma(1)}) + \underset{(t,r)}{\overset{(2)}{f}}(k_{\sigma(1)},l_1,l_2) \bigg)\underset{(l_2,l_1)}{\overset{(1)}{f}}(k_{\sigma(2)},k_{\sigma(3)})
    = \textit{rhs}.
\end{align*} \endgroup

Notice that here, and below, the $j$-terms directly cancel each other on the left and right hand sides of the equation due to the sum over permutations. 

\subsubsection{A check of $b$ at 4 Loops}
\noindent The specialized $m=4$ equation is
\begingroup \allowdisplaybreaks \begin{align*}
    \underset{\sigma \in S_4}{\sum} \underset{(t,r)}{\overset{(4)}{b}}(k_{\sigma(1)},&k_{\sigma(2)},k_{\sigma(3)},k_{\sigma(4)}) = 
    \bigg( \frac{1}{48 \pi^2} \bigg) \underset{\sigma \in S_4}{\sum} \underset{(t,r)}{\overset{(3)}{i}}(k_{\sigma(1)},k_{\sigma(2)},k_{\sigma(3)},k_{\sigma(4)}) 
    \\
    &+ \overset{\sum l_j = k_{\sigma(1)}}{\underset{\sigma \in S_{4}}{\sum}} \bigg(
    \underset{(t,r)}{\overset{(4)}{f}}(l_1,l_2,k_{\sigma(2)},k_{\sigma(3)},k_{\sigma(4)})
    + \underset{(t,r)}{\overset{(4)}{f}}(k_{\sigma(2)}, l_1,l_2,k_{\sigma(3)},k_{\sigma(4)})
    \\
    &+ \underset{(t,r)}{\overset{(4)}{f}}(k_{\sigma(2)},k_{\sigma(3)},l_1,l_2,k_{\sigma(4)})
    + \underset{(t,r)}{\overset{(4)}{f}}(k_{\sigma(2)},k_{\sigma(3)},k_{\sigma(4)},l_1,l_2)\bigg)
    \\
    &+\overset{\sum l_j = \overline{p}_{\sigma(i)}}{ \underset{\sigma \in S_{4}}{\sum}} \underset{(t,r)}{\overset{(1)}{f}}(l_1,l_2) \underset{(l_2,l_1)}{\overset{(3)}{f}}(k_{\sigma(1)},k_{\sigma(2)},k_{\sigma(3)},k_{\sigma(3)})
    \\
    &+\overset{\sum l_j = \overline{p}_{\sigma(i)}}{ \underset{\sigma \in S_{4}}{\sum}} \bigg(\underset{(t,r)}{\overset{(2)}{f}}(l_1,l_2,k_{\sigma(1)})  + \underset{(t,r)}{\overset{(2)}{f}}(k_{\sigma(1)},l_1,l_2) \bigg)
    \underset{(l_2,l_1)}{\overset{(2)}{f}}(k_{\sigma(2)},k_{\sigma(3)},k_{\sigma(4)})
    \\
    &+\overset{\sum l_j = \overline{p}_{\sigma(i)}}{ \underset{\sigma \in S_{4}}{\sum}} \bigg(\underset{(t,r)}{\overset{(3)}{f}}(l_1,l_2,k_{\sigma(1)},k_{\sigma(2)})  + \underset{(t,r)}{\overset{(3)}{f}}(k_{\sigma(1)},l_1,l_2,k_{\sigma(2)}) \\
    &+ \underset{(t,r)}{\overset{(3)}{f}}(k_{\sigma(1)},k_{\sigma(2)}l_1,l_2) \bigg)
    \underset{(l_2,l_1)}{\overset{(1)}{f}}(k_{\sigma(3)},k_{\sigma(4)})
\end{align*} \endgroup
Plugging in our expression for $b$ the left-hand-side becomes
\begingroup \allowdisplaybreaks \begin{align*}
    \underset{\sigma \in S_4}{\sum} \underset{(t,r)}{\overset{(4)}{b}}(k_{\sigma(1)},&k_{\sigma(2)},k_{\sigma(3)},k_{\sigma(4)}) = 
    \bigg( \frac{1}{96\pi^2} \bigg) \underset{\sigma \in S_4}{\sum} \bigg( \underset{(t,r)}{\overset{(3)}{i}}(k_{\sigma(4)},k_{\sigma(3)},k_{\sigma(2)},k_{\sigma(1)}) +
    \underset{(t,r)}{\overset{(3)}{i}}(k_{\sigma(1)},k_{\sigma(4)},k_{\sigma(3)},k_{\sigma(2)}) \bigg)
    \\
    & + \underset{\sigma \in S_4}{\overset{\sum l_j = k_{\sigma(4)}}{\sum}} \underset{(t,r)}{\overset{(4)}{f}}(l_1,l_2,k_{\sigma(3)},k_{\sigma(2)},k_{\sigma(1)})
    + \underset{\sigma \in S_4}{\overset{\sum l_j = k_{\sigma(3)}}{\sum}} \underset{(t,r)}{\overset{(4)}{f}}(k_{\sigma(4)},l_1,l_2,k_{\sigma(2)}, k_{\sigma(1)}) 
    \\
    & + \underset{\sigma \in S_4}{\overset{\sum l_j = k_{\sigma(2)}}{\sum}} \underset{(t,r)}{\overset{(4)}{f}}(k_{\sigma(4)},k_{\sigma(3)},l_1,l_2, k_{\sigma(1)})
    + \underset{\sigma \in S_4}{\overset{\sum l_j = k_{\sigma(1)}}{\sum}} \underset{(t,r)}{\overset{(4)}{f}}(k_{\sigma(4)},k_{\sigma(3)}, k_{\sigma(2)},l_1,l_2)
    \\
    &+ \frac{1}{2} \underset{\sigma \in S_4}{\overset{\sum l_j = \overline{p}_{\sigma(i)}}{\sum}} 
    \bigg( \underset{(t,r)}{\overset{(1)}{f}}(l_1,l_2) + \underset{(r,t)}{\overset{(1)}{f}}(l_1,l_2) \bigg) \underset{(l_2,l_1)}{\overset{(3)}{f}}(k_{\sigma(4)},k_{\sigma(3)},k_{\sigma(2)},k_{\sigma(1)})
    \\
    &+\frac{1}{2} \underset{\sigma \in S_4}{\overset{\sum l_j = \overline{p}_{\sigma(i)}}{\sum}} \bigg( \underset{(t,r)}{\overset{(2)}{f}}(l_1,l_2,k_{\sigma(1)}) + \underset{(r,t)}{\overset{(2)}{f}}(l_1,l_2,k_{\sigma(1)}) \bigg)\underset{(l_2,l_1)}{\overset{(2)}{f}}(k_{\sigma(4)}, k_{\sigma(3)},k_{\sigma(2)}) 
    \\
    &+\frac{1}{2} \underset{\sigma \in S_4}{\overset{\sum l_j = \overline{p}_{\sigma(i)}}{\sum}} \bigg( \underset{(t,r)}{\overset{(2)}{f}}(k_{\sigma(4)},l_1,l_2) + \underset{(r,t)}{\overset{(2)}{f}}(k_{\sigma(4)},l_1,l_2) \bigg)\underset{(l_2,l_1)}{\overset{(2)}{f}}( k_{\sigma(3)},k_{\sigma(2)},k_{\sigma(1)})
    \\
    &+\frac{1}{2} \underset{\sigma \in S_4}{\overset{\sum l_j = \overline{p}_{\sigma(i)}}{\sum}} \bigg( \underset{(t,r)}{\overset{(3)}{f}}(l_1,l_2,k_{\sigma(2)},k_{\sigma(1)}) + \underset{(r,t)}{\overset{(3)}{f}}(l_1,l_2,k_{\sigma(2)},k_{\sigma(1)}) \bigg)\underset{(l_2,l_1)}{\overset{(1)}{f}}(k_{\sigma(4)}, k_{\sigma(3)})
    \\
    &+\frac{1}{2} \underset{\sigma \in S_4}{\overset{\sum l_j = \overline{p}_{\sigma(i)}}{\sum}} \bigg( \underset{(t,r)}{\overset{(3)}{f}}(k_{\sigma(4)},l_1,l_2,k_{\sigma(1)}) + \underset{(r,t)}{\overset{(3)}{f}}(k_{\sigma(4)},l_1,l_2,k_{\sigma(1)}) \bigg)\underset{(l_2,l_1)}{\overset{(1)}{f}}(k_{\sigma(3)}, k_{\sigma(2)})
    \\
    &+\frac{1}{2} \underset{\sigma \in S_4}{\overset{\sum l_j = \overline{p}_{\sigma(i)}}{\sum}} \bigg( \underset{(t,r)}{\overset{(3)}{f}}(k_{\sigma(4)},k_{\sigma(3)},l_1,l_2,) + \underset{(r,t)}{\overset{(3)}{f}}(k_{\sigma(4)},k_{\sigma(3)},l_1,l_2) \bigg)\underset{(l_2,l_1)}{\overset{(1)}{f}}(k_{\sigma(2)}, k_{\sigma(1)}) 
    \\ \\
    &= \bigg( \frac{1}{48\pi^2} \bigg) \underset{\sigma \in S_4}{\sum}
    \underset{(t,r)}{\overset{(3)}{i}}(k_{\sigma(1)},k_{\sigma(2)},k_{\sigma(3)},k_{\sigma(4)})
    \\
    & + \underset{\sigma \in S_4}{\overset{\sum l_j = k_{\sigma(1)}}{\sum}} \bigg( \underset{(t,r)}{\overset{(4)}{f}}(l_1,l_2,k_{\sigma(2)},k_{\sigma(3)},k_{\sigma(4)})
    + \underset{(t,r)}{\overset{(4)}{f}}(k_{\sigma(2)},l_1,l_2,k_{\sigma(3)}, k_{\sigma(4)}) 
    \\
    & + \underset{(t,r)}{\overset{(4)}{f}}(k_{\sigma(2)},k_{\sigma(3)},l_1,l_2, k_{\sigma(4)})
    + \underset{(t,r)}{\overset{(4)}{f}}(k_{\sigma(2)},k_{\sigma(3)}, k_{\sigma(4)},l_1,l_2) \bigg)
    \\
    &+ \underset{\sigma \in S_4}{\overset{\sum l_j = \overline{p}_{\sigma(i)}}{\sum}} 
    \underset{(t,r)}{\overset{(1)}{f}}(l_1,l_2) \underset{(l_2,l_1)}{\overset{(3)}{f}}(k_{\sigma(1)},k_{\sigma(2)},k_{\sigma(3)},k_{\sigma(4)})
    \\
    &+\underset{\sigma \in S_4}{\overset{\sum l_j = \overline{p}_{\sigma(i)}}{\sum}} \bigg( \underset{(t,r)}{\overset{(2)}{f}}(l_1,l_2,k_{\sigma(1)}) + \underset{(t,r)}{\overset{(2)}{f}}(k_{\sigma(1)},l_1,l_2) \bigg) \underset{(l_2,l_1)}{\overset{(2)}{f}}(k_{\sigma(2)}, k_{\sigma(3)},k_{\sigma(4)})
    \\
    &+\underset{\sigma \in S_4}{\overset{\sum l_j = \overline{p}_{\sigma(i)}}{\sum}} \bigg( \underset{(t,r)}{\overset{(3)}{f}}(l_1,l_2,k_{\sigma(1)},k_{\sigma(2)}) + \underset{(t,r)}{\overset{(3)}{f}}(k_{\sigma(1)},l_1,l_2,k_{\sigma(2)}) 
    \\
    &+ \underset{(r,t)}{\overset{(3)}{f}}(k_{\sigma(2)},k_{\sigma(1)},l_1,l_2) \bigg)\underset{(l_2,l_1)}{\overset{(1)}{f}}(k_{\sigma(3)}, k_{\sigma(4)}) = \textit{rhs}.
\end{align*} \endgroup

\subsubsection{Solving for $c$}
In general $\overset{(m)}{c}$ decomposes into a sum over $m$ terms, contracted with different combinations of structure constants as in equation eq.\eqref{eq:cdecomp}, and so term-by-term they admit a less symmetrical and compact solution than $\overset{(m)}{b}$. \\

The associativity equations still uniquely determine all of these contributions, upon making clever specializations of the gauge indices; we also believe that, like the $\overset{(m)}{b}$, $\overset{(m)}{c}$ can be given a neat closed form solution and we hope to improve on this point in the future. \\

For a given loop order, the following less-elegant strategy suffices. Due to the relationship eq.\eqref{eq:c_m-to-b_m} and exact solution for $\overset{(m)}{b}$ one needs to solve only for a subset, e.g. $\overset{(m)}{c_j}$,  $j = 1, \ldots, \lceil m/2 \rceil$. This in turn is always possible. There is always the specialization 
\begin{equation}
    a = c = d_j \quad j > 1 
    \quad \quad 
    b \neq a 
    \quad \quad 
    d_1 = \begin{cases}
        d, &m \text{ odd} \\
        b, &m \text{ even}
    \end{cases}
\end{equation}
that isolates $\overset{(m)}{c_1}$. Here $d \notin \{ a, b \}$. \\ \\
To see this, consider the general axion-only $m$ equation with $\mathfrak{g} = \mathfrak{su}(2)$. Focusing only on the $c$ terms, we find after moving them all to one side, setting $a=c$ and fixing $k_j$ such that $\sum k_j = p_m$
\begingroup \allowdisplaybreaks \begin{align*}
    2 h^\vee \underset{\sigma \in S_{m-1}}{\sum} \bigg( &\underset{(t,r)}{\overset{(m)}{c_1}}(k_1, k_{\sigma(2,m)}) K_{a b}^{e_1 d_{\sigma(2)} \cdot \cdot \cdot d_{\sigma(m)}} f_{a e_1}^{d_1} +...+
    \underset{(t,r)}{\overset{(m)}{c_m}}(k_1, k_{\sigma(2, m)}) K_{a b}^{ d_{\sigma(2)} \cdot \cdot \cdot d_{\sigma(m)} e_1} f_{a e_1}^{d_1} 
    \\
    +&\underset{(t,r)}{\overset{(m)}{c_1}}(k_1, k_{\sigma(2,m)}) K_{a b}^{d_1 e_1 d_{\sigma(3)} \cdot \cdot \cdot d_{\sigma(m)}} f_{a e_1}^{d_\sigma(2)} +...+
    \underset{(t,r)}{\overset{(m)}{c_m}}(k_1, k_{\sigma(2,m)}) K_{a b}^{e_1 d_{\sigma(3)} \cdot \cdot \cdot d_{\sigma(m)} d_1} f_{a e_1}^{d_{\sigma(2)}}
    \\
    +...+&\underset{(t,r)}{\overset{(m)}{c_1}}(k_1, k_{\sigma(2,m)}) K_{a b}^{d_1 d_{\sigma(2)} \cdot \cdot \cdot d_{\sigma(m-1)} e_1} f_{a e_1}^{d_\sigma(m)} +...+
    \underset{(t,r)}{\overset{(m)}{c_m}}(k_1, k_{\sigma(2,m)}) K_{a b}^{ d_{\sigma(2)} \cdot \cdot \cdot d_{\sigma(m-1)} e_1 d_1} f_{a e_1}^{d_{\sigma(m)}}
    \\
    +&\underset{(t,r)}{\overset{(m)}{c_1}}(k_1, k_{\sigma(2,m)}) K_{a e_1}^{d_1 d_{\sigma(2)} \cdot \cdot \cdot d_{\sigma(m)}} f_{b a}^{e_1} +...+
    \underset{(t,r)}{\overset{(m)}{c_m}}(k_1, k_{\sigma(2,m)}) K_{a e_1}^{d_{\sigma(2)} \cdot \cdot \cdot d_{\sigma(m)} d_1} f_{b a}^{e_1} \bigg)
    \\
    -2 h^\vee \underset{\sigma \in S_{m}}{\sum} \bigg( &\underset{(t,r)}{\overset{(m)}{c_1}}(k_{\sigma(1,m)}) K_{a b}^{e_1 d_{\sigma(2)} \cdot \cdot \cdot d_{\sigma(m)}}
    +...+
    \underset{(t,r)}{\overset{(m)}{c_m}}(k_{\sigma(1,m)}) K_{a b}^{d_{\sigma(2)} \cdot \cdot \cdot d_{\sigma(m)} e_1} \bigg) f_{a e_1}^{d_\sigma(1)}
\end{align*} \endgroup
where 
\begin{equation*}
    k_{\sigma(i,j)} = k_{\sigma(i)}, k_{\sigma(i+1)},...,k_{\sigma(j)}.
\end{equation*}
Fixing $d_1 \neq a$ and $d_j = a$ for $j > 1$, all the terms with $f_{a e_1}^{d_j}$ or $K_{a \cdot}^{d_j \cdot \cdot \cdot}$ vanish and we are left with
\begin{equation*}
    4 \underset{\sigma \in S_{m-1}}{\sum}
    \underset{(t,r)}{\overset{(m)}{c_1}}(k_1, k_{\sigma(2,m)}) K_{a d}^{d_1 a_2 \cdot \cdot \cdot a_m} \epsilon_{b a d} \neq 0.
\end{equation*}
We thus see that this specialization isolates $c_1$, and so by keeping track of the other terms
\begin{equation}
    \overset{(m-1)}{i} 
    \quad \quad 
    \overset{(m-1)}{j} 
    \quad \quad
    \overset{(m)}{f} 
    \quad \quad
    \overset{(n)}{f} \overset{(m-n)}{f}
\end{equation}
we can solve for $c_1$.
\\ \\
Additional specializations, with e.g. $a \neq c$, lead to linear systems of equations, allowing for the successive solution of the required $\overset{(m)}{c_j}$. In fact, at $m=3, 4$ this is immediate, given our exact solution for $b$: $c_1$ can be isolated, determining $c_m$ (see eq.\eqref{eq:c_m-to-b_m}) and then any additional specialization can serve to determine $c_2$ (and thus $c_3$ for the case $m=4$) (which completes the solution for $c$).

\section{One-loop coefficients from Koszul Duality}\label{app:koszul}
\begin{figure}[t]
    \centering
    \begin{subfigure}[b]{0.3\textwidth}
    \centering
    \includegraphics[scale=0.4]{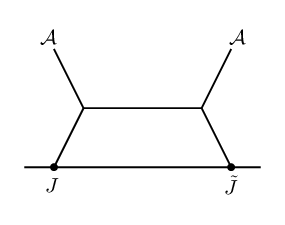}
    \end{subfigure}
    \begin{subfigure}[b]{0.3\textwidth}
\centering
         \includegraphics[scale=0.4]{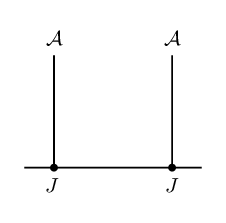}
    \end{subfigure}
     \caption{ The cancellation of the gauge anomaly of these diagrams results in the one-loop corrections to the OPEs between $J$ and itself.}
     \label{figure_appendix_c}
\end{figure}
\noindent In this section, we briefly review how to obtain the coefficients $\overset{(1)}{f}$ directly from Koszul duality. For more details, see \cite{victor}. \\ 

The $J_a[t]J_b[r]$ OPE correction at order $\hbar$ comes from demanding that the gauge anomaly (more precisely, the BRST variation) of the diagrams in Figure \ref{figure_appendix_c} cancel. \\ 

Following \cite{CP, victor}, we denote the location in twistor space of the defect operators of the left-most diagram as $Z=(z,0^{\dot{\alpha}})$ and $W=(w,0^{\dot{\alpha}})$, and let the vertices be located at $X=(x^0,v_x^{\dot{\alpha}})$ and $Y=(y^0,v_y^{\dot{\alpha}})$. In this notation, $(z, v^{\dot{\alpha}})$ label the coordinates on the ``celestial sphere'' $\mathbb{CP}^1$ (the zero section of the twistor fibration), and the coordinates on the $\mathcal{O}(1) \oplus \mathcal{O}(1)$ fibres are $v^{\dot{\alpha}}$. \\ 

We require that $\vert z-w \vert \geq \epsilon$, where $\epsilon$ is a small point-splitting regulator, and define $\underset{(k)}{D}^j \equiv \frac{1}{k^1! k^2!} \partial^{k^1}_{v^1_j} \partial^{k^2}_{v^2_j}$.\\  

Taking the linearized BRST variation of this diagram results in two terms. The contribution of the term where we took the gauge variation of the left external leg is given by: 
\begin{equation}
    \bigg(\frac{1}{2 \pi i} \bigg)^2 \underset{\mathbb{C}^2}{\int} J_{i_1}[k](z) \tilde{J}_{i_2}[l](w) K^{i_1 i_2}_{ab} \underset{(k,l)}{\mathcal{M}}(z,w;\overline{\partial}\chi^a,\mathcal{A}^b) dz dw
\end{equation} 
where 
\begingroup \allowdisplaybreaks
\begin{align} 
     \underset{(k,l)}{\mathcal{M}}(z,w;\overline{\partial}\chi^a,\mathcal{A}^b) &= \bigg( \frac{1}{4 \pi i}\bigg)^2 \underset{\mathbb{C}^2}{\int} \underset{(k)}{D}^z \underset{(Z \to X)}{P} \overline{\partial}\chi^a d^3X \underset{(X \to Y)}{P} d^3Y \mathcal{A}^b \underset{(l)}{D}^w \underset{(Y \to W)}{P} \label{eq:appendix_c_M} \\ 
     \underset{(Z \to W)}{P} &= \bigg(\frac{1}{2 \pi^2}\bigg)\epsilon_{\overline{a}\overline{b}\overline{c}} \frac{(\overline{Z}-\overline{W})^{\overline{a}} (\overline{Z}-\overline{W})^{\overline{b}} (\overline{Z}-\overline{W})^{\overline{c}}}{\vert \vert Z-W \vert \vert^6}
\end{align} \endgroup
Note that we first take the derivatives by letting $Z=(z,v_z^{\dot{\alpha}})$ and $W=(w,z_w^{\dot{\alpha}})$, and then we fix $v_z^{\dot{\alpha}}=0$ and $v_w^{\dot{\alpha}}=0$. \\ \\
Integrating by parts, we find that only the boundary term $\vert z-w \vert = \epsilon$ is non-vanishing, and its contribution is:
\begin{equation}
    -\bigg(\frac{1}{2 \pi i} \bigg)^2 \underset{\mathbb{C}}{\int} dz_0 \underset{\vert z-w \vert = \epsilon}{\oint} J_{i_1}[k](z) \tilde{J}_{i_2}[l](w) K^{i_1 i_2}_{ab} \underset{(k,l)}{\mathcal{M}}(z,w;\chi^a,\mathcal{A}^b) d(z-w) \label{eq:appendix_c_contour}
\end{equation}
where we have defined $z_0=\frac{1}{2}(z+w)$, and we restrict $d\overline{z}=d\overline{w}=d\overline{z}_0$. We can obtain the contribution of term corresponding to taking the gauge variation of the right external leg by exchanging $J_{i_1}[k] \leftrightarrow \tilde{J}_{i_2}[l]$ before performing the contour integral. \\

We take the external legs to be test functions of the form $\chi = x^0 (v_x^1)^{t^1} (v_x^2)^{t^2} \mathbf{t}_a$ and $\mathcal{A} =  (v_y^1)^{r^1} (v_y^2)^{r^2} d \overline{y}^0 \mathbf{t}_b$. Inserting this into \ref{eq:appendix_c_M} and performing the integrals in \ref{eq:appendix_c_contour}, we find that the total contribution is:\\
\begin{equation} \label{eq:appendix_c_contour_res}
    \bigg(\frac{1}{2 \pi i}\bigg) \underset{\mathcal{C}}{\int} d^2z_0 z_0  \bigg(\frac{1}{16 \pi^2}\bigg) \bigg(\underset{{(t,r)}}{\mathcal{M}}(k,l) K^{i_1 i_2}_{ab}+\underset{{(t,r)}}{\mathcal{M}}(l,k) K^{i_2 i_1}_{ab} \bigg) :J_{i_1}[k] \tilde{J}_{i_2}[l]:
\end{equation}
where $\underset{{(t,r)}}{\mathcal{M}}(k,l)$ is defined as follows:

\begin{equation}
     \underset{{(t,r)}}{\mathcal{M}}(k,l) = \delta(t+r-1-k-l) \overset{(t)}{\sum_{a,b}}  \underset{{(t,r)}}{\mathcal{M}}^{ab}(k,l)  
\end{equation}

\begin{equation}
    \underset{{(t,r)}}{\mathcal{M}}^{ab}(k,l) = \frac{(r^1)_a (r^2)_b (a+b-1)!(t^2+r^2-b)!(1+k^1+k^2)!(1+l^1+l^2-a-b)!}{(r^1+r^2)_{a+b}(1+t^1+t^2+r^1+r^2-a-b)_{1+t^2+r^2-b}}
\end{equation}
\begin{equation}
   \overset{(t)}{\sum_{a,b}} \equiv \sum_{a=1}^{\text{Min}[r^1,l^1+1]} \sum_{b=0}^{\text{Min}[r^2,l^2]} {l^1 \choose {a-1}} {l^2 \choose b} - \sum_{a=0}^{\text{Min}[r^1,l^1]} \sum_{b=1}^{\text{Min}[r^2,l^2+1]} {l^1 \choose {a}} {l^2 \choose b-1}
\end{equation}
\begin{equation}
    (x)_a \equiv \frac{x!}{(x-a)!}
\end{equation}
Note that in \ref{eq:appendix_c_contour_res} we are implicitly assuming that $k$ and $l$ are fixed. To account for all possible choices of $k$ and $l$, we need to sum over all $k$ and $l$ such that $k^i,l^i \geq 0$.  \\

We now demand that this be cancelled by the linearized BRST variation of the right-most diagram. With this choice of external legs, this contribution simplifies to:
\begin{equation}
    -\bigg(\frac{1}{2 \pi i}\bigg) \underset{\mathcal{C}}{\int} d^2z_0 z_0 \underset{z \to w}{\text{Res}}\bigg(J_a[t](z) J_b[r](w)\bigg) 
\end{equation}
We thus find that the one-loop OPE corrections to the $J_a[t]J_b[r]$ OPE that go like $J_{i_1}[k]\tilde{J}_{i_2}[l]$ are given by:
\begin{equation}
    J_a[t](z)J_b[r] \sim \frac{1}{z} \overset{k+l=t+r-1}{\sum_{k^j,l^j \geq 0}} \bigg( \frac{1}{16 \pi ^2} \bigg) \bigg(\underset{{(t,r)}}{\mathcal{M}}(k,l) K^{i_1 i_2}_{ab}+\underset{{(t,r)}}{\mathcal{M}}(l,k) K^{i_2 i_1}_{ab} \bigg) :J_{i_1}[k] \tilde{J}_{i_2}[l]: 
\end{equation}
and we find that $\overset{(1)}{f}$ is:
\begin{equation}
    \underset{(t,r)}{\overset{(1)}{f}}(k,l) =\bigg( \frac{1}{16 \pi ^2} \bigg) \theta(r^i) \theta(t^i) \theta(k^i) \theta(l^i) \underset{{(t,r)}}{\mathcal{M}}(k,l).
\end{equation}
One can readily verify in Mathematica that this expression is equivalent to eq.(\ref{eq:main_m1}).
\bibliographystyle{JHEP}
\bibliography{assoc}
\end{document}